\documentclass[
reprint,
superscriptaddress,
 amsmath,
 amssymb,
 aps,
]{revtex4-2}

\usepackage{graphicx} 
\usepackage{dcolumn} 
\usepackage{bm} 
\usepackage{amsmath} 
\usepackage[dvipsnames]{xcolor} 
\usepackage{dsfont} 
\usepackage{hyperref} 
\usepackage{cleveref}
\usepackage{ulem} 
\crefname{equation}{Eq.}{Eqs.}
\Crefname{equation}{Equation}{Equations}
\crefname{figure}{Fig.}{Figs.}
\Crefname{figure}{Figure}{Figures}
\crefname{section}{Sec.}{Secs.}
\Crefname{section}{Section}{Sections}
\crefname{appendix}{Appendix}{Apps.}
\Crefname{appendix}{Appendix}{Apps.}
\crefname{paragraph}{Sec.}{Secs.}
\crefname{table}{Table}{Tables}

\begin{document}

\title{Extensible circuit-QED architecture via amplitude- and frequency-variable microwaves}

\author{Agustin Di Paolo}
\email{adipaolo@mit.edu}
\affiliation{Research Laboratory of Electronics, Massachusetts Institute of Technology, Cambridge, MA 02139, USA}
\author{Catherine Leroux}
\affiliation{Institut quantique \& D\'epartement de Physique, Universit\'e de Sherbrooke, Qu\'ebec J1K 2R1, Canada}
\author{Thomas M. Hazard}
\affiliation{Lincoln Laboratory, Massachusetts Institute of Technology, Lexington, MA 02421-6426, USA}
\author{Kyle Serniak}
\affiliation{Lincoln Laboratory, Massachusetts Institute of Technology, Lexington, MA 02421-6426, USA}
\author{Simon Gustavsson}
\affiliation{Research Laboratory of Electronics, Massachusetts Institute of Technology, Cambridge, MA 02139, USA}
\author{Alexandre Blais}
\affiliation{Institut quantique \& D\'epartement de Physique, Universit\'e de Sherbrooke, Qu\'ebec J1K 2R1, Canada}
\affiliation{Canadian Institute for Advanced Research, Toronto, Ontario M5G 1M1, Canada}
\author{William D. Oliver}
\email{william.oliver@mit.edu}
\affiliation{Research Laboratory of Electronics, Massachusetts Institute of Technology, Cambridge, MA 02139, USA}
\affiliation{Lincoln Laboratory, Massachusetts Institute of Technology, Lexington, MA 02421-6426, USA}
\affiliation{Department of Physics, Massachusetts Institute of Technology, Cambridge, MA 02139, USA}
\affiliation{Department of Electrical Engineering and Computer Science, Massachusetts Institute of Technology, Cambridge, MA 02139, USA}

\date{\today}

\begin{abstract}

We introduce a circuit-QED architecture combining fixed-frequency qubits and microwave-driven couplers. 
In the appropriate frame, the drive parameters appear as tunable knobs enabling selective two-qubit coupling and coherent-error suppression. 
We moreover introduce a set of controlled-phase gates based on drive-amplitude and drive-frequency modulation. 
We develop a theoretical framework based on Floquet theory to model microwave-activated interactions with time-dependent drive parameters, which we also use for pulse shaping. 
We perform numerical simulations of the gate fidelity for realistic circuit parameters, and discuss the impact of drive-induced decoherence.
We estimate average gate fidelities beyond~99.9\% for all-microwave controlled-phase operations with gate times in the range~$50-120\,\mathrm{ns}$.
These two-qubit gates can operate over a large drive-frequency bandwidth and in a broad range of circuit parameters, thereby improving extensibility.
We address the frequency allocation problem for this architecture using perturbation theory, demonstrating that qubit, coupler and drive frequencies can be chosen such that undesired static and driven interactions remain bounded in a multi-qubit device.
Our numerical methods are useful for describing the time-evolution of driven systems in the adiabatic limit, and are applicable to a wide variety of circuit-QED setups.

\end{abstract}

\maketitle

\section{Introduction}
\label{sec:Introduction}

The field of quantum information processing with superconducting qubits is transitioning from small- to intermediate-scale devices. 
This is in part due to the modularity of circuit QED which has made it possible to extend few-qubit designs to multiple qubits.
However, as the number of qubits increases, average qubit coherence and two-qubit gate fidelities often tend to degrade. 
Among the leading causes are the presence of spurious interactions, spectator-qubit effects, crosstalk, and frequency crowding~\cite{takita2016demonstration,brink2018device,ku2020suppression,krinner2020benchmarking}. 

There exist two broad categories of extensible transmon-qubit-based architectures in development. 
One uses fixed-frequency qubits that are capacitively coupled, and all-microwave gates~\cite{chow2012universal,takita2016demonstration,kandala2021demonstration,chow2013microwave,krinner2020demonstration,mitchell2021hardware,kandala2021demonstration}.
Fixed-frequency layouts help preserve qubit coherence, and direct two-qubit coupling reduces hardware overhead. 
However, the need for maximizing desired gate interactions over undesirable, residual couplings, leads to tight frequency-placement constraints.
These conditions render this architecture prone to frequency collisions and exacerbate the impact of circuit-element disorder~\cite{brink2018device,berke2020transmon}.
To some extent, these issues can be mitigated by improving fabrication targeting and reducing qubit connectivity~\cite{hertzberg2021laser}.

The second approach uses tunable-frequency qubits coupled directly or via a tunable coupler. 
In this case, two-qubit gates are implemented by modulating the qubits and/or the coupler frequencies using baseband flux pulses~\cite{dicarlo2009demonstration,neeley2010generation,chen2014qubit,barends2019diabatic,mckay2016universal,caldwell2018parametrically}.
Such gates are typically faster than all-microwave gates.
Furthermore, tunable qubits and couplers help to reduce crosstalk and frequency crowding, allowing for high connectivity~\cite{arute2019quantum}. 
However, tunability leads to increased footprint, hardware overhead, and extra calibration steps~\cite{rol2020time,arute2019quantum}, in addition to sensitivity to flux noise~\cite{hutchings2017tunable}. 

Here, we theoretically investigate a transmon-qubit-based architecture that combines the advantages of fixed-frequency qubits with microwave-driven tunable couplers.
The parameters of the coupler mode and its potentially always-on microwave drive are chosen to minimize the ZZ interaction while the qubits idle. 
To perform two-qubit gates, the amplitude and frequency of the coupler drive are modulated to enhance the desired two-qubit ZZ interaction for a predetermined amount of time, leading to the accumulation of a conditional phase. 
The drive parameters are changed in time according to a pulse schedule that minimizes leakage by leveraging knowledge of the driven system Hamiltonian. 

While microwave-activated interactions are typically weaker than those implemented by direct two-qubit coupling, our proposed two-qubit gates are fast ($50-120\,\mathrm{ns}$) and have predicted average gate fidelities greater than~$99.9\%$ including dissipation. 
Moreover, because the microwave-activated ZZ interaction is largely tunable over a broad frequency range, it can be used to alleviate frequency crowding and counteract the impact of circuit-element disorder.

To address the problem at hand with generality, we present a comprehensive treatment of driven interactions in circuit QED using two complementary methods for the perturbative and the nonperturbative regimes of the drive amplitude and coupling strengths. Furthermore, we describe the two-qubit gate operation developing a version of Floquet theory where the `slow' time-dynamics of the drive amplitude and frequency can be analyzed independently of the `fast' time-dynamics of the drive phase.

The manuscript is organized as follows. 
In~\cref{sec:Two-qubit-coupler architecture}, we introduce the two-qubit architecture and study the ZZ interaction rate using a diagramatic method, also developed in this work.
We compare the results from perturbation theory obtained for a simplified model of the circuit Hamiltonian against numerical results based on Floquet theory, demonstrating an excellent quantitative agreement between these two techniques and validating our diagramatic approach. 
In~\cref{sec:Adiabatic microwave control}, and inspired by previous works, we develop a version of Floquet theory for the time-evolution operator valid for sufficiently slow changes of the drive parameters with respect to the drive frequency.
Equipped with this framework, we introduce a pulse-shaping strategy incorporating knowledge of the Floquet quasienergy spectrum to prevent nonadiabatic transitions between Floquet states that can cause leakage. 

In~\cref{sec:Controlled-phase gates}, we use Floquet theory to describe the working principles of a number of controlled-phase gates based on drive-amplitude and/or drive-frequency modulation.
We take advantage of our pulse-shaping strategy to derive a convenient parametrization for the two-qubit gate pulses, and perform time-domain simulations with and without dissipation.
We moreover define average-gate-fidelity and leakage metrics in presence of always-on microwave drives, and compute these quantities for our different two-qubit gate implementations. 
We show that the average gate fidelity for controlled-phase rotations based on drive-amplitude and/or frequency modulation can exceed 99.9\% for realistic circuit parameters. 
Finally, in~\cref{sec:Extensibility analysis}, we analyze the extensibility of our architecture to multi-qubit devices, treating the frequency allocation problem with the help of perturbation theory, and discussing microwave multi-qubit control.
We conclude in~\cref{sec:Conclusion}. 

\section{Two-qubit-coupler architecture}
\label{sec:Two-qubit-coupler architecture}

In this section, we introduce our circuit-QED architecture and describe two techniques to compute both spurious and gate interaction rates. 
The first method relies on a perturbative expansion and is useful to understand the low-power behavior of the ZZ coupling. 
The second method uses Floquet theory and is nonperturbative. 
We compare our perturbation-theory estimations against the exact numerical result provided by Floquet theory, finding an excellent agreement. 
Next, we use perturbation theory to understand the dominant processes that explain the ZZ interaction in the presence of a drive. 

\subsection{Circuit Hamiltonian}
\label{subsec:Circuit Hamiltonian}

\begin{figure}[t!]
    \includegraphics[scale=1.]{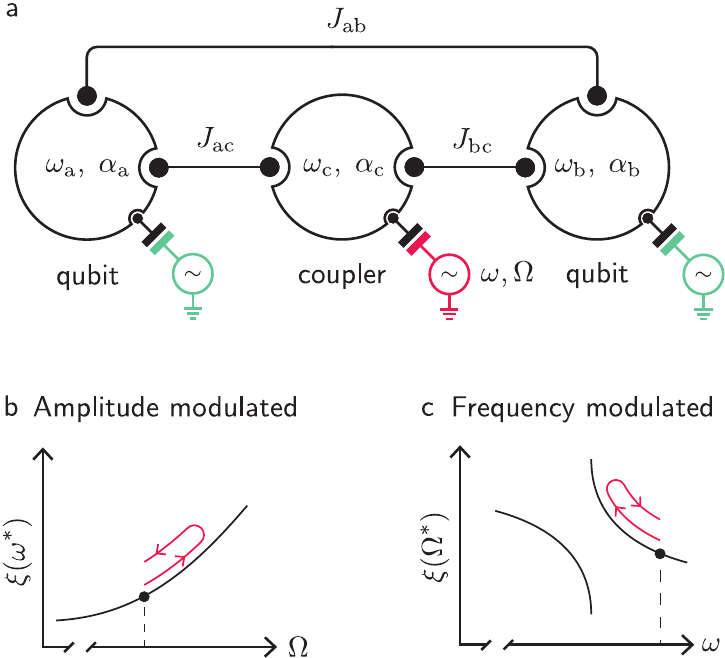}
    \caption{\label{fig:2QB device} 
    \textsf{a} Schematic of the two-qubit-coupler architecture. Each circuit mode is represented by a Kerr-nonlinear oscillator (KNO) of frequency~$\omega_{\mu}$ and anharmonicity~$\alpha_{\mu}$, with~${\mu}=\mathrm{a,b}$ for the qubits and~${\mu}=\mathrm{c}$ for the coupler. $J_{{\mu}{\nu}}$ represents the coupling between modes~${\mu}$ and~${\nu}$. 
    The coupler mode is driven by a microwave voltage source (red) of amplitude~$\Omega$ and frequency~$\omega$. 
    The qubits are individually driven by independent voltage sources (green) to perform single-qubit operations.
    \textsf{b-c} Schematic. Drive-activated ZZ interaction, $\xi$, and controlled-phase gates. The black dot indicates the idle operating point of the system.
    \textsf{b} ZZ interaction as a function of drive amplitude at fixed frequency~$\omega^*$. A controlled-phase gate is performed by modulating the drive amplitude.
    \textsf{c} ZZ interaction as a function of drive frequency for fixed drive amplitude~$\Omega^*$. A multi-photon resonance close to the operating point leads to a discontinuity of the ZZ coupling strength as a function of drive frequency. A controlled-phase gate is performed by modulating the drive frequency.}
    \end{figure}
\Cref{fig:2QB device}\textsf{a} shows a schematic where two qubits~(a and b) and a coupler~(c) are coupled by generic two-body interactions. 
The qubit modes are driven via independent voltage sources (green) used for single-qubit operations. 
The coupler mode is driven by an additional voltage source (red) to enhance and/or suppress the ZZ interaction between the qubits. 

In this work, we focus on two-qubit gates enabled by the coupler drive rather than on drive-activated ZZ cancellation.
Although we develop the two-qubit gate theory with generality concerning the circuit Hamiltonian, our numerical simulations consider an implementation where the qubit modes are fixed-frequency transmons, the coupler is a tunable transmon, and the two-mode couplings are implemented by capacitors. In absence of drives, the circuit Hamiltonian takes the form
\begin{equation}
    \hat{H}_\mathrm{s} = \sum_{{\mu}} 4E_{\mathrm{C}_{\mu}} \hat{n}_{\mu}^2-E_{\mathrm{J}_{\mu}}\cos\hat{\phi}_{\mu} + \sum_{{\mu},{\nu}} \hbar g_{{\mu}{\nu}} \hat{n}_{\mu}\hat{n}_{\nu},
    \label{eq:circuit Hamiltonian}
\end{equation}
where the first (second) sum runs over all (pairs of) circuit modes. 
$E_{\mathrm{C}_{\mu}}$ and~$E_{\mathrm{J}_{\mu}}$ refer to the charging energies and Josephson energies, respectively, of the circuit mode labelled by~${\mu}$, and~$g_{{\mu}{\nu}}$ is the effective capacitive coupling between modes~$({\mu},{\nu})$. 
Since the coupler mode is a tunable transmon, $E_{\mathrm{J}_\mathrm{c}}\to E_{\mathrm{J}_\mathrm{c}}(\Phi_\mathrm{ext})$, where~$\Phi_\mathrm{ext}$ is the external flux threading the coupler's SQUID loop.
The microwave drive applied to the coupler is described by the Hamiltonian
\begin{equation}
    \hat{H}_\mathrm{drive}(t) = 2eV_\mathrm{c}(t)\sin[\omega(t) t + \theta_0]\hat{n}_\mathrm{c} 
    \label{eq:coupler drive Hamiltonian}
\end{equation}
where~$V_\mathrm{c}(t)$ and~$\omega(t)$ are the drive-voltage amplitude and frequency, and~$\theta_0$ is a reference phase. 
We moreover define the drive phase~$\theta(t)=\omega(t) t + \theta_0$ and set~$\theta_0=0$. 
It will become clear below that this simplification does not affect the generality of our results. 

Two-qubit gates are implemented by irradiating the coupler mode with a microwave drive of large amplitude, resulting in a nonzero ZZ interaction.
Similarly, given qubit and coupling parameters, we choose the coupler mode frequency such as to minimize the ZZ interaction between the qubits.
In cases where static ZZ cancellation is not possible, we consider applying an off-resonant coupler drive to counteract the spurious ZZ coupling. 

We consider two possible pulse schedules for the two-qubit gate.
The first one relies on amplitude modulation of a fixed-frequency drive, and is illustrated in~\cref{fig:2QB device}\textsf{b}.
The second leverages the dispersion of the driven ZZ interaction against drive frequency, for fixed drive amplitude (see~\cref{fig:2QB device}\textsf{c}).
Below, we show how these seemingly distinct pulse schedules can be treated using a common formalism where the drive amplitude and frequency are treated similarly.
We also engineer the two-qubit gate controls in a way that can in principle tolerate multiple always-on drives.

\subsection{Stationary two-qubit interaction rates}
\label{subsec:Stationary two-qubit interaction rates}

Before describing the two-qubit gates, we study the amplitude of microwave-activated interactions for constant drive parameters.
This is a necessary step for the engineering of two-qubit gates based on these interactions.

\subsubsection{Perturbation theory}
\label{subsubsec:Perturbation theory}

We use perturbation theory to understand the effects of the microwave drive at low power. 
To this end, we move to a frame rotating at the drive frequency where the problem becomes time-independent under a rotating-wave approximation. 
Introducing the bosonic annihilation operators~$\hat{a},\hat{b},$ and~$\hat{c}$ for the circuit modes~a,b, and c, respectively, the system Hamiltonian is approximated by a Kerr-nonlinear-oscillator model (KNO) taking the form
\begin{equation}
    \begin{split}
    \frac{\hat{H}}{\hbar} &= \sum_{{\mu}} \Delta_{{\mu}}\hat{\mu}^\dagger \hat{\mu} + \frac{\alpha_{{\mu}}}{2}\hat{\mu}^{\dagger 2} \hat{\mu}^2 + \sum_{{\mu},{\nu}} J_{{\mu}{\nu}} (\hat{\mu} \hat{\nu}^\dagger + \hat{\mu}^\dagger \hat{\nu})\\
    & + \frac{\Omega}{2}(\hat{c}+\hat{c}^\dagger),
    \end{split}
    \label{eq:KNO Hamiltonian}
\end{equation}
where~$\hat{\mu},\hat{\nu}\in\{\hat{a},\hat{b},\hat{c}\}$, and the first (second) sum runs over all (pair of) modes. 
In this model, $\Delta_{\mu}=\omega_{\mu}-\omega$ is the detuning between the mode frequency~$\omega_{\mu}$ and the drive frequency, 
$\alpha_{\mu}$ is the mode anharmonicity, 
$J_{{\mu}{\nu}}$ is the two-mode coupling rate, 
and~$\Omega$ is the coupler-drive amplitude. 

We write~\cref{eq:KNO Hamiltonian} as~$\hat{H}=\hat{H}^0+\eta \hat{V}$, where~$\hat{H}^0$ is the noninteracting part and~$\eta \hat{V}$ groups the two-mode interactions and the drive Hamiltonian.
We denote the eigenstates (eigenvalues) of~$\hat{H}^0$ by~$|\Phi^0_\alpha\rangle$ ($\epsilon_\alpha^0$). 
Likewise, we denote the corresponding eigenstates and eigenvalues of~$\hat{H}$ by~$|\Phi_\alpha\rangle$  and~$\epsilon_\alpha$, respectively. 

In~\cref{sec:Perturbation theory calculations}, we introduce a resummation technique to approximate the self-energy~$\Sigma_\alpha=\epsilon_\alpha-\epsilon_\alpha^0$, which is given by a self-consistent infinite series.
Our resummation technique, which we refer to as~SCPT for Self-Consistent Perturbation Theory, enables us to efficiently derive equations for the self-energies of the computational states with bounded order.
The resulting semi-analytical expressions for the computational-state energies are useful to understand the origin of the drive-activated ZZ interactions. 

In addition, the implicit nature of SCPT prevents divergences due to degeneracies of~$\hat{H}^0$ for exact resonance conditions, where finite-order perturbation theory based on unitary generators can diverge.
Such a regularization is critical to predict the ZZ interaction near multi-photon resonances of the form~$|{\epsilon}_\alpha^0-{\epsilon}_\beta^0|\approx m\omega$, with~$m$ an integer.

\subsubsection{Floquet theory for time-periodic driving}
\label{subsubsec:Nonperturbative estimations}

We use Floquet theory to numerically compute the ZZ interaction rate and benchmark our perturbative approach.
For constant drive frequency, the Hamiltonian~$\hat{H}_\mathrm{s} + \hat{H}_\mathrm{drive}(t)$ is invariant under time-translations~$t\to t +T$, where~$T=2\pi/\omega$ is the period of the drive.
As a result, there exist linearly independent solutions to the Schr\"odinger equation of the form~$|\psi_\alpha(t)\rangle = \exp(-i\varepsilon_\alpha t)|u_\alpha(t)\rangle$~\cite{grifoni1998driven}.
Here, $\hbar\varepsilon_\alpha$ and~$|u_\alpha(t)\rangle$ are the \textit{quasienergy} and \textit{Floquet mode} associated with the \textit{Floquet state}~$|\psi_\alpha(t)\rangle$, respectively.
Under driven time-evolution, an initial state~$|\psi(0)\rangle$ propagates as  
\begin{equation}
    |\psi(t)\rangle = \sum_\alpha c_\alpha\exp(-i\varepsilon_\alpha t)|u_\alpha(t)\rangle,
    \label{eq:Floquet time-evolution}
\end{equation}
where~$c_\alpha=\langle\psi(0)|u_\alpha(0)\rangle$.

For a driven qubit, the Floquet modes are more commonly referred to as \textit{dressed eigenstates} of the qubit and the driving field.
In our two-qubit-coupler system, the Floquet modes define a time-dependent computational basis~$\{|u_{\alpha_{ij0}}(t)\rangle\}$, where~$\alpha_{ij0}$ indexes the Floquet mode adiabatically connected to the system eigenstate~$|\Phi_{ij0}\rangle$, which includes qubit-qubit and qubit-coupler couplings~\cite{mundada2020floquet,huang2021engineering}.
Here, $ij$ denotes the two-qubit state and `0' indicates the coupler to be in its ground state.
The ZZ interaction follows from the quasienergies~\cite{petrescu2021accurate}
\begin{equation}
    \xi = \varepsilon_{110} + \varepsilon_{000} - \varepsilon_{100} - \varepsilon_{010}.
    \label{eq:ZZ definition}
\end{equation}
Note that the assumption of a one-to-one mapping between the static and driven computational bases is implicit in this definition. 
This is not the case when the drive frequency is resonant with an energy transition of~$\hat{H}_\mathrm{s}$, and will be revisited below. 
We provide the implementation details of Floquet numerics in~\cref{sec:Adiabatic microwave control}.

\begin{table}[t]
    \begin{ruledtabular}
    \begin{tabular}{ccccccccc}
    $\frac{\omega_\mathrm{a}}{2\pi}$ & $\frac{\omega_\mathrm{b}}{2\pi}$ & $\frac{\omega_\mathrm{c}^*}{2\pi}$ & $\frac{\alpha_\mathrm{a}}{2\pi}$ & $\frac{\alpha_\mathrm{b}}{2\pi}$ & $\frac{\alpha_\mathrm{c}}{2\pi}$ & $\frac{J_\mathrm{ac}}{2\pi}$ & $\frac{J_\mathrm{bc}}{2\pi}$ & $\frac{J_\mathrm{ab}}{2\pi}$ \\
    \\[-0.75em]
    \hline
    \\[-0.9em]
    5.1 & 5.6 & 5.464 & -0.26 & -0.28 & -0.34 & 0.095 & 0.105 & 0.010\\
    \end{tabular}
    \end{ruledtabular}
    \caption{\label{tab:Circuit parameters} Mode and coupling parameters for the Kerr-nonlinear oscillator model. 
    All values are provided in GHz. 
    $\omega_\mathrm{c}^*$ denotes the coupler frequency for which the static ZZ interaction cancels out.}
\end{table}

\subsubsection{Understanding the ZZ interaction}
\label{subsubsec:System parameters and ZZ interaction}

We consider the example parameter set in~\cref{tab:Circuit parameters} for the model in~\cref{eq:KNO Hamiltonian}.
The detuning between the qubits is~$500\,\mathrm{MHz}$ (outside of the straddling regime) and the qubit-coupler coupling strength is~$100\,\mathrm{MHz}$ on average, with a variation of the order of~$10\%$.
We also assume a direct two-qubit coupling of~$10\,\mathrm{MHz}$, representing a spurious interaction.
\begin{figure}[t!]
    \includegraphics[scale=1]{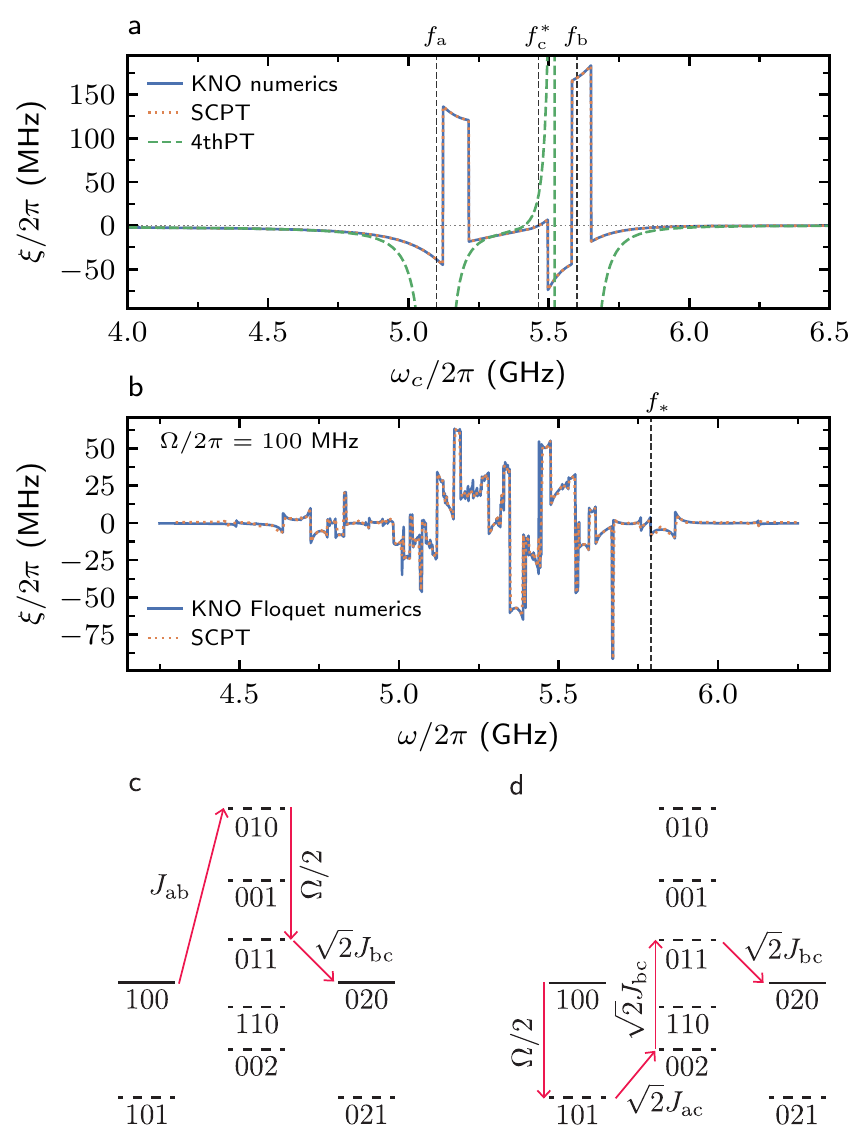}
    \caption{\label{fig:ZZ interaction} Two-qubit ZZ coupling strength. 
            \textsf{a-b} Static ZZ interaction as a function of coupler frequency. 
            \textsf{a} Comparison between numerics (KNO numerics), our perturbation theory (SCPT) and fourth-order perturbation theory (4thPT).
            \textsf{b} Driven ZZ interaction as a function of drive frequency~$\omega/2\pi$ for~$\Omega/2\pi=100\,\mathrm{MHz}$. The coupler frequency is set to~$\omega_\mathrm{c}^*$ in~\cref{tab:Circuit parameters}, for which the static ZZ interaction is zero. 
            We show the comparison between KNO Floquet numerics and SCPT.
            \textsf{c-d} Energy level diagrams to determine the coupling between states~$|\Phi_{100}^0\rangle$ and~$|\Phi_{020}^0\rangle$. The drive frequency is set to~$f_*$, as indicated by the dashed black line in panel~\textsf{b}.
            \textsf{c} Example of a third-order process. 
            \textsf{d} Example of a fourth-order process.
            }
\end{figure}

\textit{Static ZZ interaction--}
\Cref{fig:ZZ interaction}\textsf{a} shows the static ZZ interaction as a function of coupler frequency.
Because of the excitation-number-conserving symmetry of the KNO Hamiltonian, the SCPT result (SCPT) agrees with the numerical result (KNO numerics) up to numerical accuracy.
While this is in itself a remarkable fact, it is also expected, as we explain in~\cref{subsubsec:Estimating the ZZ interaction}.
To highlight the accuracy of SCPT, we contrast the result against standard fourth-order perturbation theory (4thPT).
The latter fails when the qubit-coupler detuning is small compared to the coupling strengths.
In particular, 4thPT fails for coupler frequencies approaching ZZ-cancellation condition for the current parameter set, which we consider below for two-qubit gate simulations.

\textit{Driven ZZ interaction--}
Next, with the coupler frequency set to~$f_\mathrm{c}^*$ (see panel~\textsf{a}), where the static ZZ interaction is zero, \cref{fig:ZZ interaction}\textsf{b} shows the ZZ interaction predicted for the KNO model as a function of drive frequency~$\omega$, for~$\Omega/2\pi=100\,\mathrm{MHz}$.
We compare SCPT against Floquet numerics, observing an excellent agreement between these two methods.
Indeed, perturbation theory not only estimates the ZZ interaction quantitatively, but also correctly captures the drive frequencies at which the ZZ coupling appears discontinuous due to multi-photon transitions.   
Note that the driven ZZ interaction can be nonzero in a large frequency bandwidth for strong drives and coupling strengths~\cite{chow2013microwave}.
This widely tunable ZZ coupling is the basis for our two-qubit gates.

\textit{Understanding the ZZ coupling--}
Given the excellent agreement between our perturbative approach (SCPT) and the numerical results, we use SCPT to gain insights into the origin of the drive-activated ZZ interaction.
Our strategy is to reduce the problem to an effective two-state subspace close to a multi-photon resonance. 
As an example, we consider the drive frequency~$f_*\approx 5.8\,\mathrm{GHz}$ in~\cref{fig:ZZ interaction}\textsf{b}, for which the eigenstates~$|\Phi_{100}\rangle$ and~$|\Phi_{020}\rangle$ of \cref{eq:KNO Hamiltonian} are nearly resonant, and derive an effective coupling strength between the corresponding bare states.
To compute the self-energy, we consider processes which connect~$|\Phi_{100}^0\rangle$ and~$|\Phi_{020}^0\rangle$ up to fourth order, and which involve states in a~$1\,\mathrm{GHz}$ bandwidth centered around~${\epsilon}_{100}^0/h$.
\Cref{fig:ZZ interaction}\textsf{c-d} show two (of eight) processes contributing at third- and fourth-order to the effective coupling~$J^{(4)}$, respectively.
We provide the expression for~$J^{(4)}$ in~\cref{subsubsec:Low-order estimations}.
In the same appendix, we also show that the self-energy of the state~$|\Phi_{100}\rangle$ can be approximated to eighth order in the couplings and the drive amplitude as 
\begin{equation}
    \Sigma_{100}^{(8)}\approx \frac{\Delta}{2}\left(1-\sqrt{1+\left|\frac{2 J^{(4)}}{\Delta}\right|^2}\right),
    \label{eq:self-energy 100 2}
\end{equation}
where~$\Delta={\epsilon}_{100}^0-{\epsilon}^0_{020}-\Lambda_{020}^{(2)}$, and~$\Lambda_{020}^{(2)}$ is a second-order energy shift on the state~$|\Phi_{020}\rangle$ due to the drive.  
In this approximation, the driven ZZ rate can be estimated as~$-\Sigma_{100}^{(8)}$, revealing, for instance, the scaling of this interaction rate with the different system parameters.

\textit{Comparison between the KNO and full-circuit models--}
Finally, \cref{subsec:KNO vs. full-circuit comparison} compares the numerical ZZ interaction predicted for the KNO Hamiltonian against that computed for the full-circuit model.
For such a comparison, the energies entering in~\cref{eq:circuit Hamiltonian} and provided in~\cref{tab:Full-circuit enegry parameters}, are chosen to match mode frequencies and anharmonicities of the KNO Hamiltonian in absence of two-body couplings. 
We find that the KNO model is overall a good approximation to the full-circuit model, and useful to understand static and driven ZZ interactions. 
It also justifies our SCPT approach, which is remarkably accurate in the KNO limit.  

\section{Parametric microwave control}
\label{sec:Adiabatic microwave control}

Equipped with theoretical tools to understand interactions during driven time-evolution, we now turn to the problem of engineering a controlled-phase gate using the driven ZZ interaction.
To make our description general and valid for any drive amplitude, we approach this problem using Floquet theory.

\subsection{Deconstructing a microwave pulse}
\label{subsec:Deconstructing microwave pulse}

We begin by deconstructing a pulse envelope into three sections: rise, hold, and fall.
\begin{figure*}[t!]
    \includegraphics[scale=1]{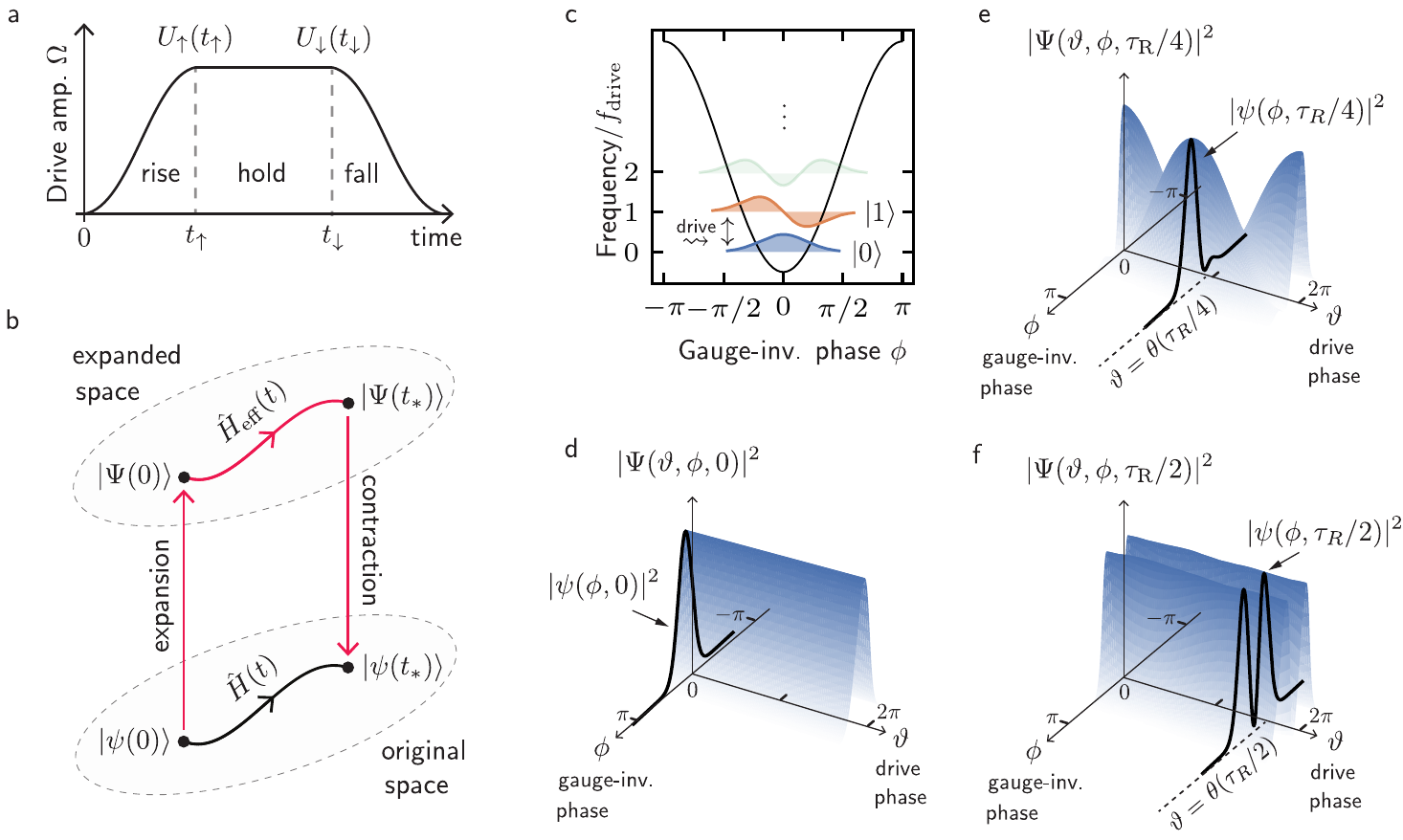}
    \caption{\label{fig:Adiabatic Floquet theory} 
    \textsf{a} Pulse envelope for microwave two-qubit gates, comprised of the three sections: `rise', `hold' and `fall'.
    $U_{\uparrow}(t_\uparrow)$ and~$U_{\downarrow}(t_\downarrow)$ correspond to the unitary operations ideally implemented by the rise and fall sections, respectively.
    \textsf{b} Solving the time-dependent Schr\"odinger equation in an expanded Hilbert space.
    See main text for details.
    \textsf{c-f} Illustration of the expansion and contraction operations, respectively, for a driven transmon qubit.
    \textsf{c} Transmon-qubit wavefunctions as a function of the gauge-invariant phase~$\phi$. The energy spectrum is inscribed in the Josephson potential and shown in units of the drive frequency, which is resonant with the qubit transition frequency.
    \textsf{d} Expansion operation. The initial wavefunction~$\psi(\phi,0)$ corresponding to the transmon in state~$|0\rangle$ (black line) is promoted to the expanded-space wavefunction~$\Psi(\vartheta,\phi,0)$ (entire blue-shaded region).
    \textsf{e-f} Time-evolution and contraction operation. 
    The expanded-space wavefunction (entire blue-shaded region) is calculated using~$\hat{H}_\mathrm{eff}(t)$ in the expanded space at time~$t_*=\tau_R/4$ (panel \textsf{e}) and~$t_*=\tau_R/2$ (panel \textsf{f}), corresponding to a~$\pi/2$ and~$\pi$ pulse, respectively.
    Here, $\tau_R=2\pi/\Omega$ is the Rabi period associated with the Rabi frequency~$\Omega$.
    The solution (black line) in the original space is found by taking the inner product between the expanded-space solution and a~$\hat{\vartheta}$-eigenstate corresponding to the phases~$\vartheta\to\omega t_*$, such that~$\vartheta\to\omega \tau_R/4=\theta(\tau_R/4)$ (panel \textsf{e}) and $\vartheta\to\omega \tau_R/2=\theta(\tau_R/2)$ (panel \textsf{f}), where~$\omega$ is the drive frequency. 
    The phase of the Rabi rotation in general differs from the drive phases~$\theta(\tau_R/4)$ and $\theta(\tau_R/2)$, because the Rabi and the drive frequencies are unrelated.}
    \end{figure*}
First, during the rise section, the system transitions are dressed by increasing the drive amplitude to a maximum value at time~$t_\uparrow$. 
The unitary~$U_\uparrow(t_\uparrow)=\sum_{ij}e^{-i\zeta_{ij}}|u_{\alpha_{ij0}}(t_\uparrow)\rangle\langle \Phi_{ij0}|$, where~$\zeta_{ij}(t_\uparrow)$ is a state-dependent phase, describes the ideal time-evolution during that section up to time~$t_\uparrow$.
Second, for the hold section, the drive remains `on' for a duration~$t_\downarrow-t_\uparrow$.
Time-evolution under the driven Hamiltonian -- in conjunction with small (but non-negligible) contributions from the rise and fall sections of the pulse -- implements the desired gate operation.
Finally, during the fall section of the pulse, the drive amplitude is returned to zero, ideally undressing the system and completing the gate.
The unitary~$U_\downarrow(t_\downarrow)=\sum_{ij}e^{i\zeta_{ij}'}|\Phi_{ij0}\rangle\langle u_{\alpha_{ij0}}(t_\downarrow)|$ describes this final step, where~$\zeta_{ij}'$ is again a state-dependent phase.

The above deconstruction assumes the stabilization of the Floquet modes~$\{|u_{\alpha_{ij0}}(t)\rangle\}$, which rapidly oscillate at the fundamental and integer multiples of the drive frequency.
Tools for quantum optimal control can be used for stabilizing the Floquet modes with high-fidelity.
However, solutions to this control problem are not guaranteed to be well-behaved, due to the time-dependent nature of the Floquet modes during the rise and fall sections of the pulse envelope ($U_{\uparrow,\downarrow}$). 
For instance, due to the rapid oscillations, a timing error~$\delta t$, such that~$t_\uparrow\to t_\uparrow +\delta t$, can lead to a substantial change in the target unitary~$U_\uparrow$, rendering the original control solution ineffective. 
Furthermore, even for a symmetric rise and fall of the pulse envelope, 
due to the time-dependence of the Floquet modes, $U_\uparrow(t_\uparrow)$ and~$U_\downarrow(t_\downarrow)$ are not conjugate operations, as one might expect. %
Rather, $U_\downarrow (t_\downarrow)= [U_\uparrow(t_\uparrow)]^\dagger$ only for times~$t_\downarrow-t_\uparrow$ that are commensurate with the period of the drive.
As we will show, these facts become increasingly important when the qubit modes are subject to always-on microwave drives of large amplitude, where the computational basis is defined by Floquet modes at all times.

\subsection{Adiabaticity and response to slow changes in the drive parameters}
\label{subsec:Response to slow drive-parameter changes}

Central to our control strategy is the concept of adiabaticity in a rotating frame.
Adiabatic time-evolution mitigates the issues associated with the fast-oscillating Floquet modes described in the previous section, and it does so in a way that requires less fine-tuning of the pulse envelope. 
The price to pay for this benefit is a potentially longer gate time. 
Nonetheless, as we shall show, this ``slow-down'' can be largely overcome by designing fast `quasiadiabatic' pulses.  

In this section, we develop an approach that enlarges the system Hamiltonian to an expanded Hilbert space (see~\cref{fig:Adiabatic Floquet theory}\textsf{b}), enabling us to separate fast and slow time dynamics, and thereby address the quasiadiabatic control problem (\cref{subsec:Designing quasiadiabatic microwave pulses}). 

Our approach is inspired by the~$t-t'$ method~\cite{peskin1993solution} and Refs.~\cite{breuer1988role,breuer1988strong,breuer1989adiabatic,breuer1990transport,drese1999floquet,guerin1997complete,weinberg2017adiabatic,hone1997time} but, in contrast to these works, it provides an explicit expression for the propagator. 
This expression enables us to understand the dynamics at all times during the pulse, and it forms the foundation for a practical framework that is used to describe and engineer all-microwave two-qubit gates.

\subsubsection{Expanded Hilbert-space representation}
\label{subsubsec:Extended-Hilbert-space Hamiltonian}

We begin by rewriting the Hamiltonians of~\cref{eq:circuit Hamiltonian} and~\cref{eq:coupler drive Hamiltonian} in the form
\begin{equation}
    \hat{H}(\hat{\boldsymbol{\phi}},\hat{\boldsymbol{n}},t) = \hat{H}_\mathrm{s}(\hat{\boldsymbol{\phi}},\hat{\boldsymbol{n}}) + \hat{H}_\mathrm{drive}[\hat{\boldsymbol{n}},\theta(t),{\Omega}(t)],
    \label{eq:complete Hamiltonian}
\end{equation}
where~$\hat{H}_\mathrm{drive}[\hat{\boldsymbol{n}},\theta(t),{\Omega}(t)]$ is the drive Hamiltonian which depends on the drive phase~$\theta(t)$ -- the fast dynamics -- and the drive amplitude~${\Omega}(t)$ -- the slow dynamics. 
The fast dynamics in~\cref{eq:complete Hamiltonian} is conveniently treated by expanding the original Hilbert space to a larger, fictitious one, which we will refer to as the expanded Hilbert space.

We enlarge the Hamiltonian~$\hat{H}(t)$ to its expanded-space counterpart~$\hat{H}_\mathrm{eff}(t)$ by promoting~$\theta(t)$ to a $2\pi$-periodic quantum degree of freedom~$\hat{\vartheta}$, with conjugate momenta~$\hat{m}\to-i{\partial}_{\vartheta}$.
Accordingly, we promote the initial condition~$|\psi(0)\rangle$ to its expanded-space representation, $|\Psi(0)\rangle$, defined as 
\begin{equation}
    |\Psi(0)\rangle = \frac{1}{2\pi}\int d\vartheta |{\vartheta}\rangle\otimes|\psi(0)\rangle,
    \label{eq:unsqueezed initial condition}
\end{equation}
where~$\hat{\vartheta}|\vartheta\rangle=\vartheta|\vartheta\rangle$ 
and $\langle\vartheta'|\vartheta\rangle=2\pi\delta(\vartheta-\vartheta')$.
This is the expansion step in~\cref{fig:Adiabatic Floquet theory}\textsf{b}.

In the expanded Hilbert space, the system evolves under the effective Hamiltonian (see~\cref{sec:Extended Hilbert space Hamiltonian}) 
\begin{equation}
    \hat{H}_\mathrm{eff}(\hat{{\vartheta}},\hat{\boldsymbol{m}},\hat{\boldsymbol{\phi}},\hat{\boldsymbol{n}},t) = \hat{H}(\hat{\vartheta},\hat{\boldsymbol{\phi}},\hat{\boldsymbol{n}},t) + \hbar
    {\omega}_{\mathrm{eff}}(t)\,\hat{{m}},
    \label{eq:Extended Hilbert space Hamiltonian}
\end{equation}
where~${\omega}_{\mathrm{eff}}(t)\equiv \dot{{\theta}}(t)$. 
The fast dynamics of the original Hamiltonian $\hat{H}(t)$, represented by the drive phase $\theta(t)$, are no longer explicitly present in the expanded Hamiltonian $\hat{H}_{\mathrm{eff}}(t)$. 
Rather, $\hat{H}_{\mathrm{eff}}$(t) only inherits slow time-dynamics via the explicit time-dependence of the drive parameters~${\Omega}(t)$ and~${\omega}_{\mathrm{eff}}(t)$. 
Nonetheless, $\hat{H}_{\mathrm{eff}}(t)$ implicitly accounts for the fast dynamics through the promoted operator $\hat{\vartheta}$.

Finally, by taking the inner product
\begin{equation}
    |\psi({\boldsymbol{\phi}},t_*)\rangle = \langle{\theta}(t_*)|\Psi({{\vartheta}},{\boldsymbol{\phi}},t_*)\rangle,
    \label{eq:wavefunction prescription}
\end{equation}
for a specific value of the drive phase~$\vartheta\to\theta(t_*)$ at time~$t_*$, we recover the solution~$|\psi({\boldsymbol{\phi}},t_*)\rangle$ to the Schr\"odinger equation in the original space associated with~\cref{eq:complete Hamiltonian}. 
This is the contraction step in~\cref{fig:Adiabatic Floquet theory}\textsf{b}.

\Cref{fig:Adiabatic Floquet theory}~\textsf{b} shows a schematic of the procedure for solving the time-dependent Schr\"odinger equation using the expanded-Hilbert-space representation, including the expansion and contraction operations.  
Time-evolution under~\cref{eq:complete Hamiltonian} is equivalent to the combination of operations: expansion, time-evolution under~\cref{eq:Extended Hilbert space Hamiltonian}, and then contraction back to the original space. 

For example, for a single-mode driven system such as a resonantly driven transmon qubit with gauge-invariant phase~$\phi$ (\cref{fig:Adiabatic Floquet theory}\textsf{c}), \cref{fig:Adiabatic Floquet theory}\textsf{d} and \cref{fig:Adiabatic Floquet theory}\textsf{e-f} illustrate the expansion and contraction operations, respectively.
In panel~\textsf{d}, the original-space wavefunction~$\psi(\phi,0)$ of the transmon in its ground state at~$t=0$ (black line) is promoted to the expanded-space wavefunction~$\Psi(\vartheta,\phi,0)$ (entire blue-shaded region). 
Equivalently, the black line represents the contraction of the expanded-space wavefunction back to the original-space wavefunction at~$t=0$.
Panels~\textsf{e} and~\textsf{f} show the contraction operation during time-evolution at times corresponding to a~$\pi/2$ and~$\pi$ pulse, respectively. 
Note that the phase of the Rabi oscillation is unrelated to the drive phase~$\vartheta\to\theta(t)=\omega t \mod 2\pi$ for constant drive frequency.

\subsubsection{Parametric time-evolution}
\label{subsubsec:Parametric time-evolution}

We desire a procedure that will enable us to design pulse envelopes that minimize leakage due to nonadiabatic transitions between Floquet modes as the drive amplitude or frequency is modulated.
To arrive at such a method, we use the expanded-space representation to describe the system dynamics under changes to the drive parameters that are slow compared to the drive frequency.
This is achieved by invoking the adiabatic theorem~\cite{young1970adiabatic,hone1997time} in a series of steps: representing the expanded-space Hamiltonian in a convenient basis, deriving an expression for adiabatic time-evolution in the expanded space, and contracting the solution back to the original space.

\textit{Representing~$\hat{H}_\mathrm{eff}$ in the~$\hat{m}$-basis--}
The first step involves switching to the basis of eigenstates of the~$\hat{m}$ operator. 
Recall that~$\hat{m}$ is conjugate to the phase operator~$\hat{\vartheta}$.
As it will become clear below, the operator~$\hat{m}$ and its eigenvalue~$m$ can also be interpreted as a photon-number operator and a photon number, respectively.
This interpretation will prove useful in understanding single- and multi-photon transitions in our driven system.

To switch bases, we introduce the eigenfunctions~$\langle{\vartheta}|{m}\rangle=e^{i{m}{\vartheta}}$, where~$\hat{{m}}|{m}\rangle={m}|{m}\rangle$ with~$m$ an integer.
We then expand the effective Hamiltonian in terms of the basis functions~$\{e^{i{l}\hat{{\vartheta}}}\}$, with~${l}$ an integer.
Using the orthogonality relation~$\langle {m'}|e^{i{l}\hat{\vartheta}}|{m}\rangle=\delta_{{m'},{l}+{m}}$, where~$\delta_{{i},{j}}$ is the Kronecker delta, we arrive at the expression
\begin{equation}
    \begin{split}
        \hat{H}_\mathrm{eff}(t) &= \sum_{{m}}\Big(\hat{h}_{{0}}[{\Omega}(t)]+ \hbar{\omega}_{\mathrm{eff}}(t)\,{m}\Big)|{m}\rangle\langle {m}|\\
        & + \sum_{{l}\neq{0}}\sum_m \hat{h}_{{l}}[{\Omega}(t)] |{m}+{l}\rangle\langle {m}| + \mathrm{H.c.},
    \end{split}
    \label{eq:Factorized Hamiltonian}
\end{equation}
where~$\hat{h}_{{l}}[\boldsymbol{\Omega}(t)]$ is the coefficient associated with~$e^{i{l}\hat{{\vartheta}}}$.
Note that no approximation has been made to this point. 

\textit{Parametric eigenspectrum--}
We analyze time-evolution under~\cref{eq:Factorized Hamiltonian} in the adiabatic limit using the parametric eigenvalue equation
\begin{equation}
    \hat{H}_\mathrm{eff}(t) |\Psi_\alpha^{{m}}(t)\rangle = \hbar\varepsilon_\alpha^{{m}}(t) |\Psi_\alpha^{{m}}(t)\rangle.
    \label{eq:Parametric eigenvalue equation}
\end{equation}
The set of time-dependent eigenvalues~$\{\varepsilon_\alpha^{{m}}(t)\}$ and eigenstates~$\{|\Psi_\alpha^{{m}}(t)\rangle\}$ are labelled by the double index~$({m},\alpha)$. 
Here, $\alpha$ is an index for the system eigenstates in both the original and expanded spaces, and~$m$ represents the number of photons added or subtracted from the drive field when driving a transition. 

To simplify the analysis below, we briefly review useful properties of the parametric eigenspectrum. 
Note that there exists an infinite number of eigenstates of~\cref{eq:Parametric eigenvalue equation} at time~$t$~\cite{grifoni1998driven}.
Yet, they are generated from only~$N$ distinct eigenstates, where~$N$ is the dimension of the system Hilbert space, as follows.

We begin by taking~$|\Psi_\alpha^{0}(t)\rangle$ to be an eigenstate of~$\hat{H}_\mathrm{eff}(t)$ with quasienergy~$\hbar\varepsilon_\alpha^0(t)$.
The superscript ``0'' indicates that the system eigenstate~$|\Phi_\alpha\rangle$ maps to~$|\Psi_\alpha^0(t)\rangle$ under the condition of zero drive amplitude.
In other words, from an expanded-space perspective, $|\Psi_\alpha^{0}(t)\rangle\to |m=0\rangle|\Phi_\alpha\rangle $ as the drive amplitude~$\Omega$ adiabatically goes to zero.

Next, we introduce the ladder operators~$\hat{m}^-=\sum_{m=-\infty}^{\infty}|m-1\rangle\langle m|$ and~$\hat{m}^+=(\hat{m}^-)^\dagger$.
The corresponding commutation relations~$[\hat{m},\hat{m}^{\pm}]=\pm \hat{m}^{\pm}$ imply that~$\hat{H}_\mathrm{eff}(t)\hat{m}^{\pm}|\Psi_{\alpha}^0(t)\rangle=\hbar[\varepsilon_{\alpha}^0(t)\pm\omega_\mathrm{eff}(t)]\hat{m}^\pm|\Psi_\alpha^0(t)\rangle$.
For a specific value of~$\alpha$, the spectrum of~$\hat{H}_\mathrm{eff}(t)$ is of the form~$\varepsilon_\alpha^{{l}}(t)=\varepsilon^{{0}}_\alpha(t) + {l}\,{\omega}_\mathrm{eff}(t)$, with respective eigenstates~$|\Psi_\alpha^{{l}}(t)\rangle=[\hat{{m}}^{\mathrm{sgn}(l)}]^{{|l|}}|\Psi_\alpha^{{0}}(t)\rangle$. In other words, $|\Psi_\alpha^{{l}}(t)\rangle=(\hat{{m}}^{+})^{{l}}|\Psi_\alpha^{{0}}(t)\rangle$ for~$l>0$, while~$|\Psi_\alpha^{{l}}(t)\rangle=(\hat{{m}}^{-})^{{-l}}|\Psi_\alpha^{{0}}(t)\rangle$ for~$l<0$.

Finally, to obtain the full spectrum, $\alpha$ must span each of its~$N$ possible values corresponding to the dimension of the driven system.
In practice, we truncate the number of basis states~$|m\rangle$ to a reasonable value~$2M+1$ symmetric about $m=0$.
We leverage the structure of the spectrum of~$\hat{H}_\mathrm{eff}(t)$ to target only a few eigenvalues in a frequency range around~$\{\varepsilon_\alpha^{0}(t)\}$ with a sparse eigensolver.
The value of~$M$ is chosen using a convergence check that ensures the eigenvalues of interest are within a desired tolerance. 

\textit{Adiabatic Floquet propagator--}
Using the parametric eigenspectrum in~\cref{eq:Parametric eigenvalue equation}, we now construct the original-space time-evolution operator~$\mathcal{U}(t_*)$ that propagates the state of the system, $|\psi(t_*)\rangle=\mathcal{U}(t_*)|\psi(0)\rangle$.

We begin by enlarging the initial state~$|\psi(0)\rangle$ to~$|\Psi(0)\rangle=|{0}\rangle|\psi(0)\rangle$ using~$(2\pi)^{-1}\smallint d\vartheta|{\vartheta}\rangle=|m={0}\rangle$ in~\cref{eq:unsqueezed initial condition}. 
Next, we calculate the time-evolution in the expanded space, assuming adiabaticity.
According to the standard adiabatic theorem, in the absence of eigenvalue degeneracies and under sufficiently slow changes of the drive parameters~$\boldsymbol{\lambda}(t) = [{\Omega}(t),{\omega}_{\mathrm{eff}}(t)]^T$, 
the time-evolution operator in the expanded space is
\begin{equation}
    \mathcal{U}_\mathrm{eff}(t_*)=\sum_{\alpha,m} e^{-i\int_0^{t_*} \varepsilon_\alpha^{{m}}[\boldsymbol{\lambda}(t)]dt}|\Psi_\alpha^{{m}}[\boldsymbol{\lambda}(t_*)]\rangle\langle \Psi_\alpha^{{m}}[\boldsymbol{\lambda}(0)]|.
    \label{eq:Parametric time evolution}
\end{equation}
We choose a gauge such that the parametric eigenstates~$|\Psi_\alpha^{{m}}[\boldsymbol{\lambda}(t)]\rangle$ satisfy~$\langle \Psi_\alpha^{{m}}[{\lambda}(t)] | \nabla_{{\lambda}}\Psi_\alpha^{{m}}[\boldsymbol{\lambda}(t)]\rangle \cdot\dot{\boldsymbol{\lambda}}(t)=0$.
This choice of gauge accounts for possible geometric phases due to the parametric time-evolution. 

Finally, we contract the expanded-space solution back to the original space using the prescription in~\cref{eq:wavefunction prescription}. 
Using the~$\hat{m}$-basis representation~$\langle{\vartheta}|=\sum_{{m}}e^{i{\vartheta}{m}}\langle{m}|$, we arrive at the propagator
\begin{equation}
    \begin{split}
        {\mathcal{U}}(t_*) = \sum_{\substack{\alpha\\ m,m'}}& \langle{m'}-{m}|\Psi_\alpha^{{0}}[\boldsymbol{\lambda}(t_*)]\rangle\langle\Psi_\alpha^{{0}}[\boldsymbol{\lambda}(0)]|-{m}\rangle \\
        & \times e^{i{\theta}(t_*){m'}}e^{-i\int_0^{t_*}  \varepsilon_\alpha^{{m}}[\boldsymbol{\lambda}(t)]dt},
    \end{split}
    \label{eq:Adiabatic Floquet propagator}
\end{equation}
where we have leveraged properties of the parametric eigenspectrum to simplify the resulting expression.
Note that the eigenstates of~$\hat{H}_\mathrm{eff}(t)$ propagate the system and the Floquet modes at \textit{any time} within one period of the drive, and thus encode the solution of the time-dependent Schr\"odinger equation at \textit{all times}. 

To better understand the meaning of~\cref{eq:Adiabatic Floquet propagator}, let us consider the case of an off-resonant drive with zero drive amplitude at~$t=0$, which is of particular interest in this work.
The eigenstates of the effective Hamiltonian at~$t=0$ are~$|\Psi_\alpha^{{0}}\rangle=|{0}\rangle|\Phi_\alpha\rangle$, where~$\{|\Phi_\alpha\rangle\}$ are the system eigenstates without a drive. 
Accordingly, we have
\begin{equation}
        {\mathcal{U}}(t_*) = \sum_{\alpha} e^{-i\int_0^{t_*} \varepsilon_\alpha^{{0}}[\boldsymbol{\lambda}(t)]dt}|u_\alpha[\boldsymbol{\lambda}(t_*)]\rangle\langle\Phi_\alpha|,
    \label{eq:Adiabatic Floquet propagator - Zero power}
\end{equation}
where we have introduced the Floquet modes, defined as
\begin{equation}
        |u_\alpha[\boldsymbol{\lambda}(t)]\rangle=\sum_m e^{i\theta(t)m}\langle m|\Psi_\alpha^{{0}}[\boldsymbol{\lambda}(t)]\rangle
    \label{eq:Floquet mode}
\end{equation}
\Cref{eq:Adiabatic Floquet propagator - Zero power} implements~$U_\uparrow(t_\uparrow)$ (see~\cref{subsec:Deconstructing microwave pulse} and~\cref{fig:Adiabatic Floquet theory}) for~$t_*=t_\uparrow$, and it does so adiabatically.
Note that the definition of the Floquet mode in~\cref{eq:Floquet mode} is $2\pi$-periodic in the phase~$\theta(t)$, rather than periodic in time.
We do this to accommodate cases for which the instantaneous drive frequency during a pulse is itself time-dependent.
For constant drive frequency~$\omega$, this generalization reduces to the time-periodic Floquet modes introduced in~\cref{subsubsec:Nonperturbative estimations} through the expression~$\theta(t)=\omega t$.

To the best of our knowledge, while the adiabatic limit of Floquet theory has been analyzed in several previous works, an explicit expression for the Floquet propagator has not been provided before.
More importantly, \cref{eq:Adiabatic Floquet propagator} forms the basis of the pulse-engineering strategy that we present below, which leverages both the original- and expanded-space representations of the driven problem.
Finally, we emphasize that~\cref{eq:Adiabatic Floquet propagator} reduces to the correct Floquet propagator for constant drive amplitude and frequency.

\subsection{Designing quasiadiabatic microwave pulses}
\label{subsec:Designing quasiadiabatic microwave pulses}

Using the theory introduced in~\cref{subsec:Response to slow drive-parameter changes}, we now focus on the engineering of adiabatic microwave controls that are also reasonably fast, or quasiadiabatic.
As a first step in this direction, we define a convenient parametrization of the pulse shape. 
We are interested in operating close to the speed limit where nonadiabatic transitions cause leakage errors of the order of $10^{-4}$. 

\subsubsection{Pulse-shape parametrization}
\label{subsubsec:Pulse shape parametrization}
For exact adiabatic time-evolution, the populations of the parametric eigenstates of~$\hat{H}_\mathrm{eff}(t)$ remain constant in time. 
In practice, however, a change in the drive parameters with finite speed leads to nonadiabatic transitions between these eigenstates~\cite{breuer1988strong,drese1999floquet}.

To understand how these transitions impact the dynamics, let us consider the system initialized in the computational eigenstate~$|\psi(0)\rangle=|\Phi_\alpha\rangle$ for~$\Omega(0)=0$.
The expanded-space wavefunction at~$t=0$ is thus~$|\Psi(0)\rangle=|0\rangle|\Phi_{\alpha} \rangle$. 
We assume that the system evolves according to~\cref{eq:Adiabatic Floquet propagator - Zero power} until time~$t_*$, when the nonadiabatic transition~$(0,\alpha)\to(m,\alpha')$ takes place, representing leakage.
Adiabatic evolution follows for~$t > t_*$ and the drive amplitude is returned back to zero at time~$t_\mathrm{g}$, where~$|\Psi_{\alpha'}^{m}(t_\mathrm{g})\rangle\simeq |m\rangle|\Phi_{\alpha'}\rangle$.
As a consequence, the nonadiabatic transition at~$t=t_*$ builds population in the state~$|\Phi_{\alpha'}\rangle$ at~$t=t_\mathrm{g}$, which we assume belongs to the noncomputational subspace.

Realizing adiabatic time-evolution in the expanded space is thus a necessary condition for engineering adiabatic pulses. 
While the drive frequency does not seem to play a role in our analysis, because the minimum energy difference between eigenstates of~$\hat{H}_\mathrm{eff}(t)$ with different $\alpha$-index is upper bounded by~$\hbar\omega_\mathrm{eff}(t)$, realizing adiabatic time-evolution in the expanded Hilbert space is harder for slow drive frequencies. 
However, a rigorous analysis of nonadiabatic transitions in the expanded space is challenging at finite drive frequency~\cite{weinberg2017adiabatic}.

We thus address the problem of pulse shaping in a practical way, by first defining a pulse shape that incorporates the details of the expanded-space Hamiltonian. 
Secondly, we adjust the time-scale of the pulse such that leakage is minimized in time-dependent simulations. 
More precisely, we design the pulse shape using an estimate of the unwanted population~$|c_{\alpha'}^m|^2$ that can result from nonadiabatic transitions of the form~$({l},\alpha)\to({m},\alpha')$ in the expanded Hilbert space. 
First-order time-dependent perturbation theory leads to the expression~\cite{martinez2015fast}
\begin{equation}
    |c_{\alpha'}^{{m}}(t)| \approx \frac{1}{\hbar}\frac{|\langle \Psi_{\alpha'}^{{m}}(\boldsymbol{\lambda})|\nabla_{\boldsymbol{\lambda}}H_{\mathrm{eff}}(\boldsymbol{\lambda})|\Psi_\alpha^{{l}}(\boldsymbol{\lambda})\rangle\cdot\dot{\boldsymbol{\lambda}}|}{[\varepsilon_{\alpha'}^{{m}}(\boldsymbol{\lambda})-\varepsilon_{\alpha}^{{l}}(\boldsymbol{\lambda})]^2},
    \label{eq:Leading order nonadiabatic expression}
\end{equation}
where some of the explicit time-dependence of the r.h.s. has been omitted for clarity.
As expected, nonadiabatic transitions are more likely for densely packed quasienergy spectra and effective Hamiltonians with strong dispersion against the drive parameters~$\boldsymbol{\lambda}$.

This information can be incorporated into the pulse shape by setting~$|c_{\alpha'}^{{m}}(t)|$ to a constant much smaller than unity, and solving for~$\dot{\boldsymbol{\lambda}}(t)$.
This leads to the equation~$\dot{\boldsymbol{\lambda}}(t)=\Lambda(\boldsymbol{\lambda})$, where the function~$\Lambda$ follows from~\cref{eq:Leading order nonadiabatic expression}. 
In this way, we arrive at a `fast-quasiadiabatic' pulse shape where leakage is approximately bounded to a desired tolerance at all times~\cite{martinez2015fast}.
Moreover, the equation for the pulse shape can account for transitions out of a subspace of interest~$\{|\Psi_\alpha^{{l}}\rangle\}$, by ensuring that~$|c_{\alpha'}^{{m}}(t)|$ remains bounded for all initial conditions in~$\{|\Psi_\alpha^{{l}}\rangle\}$~\cite{garcia2020quantum}.
We find that this pulse-shaping strategy works well in all analyzed cases.

\subsubsection{Pulse-shape implementation details}
\label{subsubsec:Implementation details}
We use the fast-quasiadiabatic approach discussed in the previous subsection to calculate a suitable pulse shape for~$U_\uparrow(t_\uparrow)$ in~\cref{fig:Adiabatic Floquet theory}\textsf{a}. 
As~$\Lambda(\boldsymbol{\lambda})$ is in general nonzero at the boundaries~$\boldsymbol{\lambda}_0$ and~$\boldsymbol{\lambda}_1$, which define the range of the drive-parameter modulation, we incorporate a time-dependent filter function~$w_\tau(t)$ such that~$\Lambda(\boldsymbol{\lambda})\to w_\tau(t)\Lambda(\boldsymbol{\lambda})$.
The purpose of~$w_\tau(t)$ is to smooth the pulse envelope at~$t=0$ and~$t=t_\uparrow$, such that its time derivative is continuous at all times. 
In particular, we consider a cosine filter-function of the form~$w_\tau(t)=1$ for~$\tau/2\leq t \leq t_\mathrm{flat}+\tau/2$, while~$2w_\tau(t)=1-\cos(2\pi t/\tau)$ for~$t<\tau/2$ and~$w_\tau(t)=1-\cos[2\pi (t-t_\mathrm{flat})/\tau]$ for~$t>t_\mathrm{flat}+\tau/2$.
Here,~$\tau$ is an additional pulse parameter that can be optimized to minimize leakage, and~$t_\mathrm{flat}$ follows from the gate time as~$t_\mathrm{flat}=t_\mathrm{g}-\tau$. 
Finally, $\dot{\boldsymbol{\lambda}}(t)=w_\tau(t)\Lambda(\boldsymbol{\lambda})$ is normalized and solved such that the boundary conditions~$\boldsymbol{\lambda}(0)=\boldsymbol{\lambda}_0$ and~$\boldsymbol{\lambda}(t_\mathrm{g})=\boldsymbol{\lambda}_1$ are respected.
Drawing inspiration from the GRAPE algorithm, we refer to these waveforms as Locally constraIned MicrowavE (LIME) pulses. 

If necessary, we complement the LIME pulse with a pulse of duration~$t_\downarrow-t_\uparrow$ during which the drive parameters are constant and equal to~$\boldsymbol{\lambda}_1$ (hold section).
Moreover, since we operate in the adiabatic limit, we use the time-reversed version of the LIME pulse to implement~$U_\downarrow\approx (U_\uparrow)^\dagger$~\cite{yatsenko2004pulse}.
In total, the pulse parameters are only a few, including the span~$[\boldsymbol{\lambda}_0,\boldsymbol{\lambda}_1]$, $t_{\uparrow,\downarrow}$ and~$\tau$. 
While the LIME waveform is usually well-behaved, it can significantly change with the system and drive parameters, as it encodes details of the expanded-Hilbert-space Hamiltonian.
Some of the pulse parameters, such as the gate time, can be estimated using the parametric quasifrequency spectrum. 
However, parameters such as~$t_\uparrow$ and $\tau$, which impact the rate of nonadiabatic transitions, are numerically found by running time-dependent simulations that determine what `sufficiently slow' means in practice~\cite{drese1999floquet}. 
We discuss additional implementation details below. 

\section{Controlled-phase gates}
\label{sec:Controlled-phase gates}

In this section, we focus on the implementation of controlled-phase gates based on amplitude and frequency modulation of the coupler drive. 
We describe these gate operations using the tools developed in previous sections.
We use the expanded-space representation to derive expressions for the conditional phase and leakage-cancellation conditions.
We simulate the various two-qubit gates using the full-circuit Hamiltonian including dissipation, and discuss gate fidelity and leakage metrics. 

\subsection{New drive-amplitude adiabatic two-qubit gates}
\label{subsec:Drive-amplitude-adiabatic two-qubit gates}

Here we discuss controlled-phase gates that use drive-amplitude modulation. 
For concreteness, we describe the gate operation assuming that the drive amplitude is zero at the beginning of the pulse~($\Omega_0=0$), and reaches a maximum value~$\Omega_1$.
This assumption does not limit the applicability of our results.

To model these two-qubit gates, we consider the Hamiltonian
\begin{equation}
    \hat{H}(\hat{\boldsymbol{\phi}},\hat{\boldsymbol{n}},t) = \hat{H}_\mathrm{s}(\hat{\boldsymbol{\phi}},\hat{\boldsymbol{n}}) + \hbar\Omega(t)\sin(\omega t)\,\hat{n}_\mathrm{c}/n^\mathrm{zpf}_\mathrm{c},
    \label{eq:Power-adiabatic gate Hamiltonian}
\end{equation}
where~$\Omega(t)$ is the drive amplitude (frequency) applied to the coupler mode and~$\omega$ its frequency.
$n^\mathrm{zpf}_\mathrm{c}$ denotes the magnitude of the zero-point fluctuations of~$\hat{n}_\mathrm{c}$.

\subsubsection{General qualitative picture}
\label{subsubsec:Qualitative picture}

\begin{figure*}[t!]
    \includegraphics[scale=1]{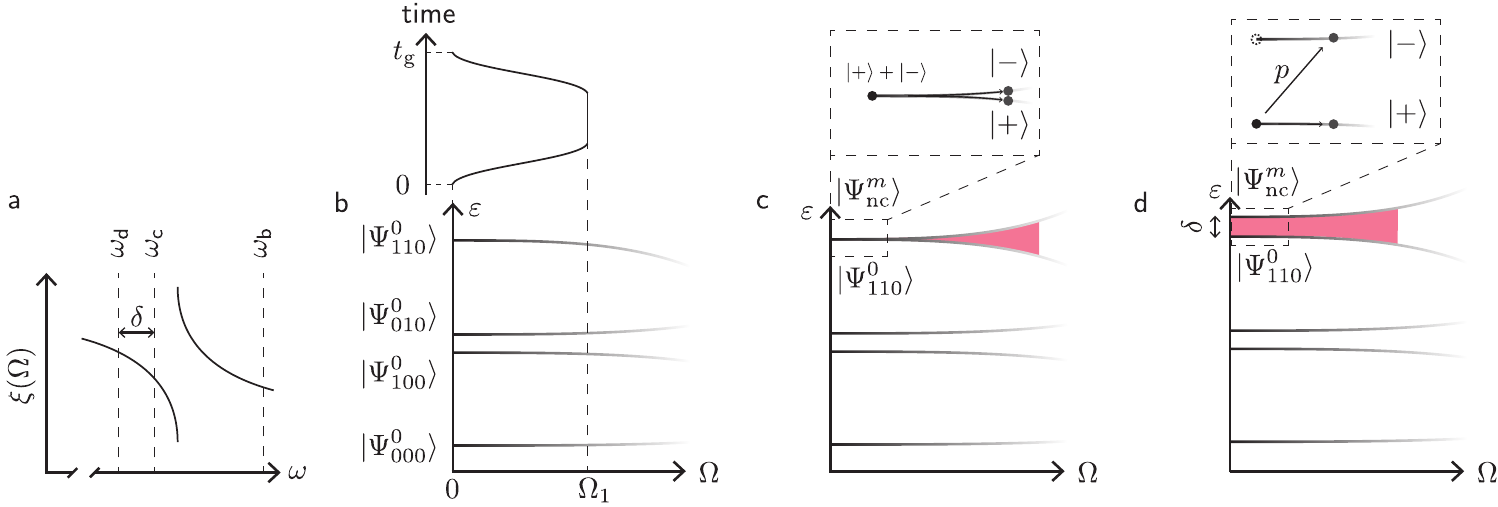}
    \caption{\label{fig:Power-adiabatic gates} Drive-amplitude adiabatic gates. 
            \textsf{a} ZZ interaction, $\xi$, as a function of drive frequency for fixed drive amplitude~$\Omega$. 
            Example frequencies (dashed lines) corresponding to the situations illustrated in panels~\textsf{b}-\textsf{d}. $\delta$ represents a detuning with respect to a resonant condition at~$\omega_\mathsf{c}$ for zero drive amplitude.
            \textsf{b} Off-resonant gate. 
            (Bottom) Logical-subspace quasifrequencies as a function of the instantaneous drive amplitude~$\Omega$. 
            (Top) Drive amplitude as a function of time. 
            \textsf{c} Resonant gate. 
            (Bottom) Level~$|\Phi_{110}\rangle$ is resonant with a noncomputational state,~$|\Phi_\mathrm{nc}\rangle$, by a~$m$-photon transition. 
            (Top) Adiabatic evolution in a two-state subspace~$\{|+\rangle,|-\rangle\}$, which are eigenstates of the expanded-Hilbert-space Hamiltonian. 
            \textsf{d} Nearly resonant gate. 
            (Bottom) Level~$|\Phi_{110}\rangle$ is nearly resonant with~$|\Phi_\mathrm{nc}\rangle$ by a~$m$-photon transition. 
            (Top) Sudden-adiabatic evolution corresponding to the off-resonant drive of an $m$-photon transition between state~$|\Phi_{110}\rangle$ and the noncomputational state~$|\Phi_\mathrm{nc}\rangle$. Here, $p$ schematically represents the probability of a nonadiabatic transition between the respective expanded-Hilbert-space eigenstates.}
\end{figure*}

\Cref{fig:Power-adiabatic gates} shows possible amplitude-modulated gates where the choice of drive frequency leads to qualitatively different dynamics.

\textit{Off-resonant drive-amplitude adiabatic gate--}
Let us first examine the case of a drive with frequency far off-resonant from all multi-photon transitions for~$\Omega(t)\in [\Omega_0,\Omega_1]$ (see~\cref{fig:Power-adiabatic gates}\textsf{a}).
The drive amplitude~$\Omega(t)$ is 0 at~$t=0$, and adiabatically reaches a maximum~$\Omega_1$ for which the ZZ interaction between the two qubits is large in magnitude.
Time-evolution under the strong drive leads to the accumulation of a conditional phase, and the gate is completed by returning the drive amplitude back to zero at~$t=t_\mathrm{g}$.
The pulse schedule is illustrated in~\cref{fig:Power-adiabatic gates}\textsf{b}, where we also show the expanded-space states that are adiabatically connected to the computational levels.

According to~\cref{eq:Adiabatic Floquet propagator - Zero power}, the initial condition~$|\psi(0)\rangle = \sum_{ij} c_{ij}|\Phi_{i j0}\rangle$ evolves to 
\begin{equation}
    |\psi(t_\mathrm{g})\rangle = \sum_{ij} c_{ij}e^{-i\int_0^{t_\mathrm{g}} \varepsilon_{ij0}^0[\Omega(t)]dt}|\Phi_{ij0}\rangle.
    \label{eq:Power-adiabatic gate wavefunction}
\end{equation}
The conditional phase accumulated during the adiabatic trajectory is given by 
\begin{equation}
    \varphi = \int_0^{t_\mathrm{g}} dt\,\xi[\Omega(t)],
    \label{eq:conditional phase offresonant gate}
\end{equation}
leading to a controlled-phase gate for~$\varphi=\pi$, up to zero-duration single-qubit~Z rotations~\cite{mckay2017efficient}.

\textit{Resonant drive-amplitude adiabatic gate--}
Next, let us consider a situation where the drive frequency is resonant with an~$m$-photon transition of the static Hamiltonian (see~\cref{fig:Power-adiabatic gates}\textsf{c}). 
The resonance condition between a given computational state~$|\Phi_{\widetilde{ij}0}\rangle$ and a noncomputational one~$|\Phi_\mathrm{nc}\rangle$ leads to a degeneracy between the expanded-space states~$|\Psi_{\widetilde{ij}0}^0\rangle=|0\rangle|\Phi_{\widetilde{ij}0}\rangle$ and~$|\Psi_\mathrm{nc}^m\rangle=|m\rangle|\Phi_\mathrm{nc}\rangle$ for~$\Omega=0$.
For nonzero drive amplitude, the bonding and antibonding superpositions~$|\pm(\Omega)\rangle = |\Psi_{\widetilde{{ij}}0}^0(\Omega)\rangle \pm |\Psi_\mathrm{nc}^m(\Omega)\rangle)/\sqrt{2}$ diagonalize~$\hat{H}_\mathrm{eff}[\Omega]$.
The initial condition~$|\psi(0)\rangle = \sum_{ij} c_{ij}|\Phi_{i j0}\rangle$ can be written in the expanded Hilbert space as 
\begin{equation}
    |\Psi(0)\rangle = \sum_{ij\neq\widetilde{ij}} c_{ij}|\Psi_{i j0}^0\rangle + \frac{c_{\widetilde{ij}}}{\sqrt{2}} \left[|+\rangle+|-\rangle\right]_{\Omega\to 0}.
    \label{eq:Initial condition resonant gate}
\end{equation}
As the drive amplitude increases, the degeneracy between the states $|+(\Omega)\rangle$ and $|-(\Omega)\rangle$ is lifted~\cite{yatsenko2004pulse}.
Assuming adiabatic evolution at all times, the expanded-Hilbert-space wavefunction at time~$t_\mathrm{g}$ takes the form
\begin{equation}
    \begin{split}
        |\Psi(t_\mathrm{g})\rangle &= \sum_{ij\neq\widetilde{ij}} c_{ij}e^{-i\int_0^{t_\mathrm{g}} \varepsilon_{ij0}^0[\Omega(t)]dt}|\Psi^0_{ij0}\rangle \\
        & + \frac{c_{\widetilde{ij}} e^{-i\frac{\Delta_{+}}{2}}}{\sqrt{2}} \left[e^{-i\frac{\Delta_{-}}{2}}|+\rangle+e^{i\frac{\Delta_{-}}{2}}|-\rangle\right]_{\Omega\to 0},
    \end{split}
    \label{eq:Extended Hilbert space wavefunction resonant gate}
\end{equation}
where
\begin{equation}
        \Delta_{\pm} =\int_0^{t_\mathrm{g}} \{\varepsilon_{+}[\Omega(t)]\pm \varepsilon_{-}[\Omega(t)]\} dt.
    \label{eq:Phases for the resonant gate}
\end{equation}
Here, $\varepsilon_{\pm}[\Omega]$ are the quasifrequencies associated with the eigenstates~$|\pm(\Omega)\rangle$, respectively.

Since at the end of the pulse the population in~$|\Psi^m_\mathrm{nc}\rangle$ must be zero to prevent leakage out of the computational manifold, the interference condition~$\Delta_{-}/2=0\mod 2\pi$ needs to be satisfied. 
In other words, the gate evolution must complete a so-called generalized~$2\pi$-pulse between the computational and noncomputational states~\cite{holthaus1994generalized}, accumulating the conditional phase
\begin{equation}
    \varphi = \sum_{ij\neq \widetilde{ij}} (-1)^{i+j}\int_0^{t_\mathrm{g}} \varepsilon^0_{ij0}[\Omega(t)]dt + (-1)^{\tilde{i}+\tilde{j}} \frac{\Delta_{+}}{2}. 
    \label{eq:Controlled phase resonant gate}
\end{equation}

\textit{Nearly resonant drive-amplitude adiabatic gate--}
Let us assume that the drive frequency is instead nearly resonant with the~$m$-photon transition of the previous example. 
For~$\Omega=0$, the quasifrequencies associated with~$|\Psi_{\widetilde{ij}0}^0\rangle$ and~$|\Psi_{\mathrm{nc}}^m\rangle$ now differ by the detuning~$\delta$.

The dynamics in the expanded space depend on the speed~$\dot{\Omega}$ at which the drive amplitude is modulated.
We focus on the limit~$\dot{\Omega}\gg\delta^2$, where a Landau-Zener-like transition takes place at~$t=0$~\cite{yatsenko2004pulse}.
This process transfers population from the computational to the noncomputational state and introduces nonadiabatic phases. 
The initial population in~$|\Psi_{\widetilde{ij}0}^0\rangle$ splits into two branches defined by the states~$|+(\Omega)\rangle$ and~$|-(\Omega)\rangle$, and we assume is followed by adiabatic time-evolution for larger drive amplitudes.
A second Landau-Zener-like process occurs at the end of the pulse, when the drive amplitude is returned back to zero.

The above description is a simplified interpretation of a rather complex process~\cite{yatsenko2004pulse,zhang2017preparing}. 
However, it allows us to formulate an approximate leakage-cancellation condition by extending the reasoning of the resonant case: the interference condition~$\Delta_{-}/2=0\mod 2\pi$ simply corresponds to a~$2\pi$-off-resonant Rabi oscillation in the two-state manifold~$\{|\Phi_{\widetilde{ij}0}\rangle,|\Phi_{\mathrm{nc}}\rangle\}$.
Furthermore, \cref{eq:Controlled phase resonant gate} remains useful to estimate the total conditional phase accumulated, disregarding nonadiabatic contributions~\cite{breuer1988strong}.  

\subsubsection{Master equation, leakage and fidelity metrics}
\label{subsubsec:Numerical simulations}

Before presenting our numerical simulations, we briefly discuss our device modeling and gate fidelity metrics. 

\textit{Lindblad master equation--}
To model the open-system dynamics, we specify the Lindblad master equation in the device eigenbasis as
\begin{equation}
    \begin{split}
        \dot{\hat{\rho}}(t) &= -i\left[{\hat{H}}(t)/{\hbar}, \hat{\rho}(t)\right] + \sum_{\alpha\neq\beta}\gamma^{1}_{\alpha\beta}\mathcal{D}[\hat{\sigma}_{\beta\alpha}]\hat{\rho}(t) \\
        & + \sum_{\alpha,\beta}\gamma^{\varphi}_{\alpha\beta}\mathcal{D}[\hat{\sigma}_{\alpha\alpha},\hat{\sigma}_{\beta\beta}]\hat{\rho}(t),
    \end{split}    
    \label{eq:Master equation}
\end{equation}
where we have introduced the operators~$\hat{\sigma}_{\alpha\beta}=|\Phi_\alpha\rangle\langle\Phi_\beta|$ and the dissipator superoperator~$\mathcal{D}[\hat{x},\hat{y}]\bullet=\hat{x}\bullet\hat{y}^\dagger-\{\hat{y}^\dagger\hat{x},\bullet\}$ with~$\mathcal{D}[\hat{x}]\bullet=\mathcal{D}[\hat{x},\hat{x}]\bullet$.
The rates~$\gamma^{1}_{\beta \alpha}$ describe multi-level relaxation and excitation processes, whereas pure-dephasing is represented by the rates~$\gamma^{\varphi}_{\alpha\beta}$. 

Dissipation is modeled using a frequency-independent quality factor~$Q$ for capacitive loss, which we assume to be same for the qubits and coupler modes.
Pure-dephasing is modeled using a white-noise approximation to $1/f$ flux noise affecting the coupler flux bias.
While we do not account for additional pure-dephasing channels for the qubit modes, our master equation properly captures flux-noise dephasing of the qubit states due to hybridization with the coupler mode.
Decoherence rates in terms of~$Q$ and the coupler pure-dephasing time~$T_\varphi$ are provided in~\cref{subsec:Lindblad master equation}.
We numerically solve~\cref{eq:Master equation} including 40-50 device eigenstates, using QuTiP's~\texttt{mesolve} function with error tolerances set by convergence checks~\cite{johansson2012qutip}.

\textit{Average gate fidelity and leakage--}
Following Ref.~\cite{wood2018quantification}, we partition the system Hilbert space~$\mathcal{X}$ into two disjoint subspaces~$\mathcal{X}=\mathcal{X}^1_t \oplus\mathcal{X}^2_t$. 
$\mathcal{X}^{1}_t$ is a~$d_1$-dimensional subspace with projector~$\mathds{1}^1_t=\sum_\alpha |u_\alpha(t)\rangle\langle u_\alpha(t)|$, where the sum runs over computational states.
$\mathcal{X}^{2}_t$ is the~$d_2$-dimensional complement of~$\mathcal{X}^1_t$, with associated projector~$\mathds{1}^2_t=\mathds{1}-\mathds{1}^1_t$.
We define the average gate fidelity as
\begin{equation}
    F_\mathrm{avg} = \int d\psi^1_t\langle\psi^1_t|U_{\mathrm{tg}}^{\dagger}\mathcal{E}_{t}\left(\Pi_{t}^{\dagger}|\psi^1_t\rangle\langle\psi^1_t|\Pi_{t}\right)U_{\mathrm{tg}}|\psi^1_t\rangle,
    \label{eq:Average gate fidelity integral}
\end{equation}
where the integral is performed over the Haar measure in~$\mathcal{X}^1_t$.
To accommodate the time-dependence of the computational states, our fidelity definition incorporates the additional operator~$\Pi_{t}$ that maps~$\mathds{1}_0^1$ to~$\mathds{1}_t^1$.
The meaning of~\cref{eq:Average gate fidelity integral} is, however, simple: $F_\mathrm{avg}=1$ if and only if the process~$\mathcal{E}=\mathcal{E}_{U_{\mathrm{tg}}^{\dagger}}\circ\mathcal{E}_{t}\circ\mathcal{E}_{\Pi_t^{\dagger}}$ maps~$\mathds{1}_t^1$ to itself, where~$\mathcal{E}_{U_{\mathrm{tg}}^{\dagger}}$ is the channel associated with the adjoint of the target operation.
The average gate fidelity takes the form
\begin{equation}
    F_\mathrm{avg} = \frac{d_1 F_\mathrm{proc}(\mathcal{E}) + 1 - L_1}{d_1+1},
    \label{eq:Average gate fidelity formula}
\end{equation}
where~$F_\mathrm{proc}(\mathcal{E})$ is the process fidelity associated with~$\mathcal{E}$.
$L_1$ quantifies leakage as the trace of the operator that results from projecting~$\mathcal{E}(\mathds{1}^1_t/d_1)$ on the complement~$\mathcal{X}^2_t$.
Incorporating~$\Pi_{t}$ in~\cref{eq:Average gate fidelity integral} is crucial for properly quantifying the gate fidelity in setups involving always-on drives and within sections of a microwave pulse (see~\cref{subsubsec:Numerical simulations frequency}). 

\subsubsection{Numerical results for selected circuit parameters}
\label{subsubsec:Results drive amplitude}

We now discuss the result of numerical simulations of the various two-qubit gates described in~\cref{subsubsec:Qualitative picture}. 
We demonstrate the proposed gates schemes using transitions that not only involve noncomputational qubit states, but also coupler excitations.

\textit{LIME pulse shape--}
To construct the LIME-pulse equation, we consider the subspace defined by the set of levels~$\{({l},\alpha)\}$ including all computational and noncomputational states that participate in the gate.
The matrix elements in~\cref{eq:Leading order nonadiabatic expression} are calculated considering the full-circuit Hamiltonian in~\cref{eq:Power-adiabatic gate Hamiltonian}.
The corresponding transition operator is~$\partial_{\Omega}\hat{H}_{\mathrm{eff}}/\hbar = \sin\hat{\theta}\,\hat{n}_\mathrm{c}/{n}^\mathrm{zpf}_\mathrm{c}$, where~$\sin\hat{\theta}=\sum_{m}(2i)^{-1}|m+1\rangle\langle m| + \mathrm{H.c.}$
For resonant (nearly resonant) driving, we exclude the matrix element between the states that undergo a (an off-resonant) Rabi rotation, but we account for transitions between these and other states. 

\textit{Off-resonant drive-amplitude adiabatic gate--}
\begin{figure*}[t!]
    \includegraphics[scale=1]{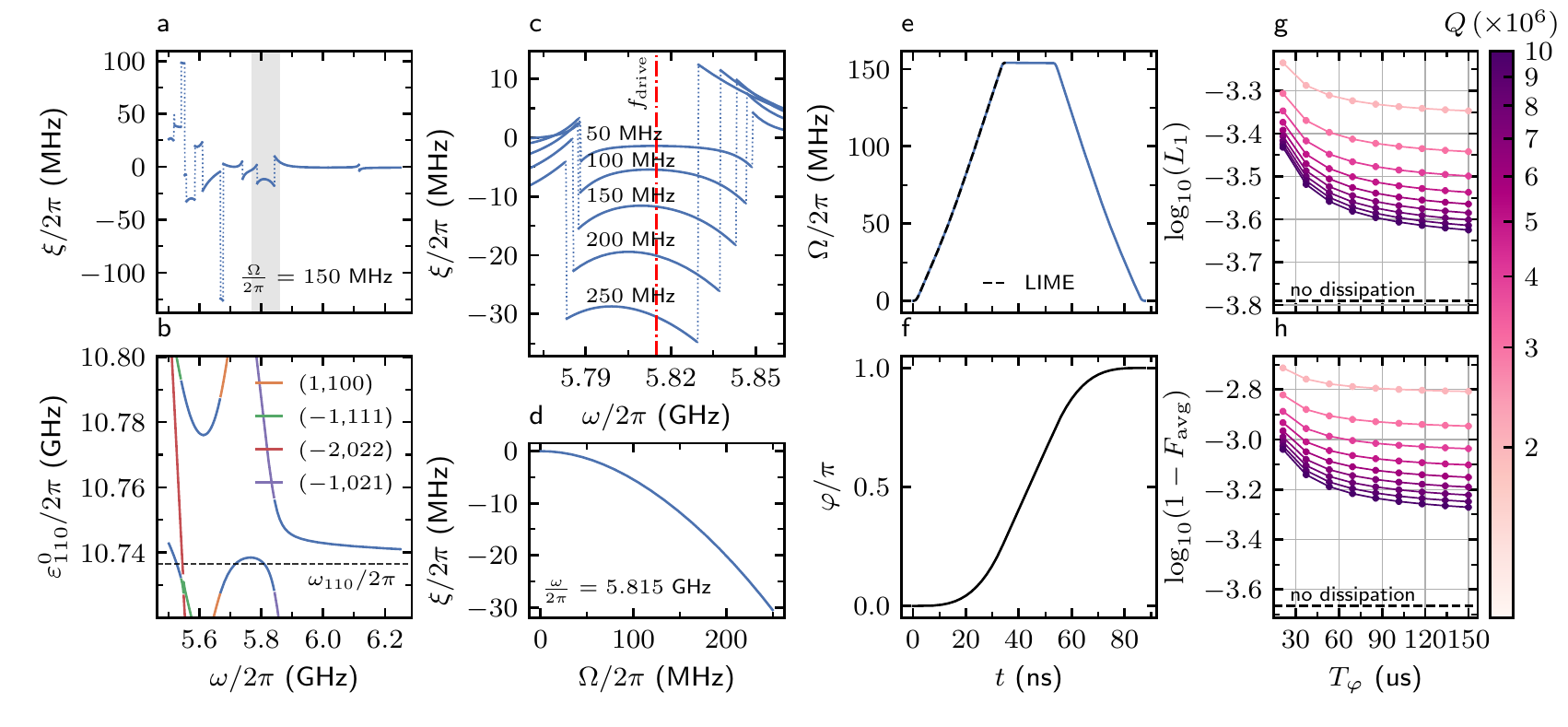}
    \caption{\label{fig:Amplitude adiabatic gate} Drive-amplitude adiabatic resonant and nearly resonant gates. 
            \textsf{a} ZZ interaction, $\zeta$, as a function of drive frequency for~$\Omega/2\pi=150\,\mathrm{MHz}$.
            \textsf{b} Quasifrequency spectrum near the static two-qubit eigenfrequency~$\omega_{110}/2\pi$ (dashed black line). $\varepsilon^0_{110}$ (solid blue line) is the quasifrequency associated with the expanded-Hilbert-space eigenstates adiabatically connected to the state eigenstate~$|\Phi_{110}\rangle$. 
            The label~$(m,ijk)$ incorporates the Floquet photon number~$m$ and the excitations~$(i,j,k)$ of qubit a, b and coupler c, respectively. 
            \textsf{c} ZZ interaction in a narrow frequency bandwidth (grayed area in panel~\textsf{a}) for drive amplitudes in the range~$50-250\,\mathrm{MHz}$.
            \textsf{d} ZZ interaction as a function of drive amplitude for the drive frequency~$f_\mathrm{drive}=5.815\,\mathrm{GHz}$.
            \textsf{e} LIME (dashed black line) and complete (solid blue line) pulse schedules as a function of time. 
            \textsf{f} Accumulated conditional phase as a function of time for the complete two-qubit gate pulse in panel~\textsf{e}.
            This pulse leads to a unitary CZ average-gate-fidelity~$F_\mathrm{avg}=99.98\%$ up to single-qubit Z rotations and leakage~$L_1=0.016\%$.
            \textsf{g-h} Leakage and average-gate-fidelity up to single-qubit Z rotations as a function of coupler pure-dephasing time~$T_\varphi$ and capacitive quality factor~$Q$.
            }
\end{figure*}
\Cref{fig:Amplitude adiabatic gate}\textsf{a} shows the driven ZZ interaction as a function of drive frequency for the drive amplitude~$\Omega/2\pi=150\,\mathrm{MHz}$.
We zoom in on drive frequencies above~$5.5\,\mathrm{GHz}$, but the driven ZZ coupling is nonzero in a larger frequency range of about~$2\,\mathrm{GHz}$ (see also~\cref{fig:ZZ interaction}\textsf{d}).
The structure of the ZZ interaction can be understood in terms of the quasifrequency spectrum~\cite{petrescu2021accurate}, as shown in panel~\textsf{b} for states close in energy to~$|\Phi_{110}\rangle$, with frequency~$\omega_{110}/2\pi$. 
There, the quasifrequency~$\varepsilon_{110}^0$ (solid blue line), associated with the Floquet mode adiabatically connected to~$|\Phi_{110}\rangle$ as~$\Omega\to 0$, appears discontinuous at specific drive frequencies due to anticrossings with noncomputational states in the expanded Hilbert space. 

Next, in panel~\textsf{c}, we examine the ZZ interaction in a narrower frequency range. 
We plot the ZZ interaction as a function of drive amplitude in the range~$0-250\,\mathrm{MHz}$, and focus on the drive frequency~$\omega/2\pi=5.815\,\mathrm{GHz}$. 
The chosen frequency is well off-resonant with respect to multi-photon transitions for all drive amplitudes in the range of interest.  
Panel~\textsf{d} shows the ZZ interaction as a function of drive amplitude for the selected drive frequency. 
The ZZ coupling reaches about~$-30\,\mathrm{MHz}$ as the drive amplitude approaches~$250\,\mathrm{MHz}$, leading to fast and high-fidelity gates. 

We use the parametric quasienergy spectrum to engineer the pulse schedule.
Panel~\textsf{e} shows the resulting LIME pulse shape (dashed black line), which we complete with a pulse of constant amplitude and the time-reverse version of the rise section (solid blue line).
The gate time is chosen such that the accumulated conditional phase is~$\pi$ radians, as shown in panel~\textsf{f}.
The conditional phase is estimated using~\cref{eq:conditional phase offresonant gate}.
We find an excellent agreement between this estimation and the conditional phase obtained in time-domain simulations (not shown).

The unitary average gate fidelity that we obtain in simulation is~$99.98\%$ up to single-qubit Z rotations [see~\cref{eq:Average gate fidelity formula}], and is limited by leakage~$L_1=0.016\%$.
Panels~\textsf{g} and \textsf{h} show~$L_1$ and~$F_\mathrm{avg}$, respectively, as a function of~$T_\varphi$ and~$Q$.
According to our model of dissipation, two-qubit gate fidelities beyond~$99.9\%$ are possible for realistic circuit parameters. 

\begin{figure*}[t!]
    \includegraphics[scale=1]{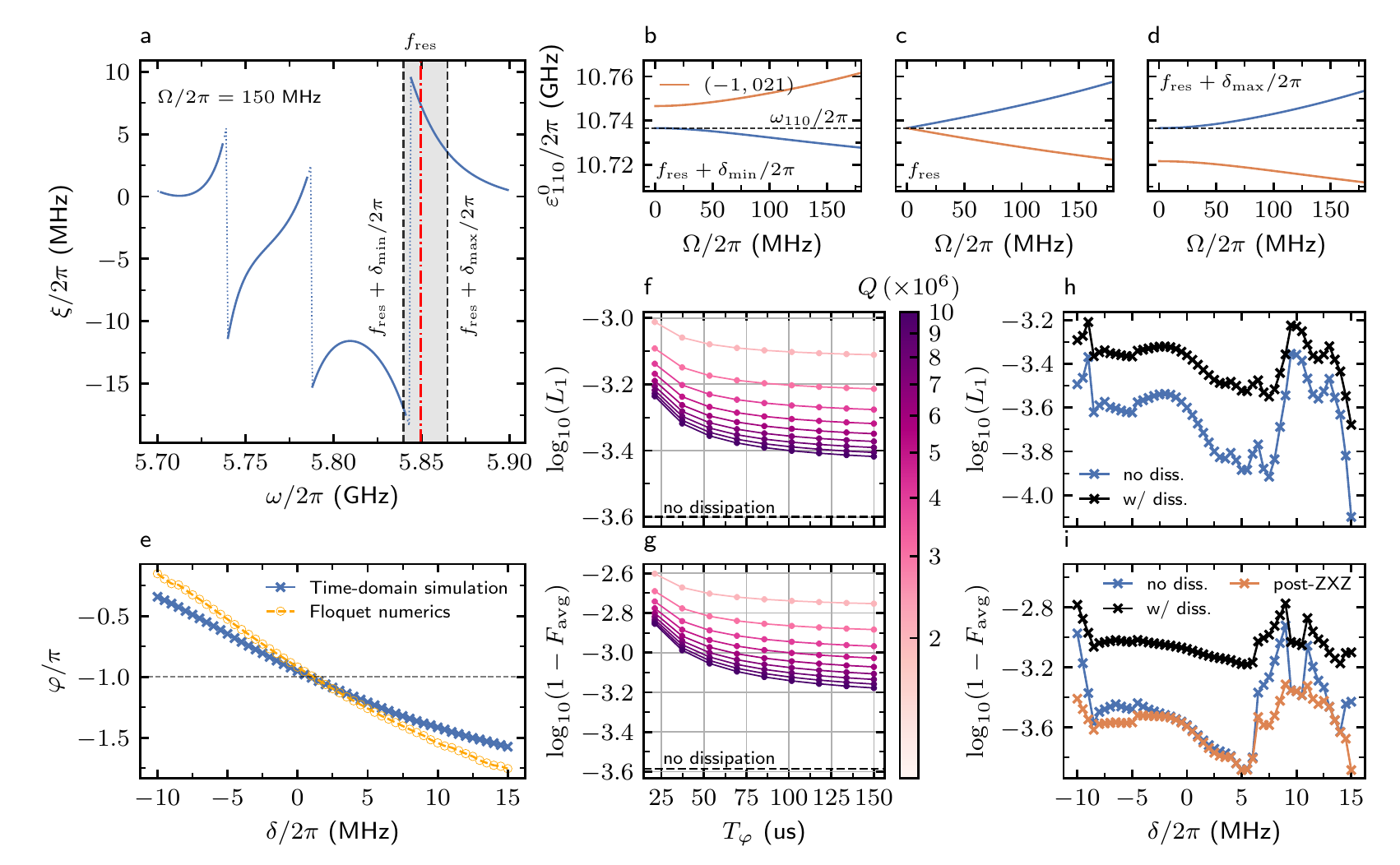}
    \caption{\label{fig:Amplitude adiabatic resonant gate} Drive-amplitude adiabatic resonant and nearly resonant gates. 
            \textsf{a} ZZ interaction as a function of drive frequency.
            The shaded area represents the frequency range in which we study the gate fidelity. 
            $f_\mathrm{res}\approx 5.850\,\mathrm{GHz}$ is the transition frequency between the full-device eigenstates~$|\Phi_{110}\rangle$ and~$|\Phi_{021}\rangle$ at zero drive power.
            \textsf{b-d} Quasienergies of the expanded-Hilbert-space eigenstates that are adiabatically connected to~$|\Phi_{110}\rangle$ and~$|\Phi_{021}\rangle$. 
            The solid blue line corresponds to the energy~$\varepsilon_{110}^0$ of the computational state.
            The frequency~$\omega_{110}/2\pi$ of the computational state~$|\Phi_{110}\rangle$ is shown for comparison (dashed black line). 
            \textsf{b} Red-detuned drive of frequency~$f_\mathrm{res}+\delta_\mathrm{min}/2\pi\approx 5.840\,\mathrm{GHz}$.
            \textsf{c} Exact resonance condition.
            \textsf{d} Blue-detuned drive of frequency~$f_\mathrm{res}+\delta_\mathrm{max}/2\pi\approx 5.865\,\mathrm{GHz}$.
            \textsf{e} Accumulated conditional phase as a function of drive detuning~$\delta$. We compare the result from time-domain simulations against predictions based on Floquet theory.
            The LIME pulse shape used in the two-qubit gate simulations is independently optimized for each drive frequency.
            The total gate time varies in the range~$65-90\,\mathrm{ns}$ as the drive frequency goes from blue- to red-detuned, and is numerically optimized to satisfy the resonant and nearly resonant leakage-cancellation conditions.
            \textsf{f-g} Resonant gate. Leakage and average gate fidelity up to single-qubit Z rotations as a function~$T_\varphi$ and~$Q$.
            \textsf{h-i} Leakage and average gate fidelity as a function of~$\delta$, with and without dissipation. 
            We select~$Q=8\times10^6$ and $T_\varphi=80\,\mu$s for the simulations that include dissipation.
            }
\end{figure*}

\textit{Resonant drive-amplitude adiabatic gate--}
In~\cref{fig:Amplitude adiabatic resonant gate}, we consider the resonant and nearly resonant gates. 
Panel~\textsf{a} shows the ZZ interaction in a narrow frequency range that includes the frequency~$f_\mathrm{res}$ of the~$|\Phi_{110}\rangle\to|\Phi_{021}\rangle$ transition at zero drive power (dash-dotted red line).
We consider drive frequencies in the range~$[f_\mathrm{res}+\delta_\mathrm{min}/2\pi,f_\mathrm{res}+\delta_\mathrm{max}/2\pi]$ (shaded area).
Panels~\textsf{b-d} show the quasifrequencies corresponding to the nearly degenerate states coupled by the drive, as a function of drive amplitude and frequency (see also~\cref{fig:Power-adiabatic gates}\textsf{c-d}). 

We first consider the case~$f_\mathrm{drive}=f_\mathrm{res}$. 
Similarly to the simulation in~\cref{fig:Amplitude adiabatic gate}, we set the maximum drive power to~$\Omega_1/2\pi=150\,\mathrm{MHz}$, and design a LIME pulse such that leakage is minimized and the total conditional phase is~$\varphi=-\pi+\delta\varphi$, with~$|\delta\varphi|\ll \pi$.
[While~$\delta\varphi$ is in general nonzero because of the contribution of the off-resonant computational states in~\cref{eq:Controlled phase resonant gate}, we discuss below how to target~$\delta\varphi\to 0$ by selecting the drive frequency.]
The resulting pulse shape is similar to that shown in~\cref{fig:Amplitude adiabatic gate}\textsf{e}, with a comparable gate time~$t_\mathrm{g}\approx 87\,\mathrm{ns}$.
Because this pulse schedule populates the noncomputational state~$|\Psi^{-1}_{021}\rangle$ in the expanded Hilbert space, the pulse time must be set to restore the initial population to~$|\Psi^{0}_{110}\rangle$ at the end of the gate.
The two-qubit gate implements a total conditional-phase~$\varphi\approx \pi$ (see~\cref{fig:Amplitude adiabatic resonant gate}\textsf{b}) with~$F_\mathrm{avg}=99.97\%$ up to single-qubit Z rotations and leakage~$L_1\approx0.03\%$.
The gate fidelity is computed against an arbitrary-phase gate that best approximates the two-qubit unitary operation. 
Panels~\textsf{f-g} show leakage and average-gate-fidelity, respectively, as a function of~$T_\varphi\in[20,150]\,\mu$s and capacitive quality factor~$Q$.
Due to the stronger coupling to the noncomputational state and a comparable gate time, the impact of dissipation is greater than for the off-resonant gate.
However, average gate fidelities above~$99.9\%$ are still possible for typical circuit parameters, according to our simulations. 

\textit{Nearly resonant drive-amplitude adiabatic gate--}
Next, we investigate the gate operation as a function of detuning~$\delta$.
We use Floquet numerics to determine suitable initial values for the pulse parameters: given~$t_\uparrow$, we estimate the required duration~$t_\downarrow-t_\uparrow$ by integrating the quasienergies such that the zero-leakage condition~$\Delta_{-}/2=0\mod 2\pi$~[see~\cref{eq:Phases for the resonant gate}] is met for each drive frequency. 
We also perform time-domain simulations to adjust~$\tau$, $t_\uparrow$ and the total duration of the pulse, such that leakage is further minimized.
Because the zero-leakage condition depends on~$\delta$, the accumulated conditional phase~$\varphi$ varies with detuning.
\Cref{fig:Amplitude adiabatic resonant gate}~\textsf{e} compares the conditional phase obtained by time-domain simulations (blue symbols) to that predicted by~\cref{eq:Controlled phase resonant gate} (orange symbols).
These two estimations are in agreement for small detunings, but deviations appear for~$|\delta/2\pi|\gtrsim 5\,\mathrm{MHz}$. 
We attribute this discrepancy to the nonadiabatic phases introduced by the off-resonant drive, which are not taken into account in~\cref{eq:Controlled phase resonant gate}. 

Panels~\textsf{h} and~\textsf{i} show, respectively, leakage and average gate fidelity against the arbitrary-phase two-qubit gate that best approximates the process as a function of detuning. 
Blue symbols (no diss.) show the result of unitary time-domain simulations, where the average gate fidelity is estimated up to single-qubit Z rotations. 
In panel~\textsf{i}, orange symbols (post-ZXZ) show the average gate fidelity up to arbitrary single-qubit rotations applied after the two-qubit gate. 
For larger detunings, the two-qubit gate fidelity improves when correcting for arbitrary single-qubit rotations.
We speculate that single-qubit gates account for nonadiabatic deviations with respect to the ideal process map due to the off-resonant drive.
Selecting~$Q=8\times 10^6$ and~$T_\varphi=80\,\mu$s, we investigate leakage and average gate fidelity in the presence of dissipation (black symbols, w/diss.).
We find that the gate fidelity can exceed~$99.9\%$ for realistic circuit parameters for~$|\delta/2\pi|\lesssim 5\,$MHz, while also offering significant tunability of the conditional phase.

The analysis of the various amplitude-adiabatic gates presented in this work is valid for any drive amplitude and frequency.
While the off-resonant gate leverages virtual coupling to noncomputational states, leading to the `bending' of the computational energy levels as a function of drive amplitude, the resonant and nearly resonant gates harness direct coupling to noncomputational levels. 
In near-term devices, the impact of dissipation can in principle be mitigated by leveraging transitions to a noncomputational state with a minimum number of qubit and coupler excitations. 

Finally, we expect our approach to pulse engineering to work complementarily with transitionless-quantum-driving schemes such as `Derivative Removal by Adiabatic Gate' (DRAG)~\cite{motzoi2009simple}, and benefit other types of two-qubit gates.
Controlled-phase gates based on direct coupling~\cite{chow2013microwave,krinner2020demonstration,mitchell2021hardware,kandala2021demonstration,ficheux2021fast}, or via a resonator mode~\cite{paik2016experimental}, are clear choices to investigate next. 
Two-qubit gates based on cross-resonance or parametric interactions are other possible candidates.

\subsection{New drive-frequency adiabatic two-qubit gates}
\label{subsec:Drive-frequency-adiabatic two-qubit gates}

So far, we have thoroughly discussed two-qubit gates that operate at fixed drive frequency and rely on drive-amplitude modulation. 
Now, we turn our attention to a different type of controlled-phase gates that are implemented by modulating the phase of the drive. 
In other words, the drive frequency is chirped. 

Using the tools developed in~\cref{sec:Adiabatic microwave control}, here we introduce the concept of a drive-frequency-variable two-qubit gate, and show that it features very unique and interesting properties. 
These two-qubit gates offer new possibilities for coherent control, and are especially well suited for driven (\textit{i.e.} Floquet) qubits. 

\subsubsection{General qualitative picture}
\label{subsubsec:Qualitative picture frequency adiabatic}

\begin{figure}[t!]
    \includegraphics[scale=1.]{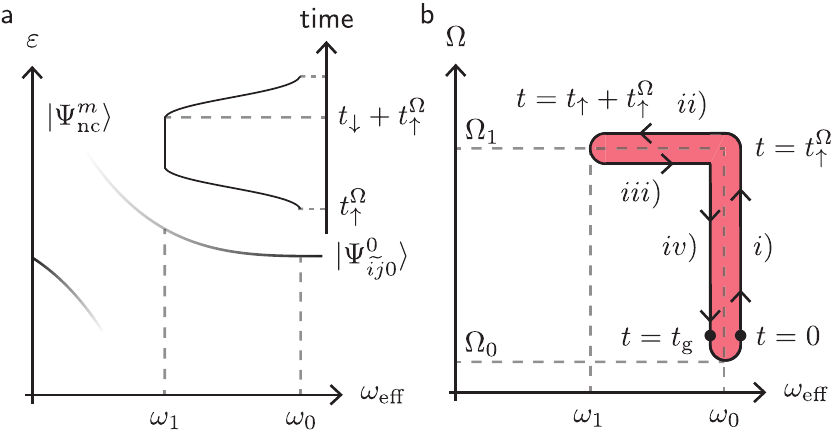}
    \caption{\label{fig:Frequency-adiabatic gates} Frequency-adiabatic two-qubit gate. 
             \textsf{a} Anticrossing between two states~$|\Psi^0_{\widetilde{ij}0}\rangle$ and~$|\Psi^m_\mathrm{nc}\rangle$ of the expanded Hilbert space connected to computational (noncomputational) states of the system, respectively. 
             The inset shows a schematic of the effective drive frequency as a function of time.
             \textsf{b} Complete pulse schedule in the~$\omega_\mathrm{eff}-\Omega$ plane. While our pulse schedule has zero area in parameter space, we schematically separate overlapping trajectories for clarity.
             }
\end{figure}

\textit{Working principle--}
The working principle of our frequency-modulated two-qubit gate is illustrated in~\cref{fig:Frequency-adiabatic gates}\textsf{a}.
There, we show the anticrossing between a pair of states~$|\Psi^0_{\widetilde{ij}0}\rangle$ (computational)  and~$|\Psi^m_\mathrm{nc}\rangle$ (noncomputational) of the expanded Hilbert space, as a function of effective drive frequency~$\omega_\mathrm{eff}$.
Due to the strong coupling between these states, the energy of the computational level shifts as the drive frequency approaches~$\omega_1$ from its initial value~$\omega_0$.
This energy shift of the computational state leads to the accumulation of a conditional phase as the drive frequency is chirped, and it forms the basis of the proposed two-qubit operation.

The accumulation of the conditional phase takes place at constant drive amplitude~$\Omega_1$.
Thus, to perform the gate, the drive amplitude is first adiabatically modified to reach~$\Omega_1$ from an initial value~$\Omega_0$, in a time~$t_\uparrow^\Omega$ [see~\cref{fig:Frequency-adiabatic gates}\textsf{b}, step \textit{i})]. 
The frequency chirp~$\omega_0\to\omega_1$ [step \textit{ii})] follows immediately after step \textit{i}).
A `hold' section (not shown) can follow after step \textit{ii}).
Step~\textit{iii}) serves to restore the drive frequency to its original value~$\omega_1\to\omega_0$. 
Finally, in step \textit{iv}), the drive amplitude is adiabatically returned to its original value~$\Omega_0$.

By engineering a closed trajectory in the~$\Omega-\omega_{\mathrm{eff}}$ plane, the dynamical phases accumulated by the computational states can be adjusted to implement a controlled-phase gate. 
More precisely, the total conditional phase accumulated during the pulse is given by
\begin{equation}
    \varphi = \int_0^{t_\mathrm{g}} \xi[\omega_\mathrm{eff}(t),\Omega(t)]\,dt,
    \label{eq:Conditional phase chirp}
\end{equation}
where the integral is taken along the closed path in~\cref{fig:Frequency-adiabatic gates}\textsf{b}.

\textit{Engineering the pulse schedule--}
We divide the pulse schedule in~\cref{fig:Frequency-adiabatic gates}\textsf{b} in two main parts, corresponding to the drive-amplitude modulation and the frequency chirp.
We design the former with the methods demonstrated in~\cref{subsubsec:Results drive amplitude} (off-resonant case), and use similar tools to engineer the frequency chirp, as follows. 
According to~\cref{eq:Leading order nonadiabatic expression}, the matrix elements of the operator~$\partial_{\lambda}\hat{H}_{\mathrm{eff}}$ for~$\lambda=\omega_\mathrm{eff}$ are needed in this case.
From~\cref{eq:Extended Hilbert space Hamiltonian}, it follows that~$\partial_{\omega_\mathrm{eff}}\hat{H}_{\mathrm{eff}}/\hbar=\hat{m}$. 

We set up the differential equation for~$\omega_\mathrm{eff}(t)$ considering all matrix elements of~$\hat{m}$ between computational and noncomputational states in the expanded space.
We moreover account for the boundary conditions~$\omega_\mathrm{eff}(t_\uparrow^\Omega)=\omega_0$ and~$\omega_\mathrm{eff}(t_\uparrow+t_\uparrow^\Omega)=\omega_1$, where~$t_\uparrow$ sets the time of the frequency chirp~$\omega_0\to\omega_1$ (see~\cref{fig:Frequency-adiabatic gates}\textsf{b}).
The chirp is concatenated with a pulse of constant frequency during which~$\omega_\mathrm{eff}(t)=\omega_1$, and completed by its time-reverse version~$\omega_1\to\omega_0$ which restores the frequency to its original value.
Finally, the pulse schedules for amplitude and frequency modulation are integrated to compose the full two-qubit-gate schedule in~\cref{fig:Frequency-adiabatic gates}\textsf{b}.

\textit{Implementation details--}
For the frequency chirp, we introduce an additional fine-tuning condition that helps mitigating the impact of the dominant nonadiabatic transition as~$\omega_\mathrm{eff}(t)\to\omega_1$.
Denoting~$\Delta$ to be the quasifrequency difference between~$|\Psi^0_{\widetilde{ij}0}\rangle$ and~$|\Psi^m_\mathrm{nc}\rangle$, first-order perturbation theory within a two-level approximation leads to the condition
\begin{equation}
    \int_{t_\uparrow^\Omega}^{t_\uparrow+t_\uparrow^\Omega} \Delta[\omega_\mathrm{eff}(t)]\,dt=0\mod2\pi,
    \label{eq:Chirp interference condition}
\end{equation}
for minimum leakage~\cite{martinez2015fast}.
In practice, given~$\omega_0$, $\omega_1$ and the drive amplitude~$\Omega_1$, we pick~$t_\uparrow$ such that~\cref{eq:Chirp interference condition} is satisfied. 
Using time-domain simulations, we further optimize the value of~$t_\uparrow$ by evaluating leakage out of the computational-state manifold at time~$t_\uparrow+t_\uparrow^\Omega$.
We find that the optimal value for~$t_\uparrow$ is generally close to that predicted by~\cref{eq:Chirp interference condition}. 
Other pulse parameters, such as the time~$\tau$ used in the cosine filter, are optimized together with~$t_\uparrow$. 

Additionally, we calculate the instantaneous drive frequency~$\omega(t)$ that is used in time-domain simulations, by solving
\begin{equation}
    \omega(t) + \dot{\omega}(t)\, t=\omega_\mathrm{eff}(t).
    \label{eq:Effective frequency}
\end{equation}
While seemingly simple, this relation has interesting consequences. 
For instance, let us consider the instant~$t_*=t_\uparrow^\Omega+t_{\uparrow}+t_{\downarrow}$, after which the frequency is no longer modulated. 
For~$t\geq t_*$, we must have~$\omega(t)=\omega_0$. 
However, $\omega(t\to t_*)\neq \omega_0$ for $t\leq t_*$. 
This apparent contradiction implies a discontinuity of the instantaneous drive frequency at time~$t=t_*$, while the drive phase~$\theta(t)=\omega(t)\,t$ remains continuous. 
We return to these details below. 

\subsubsection{Numerical results for selected circuit parameters}
\label{subsubsec:Numerical simulations frequency}

We now discuss the simulation and predicted fidelities of frequency-modulated two-qubit gates. 

\begin{figure*}[t!]
    \includegraphics[scale=1]{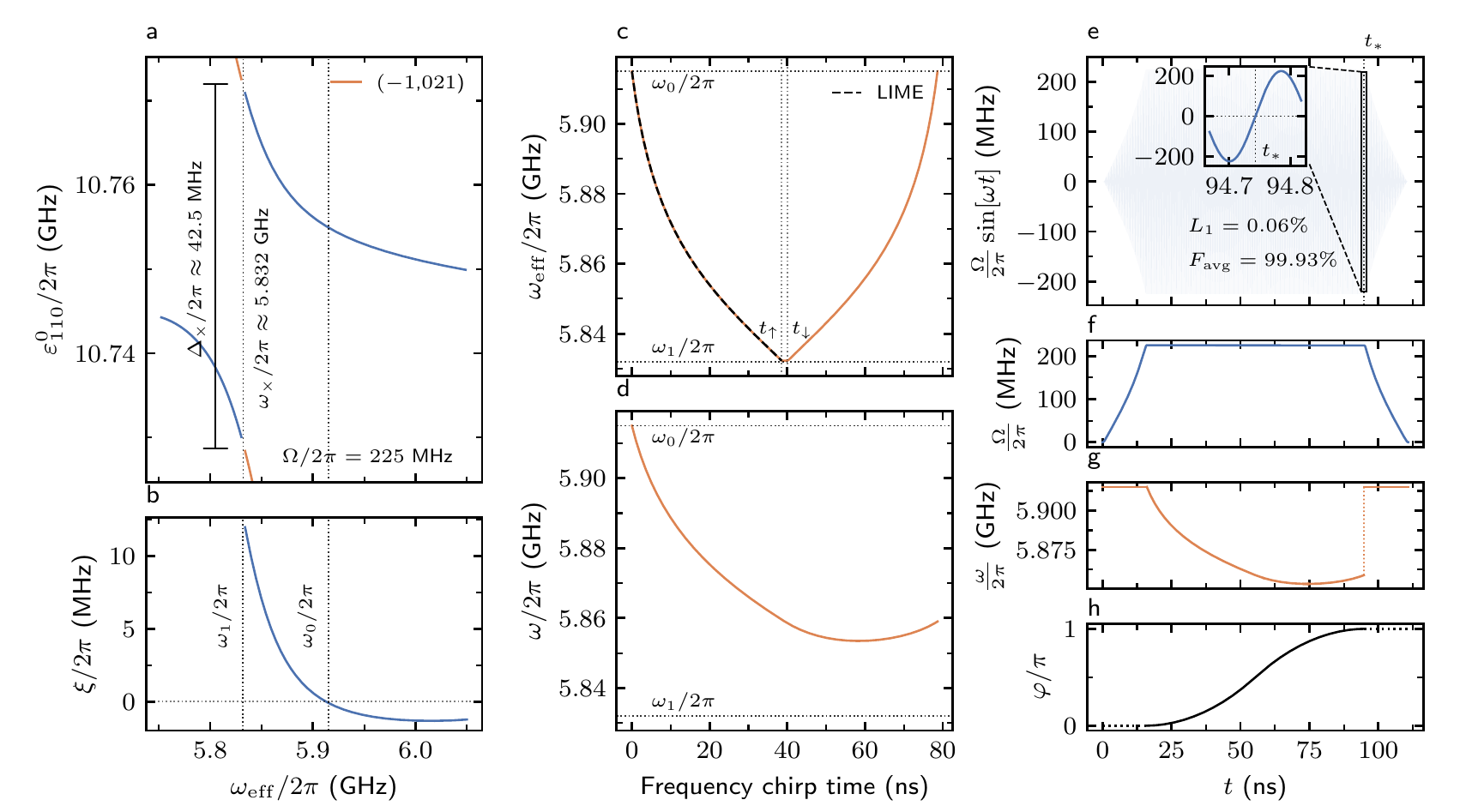}
    \caption{\label{fig:Frequency adiabatic gate} Drive-frequency adiabatic two-qubit gate.
            \textsf{a} Anticrossing between computational~(0,110) (blue) and  noncomputational~(-1,021) (orange) states of the expanded space as a function of drive frequency.
            \textsf{b} ZZ interaction as a function of drive frequency. 
            The frequencies $\omega_0$ and~$\omega_1$ define the boundary conditions for the frequency chirp.
            \textsf{c} Effective frequency pulse. 
            The LIME pulse shape that implements~$\omega_0\to\omega_1$ in a time~$t_\uparrow$ is concatenated with a pulse of duration~$t_\downarrow-t_\uparrow\approx 1.7\,\mathrm{ns}$ and then time-reversed to restore the effective drive frequency to its initial value~$\omega_0$.
            The time~$t_\downarrow-t_\uparrow$ serves as a buffer to adjust the total controlled phase, which in this case is~$\varphi=\pi$.
            \textsf{d} Instantaneous drive frequency derived from~\cref{eq:Effective frequency} for the pulse in panel~\textsf{c}. 
            \textsf{e} Complete (envelope and carrier) frequency-adiabatic pulse schedule as a function of time. 
            We show leakage and average gate fidelity obtained from a unitary time-domain simulation of the two-qubit gate.
            The inset shows a thin slice of the waveform around time~$t_*$, where the instantaneous drive frequency is discontinuous. 
            As pointed out in the main text and shown in the inset, the drive phase is however continuous at~$t=t_*$.
            \textsf{f-g} Drive amplitude and instantaneous drive frequency as a function of time for the complete pulse schedule.
            \textsf{h} Controlled phase accumulated as a function of time. 
            The dotted line represents the phase accumulated at constant frequency~$\omega_0$, which is negligible due to a vanishing small ZZ coupling.
            The full line represents the conditional phase accumulated during the frequency chirp.}
\end{figure*}

\textit{Frequency-modulated two-qubit gates--}
\Cref{fig:Frequency adiabatic gate}~\textsf{a} shows an anticrossing between two expanded-space eigenstates~$|\Psi_{110}^0\rangle$ [labelled by~$(0,110)$] and~$|\Psi_{\mathrm{nc}}^m\rangle$ [labelled by~$(-1,021)$], for the drive amplitude~$\Omega_1/2\pi=225\,\mathrm{MHz}$. 
Recall that the label~$(m,ijk)$ includes the Floquet photon number~$m$ and the excitations~$(i,j,k)$ of qubit a, b and coupler c, respectively.
The states~$(0,110)$ and~$(-1,021)$ are adiabatically connected to computational and noncomputational states of the undriven system, respectively.
The circuit parameters are provided in~\cref{tab:Full-circuit enegry parameters} (`Zero static ZZ') and already used in previous sections.

To design the frequency chirp, we numerically find the exact frequency~$\omega_\times$ where the anticrossing takes place.
Solving~$\partial_{\omega_\mathrm{eff}}\Delta(\omega_\times)=0$, where
\begin{equation}
    \partial_{\omega_\mathrm{eff}}\Delta(\omega_\mathrm{eff})=\langle\Psi_{110}^0|\hat{m}|\Psi_{110}^0\rangle - \langle\Psi_{\mathrm{nc}}^m|\hat{m}|\Psi_{\mathrm{nc}}^m\rangle,
    \label{eq:gap derivative}
\end{equation}
we find~$\omega_\times/2\pi=5.832\,\mathrm{GHz}$ and~$\Delta_\times=\Delta(\omega_\times)\approx 42.5\,\mathrm{MHz}$, as indicated in~\cref{fig:Frequency adiabatic gate}~\textsf{a}.
In addition, panel~\textsf{b} shows the ZZ interaction as a function of effective drive frequency. 
Dashed black lines represent the selected boundary conditions~$\omega_0$ and~$\omega_1$ for the frequency chirp. 
We choose the initial frequency to be~$\omega_0/2\pi=5.915$ ensuring that the ZZ interaction at this frequency does not counteract the conditional phase accumulated during the chirp. 
With our choice, the ZZ coupling is vanishing small at~$\omega_0$ in the full range~$\Omega\in[\Omega_0,\Omega_1]$.
Note that the ZZ interaction at~$\omega_0$ can more generally be used to fine-tune the total conditional phase.
Next, we choose~$\omega_1=\omega_\times$, for which the ZZ interaction reaches about $12\,\mathrm{MHz}$.

We calculate a LIME pulse schedule for the frequency chirp~$\omega_0\to\omega_1$, see panel~\textsf{c} (dashed black line). 
The rise time~$t_\uparrow$ is chosen such that~\cref{eq:Chirp interference condition} is satisfied, and then numerically optimized to minimize leakage outside the computational subspace defined by the drive parameters~$(\Omega_1,\omega_1)$ and~$t_\uparrow$.
We compute the conditional phase~$\varphi_{01}$ accumulated during this pulse by integrating the quasienergies as a function of time.
Then, we estimate the hold time~$t_\downarrow-t_\uparrow$ required to implement a controlled-phase gate, according to the relation~$\xi[\omega_1](t_\downarrow-t_\uparrow) \approx \pi-2\varphi_{01}$.
We optimize the hold time further using time-domain simulations to target~$\varphi=\pi$ accurately, arriving at~$t_\downarrow-t_\uparrow\approx 1.7\,\mathrm{ns}$.
The LIME pulse is then time-reversed and concatenated with the rise and hold sections of the frequency chirp.
The complete pulse schedule for~$\omega_\mathrm{eff}(t)$ is used in~\cref{eq:Effective frequency} to obtain the instantaneous drive frequency~$\omega(t)$, shown in panel~\textsf{d}.
Contrary to~$\omega_\mathrm{eff}(t)$, $\omega(t)$ is not symmetric with respect to~$(t_\uparrow+t_\downarrow)/2$, as one might expect.
This is one of the very unique and interesting characteristics of this gate. 

\Cref{fig:Frequency adiabatic gate}~\textsf{e} shows the complete pulse schedule for the two-qubit gate, comprised of the amplitude-modulated pulse in panel~\textsf{f} and the frequency-modulated pulse in~\textsf{g}. 
The inset shows the waveform around~$t=t_*$, where the instantaneous frequency is discontinuous: while the drive phase remains continuous, its slope as a function of time is different as~$t\to t_*$ for~$t<t_*$ or~$t>t_*$. 
The amplitude-adiabatic waveform in~\textsf{f} is a LIME pulse designed using the methods described in previous sections. 
The time for the process~$\Omega_0\to\Omega_1$ is chosen such that leakage is minimized to a level comparable to that of the frequency chirp.
The total gate time is approximately $120\,\mathrm{ns}$, of which~$80\,\mathrm{ns}$ correspond to the frequency chirp and~$30\,\mathrm{ns}$ correspond to the amplitude-modulated pulse. 

Finally, panel~\textsf{h} shows the conditional phase accumulated as a function of time, as estimated from the quasienergy spectrum. 
The conditional phase accumulated during the frequency chirp is represented by a full line.
The contribution of the amplitude-modulated pulses to~$\varphi$ (dotted line) is negligible due to a vanishing small ZZ coupling strength at~$\omega_0$.
The complete pulse achieves a unitary average gate fidelity of~$99.93\%$ up to single-qubit Z rotations, limited by leakage~$L_1=0.06\%$.
Indeed, we find that the fidelity of frequency-adiabatic two-qubit gates can exceed~$99.9\%$ for realistic circuit parameters. 

\textit{Frequency-modulated control of driven qubits--}
We now demonstrate that frequency chirps are especially useful for engineering two-qubit gates in the presence of always-on drives.
In particular, we focus on the case where a coupler drive is used to suppress spurious two-qubit interactions when the qubits idle. 
More generally, however, our simulations suggest that frequency modulation can be an versatile tool for Floquet-qubit control.

\begin{figure}[t!]
    \includegraphics[scale=1]{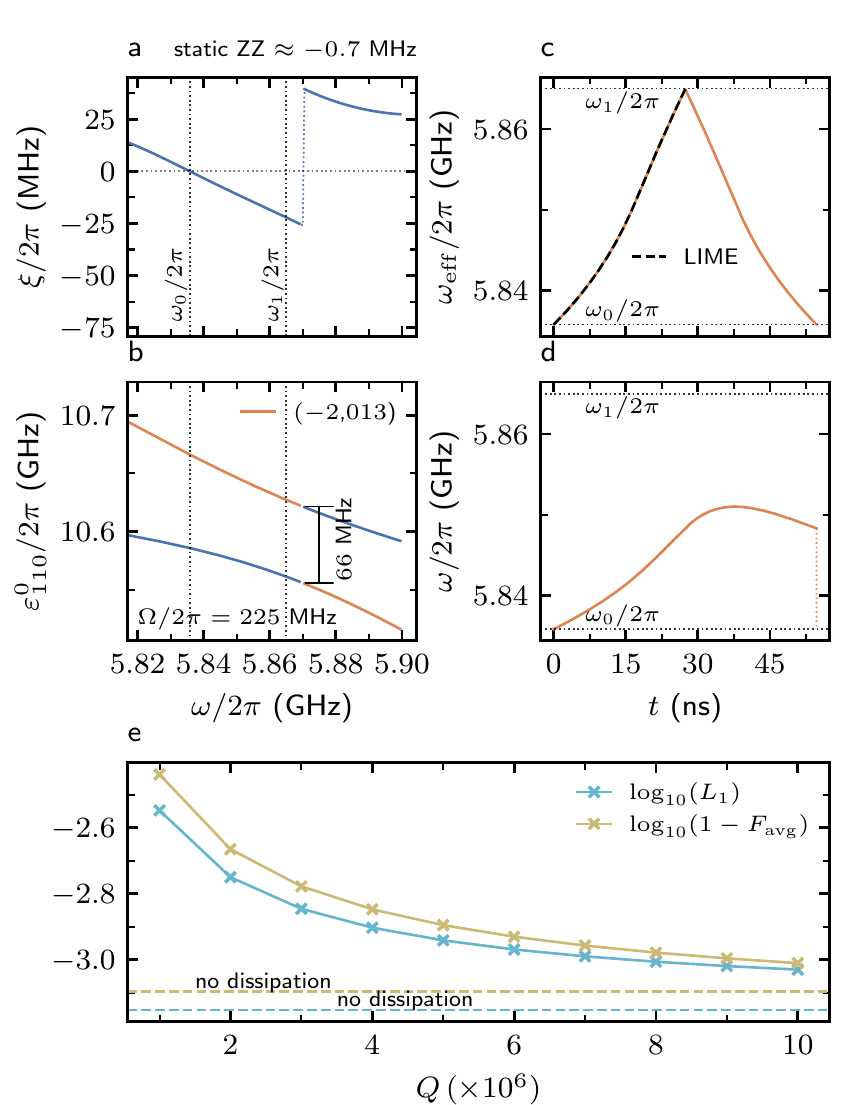}
    \caption{\label{fig:Frequency adiabatic gate 0ZZ} Drive-frequency adiabatic two-qubit gate for Floquet-transmon qubits.
            \textsf{a} ZZ interaction as a function of drive frequency for~$\Omega_0/2\pi=\Omega_1/2\pi=225\,\mathrm{MHz}$. 
            The operating frequency~$\omega_0$ is chosen to counteract the static ZZ coupling of~$-0.7\,\mathrm{MHz}$, such that the total ZZ interaction is zero.
            The frequency chirp spans the range~$[\omega_0,\omega_1]$, where~$\omega_1/2\pi=5.865$ is a frequency for which the ZZ interaction exceeds~$20\,\mathrm{MHz}$ in magnitude.  
            \textsf{a} Anticrossing between computational~(0,110) and  noncomputational~(-2,013) levels of the expanded Hilbert space.
            \textsf{c} LIME pulse for the effective drive frequency.
            \textsf{d} Instantaneous drive frequency corresponding to the pulse in~\textsf{c}.
            \textsf{e} Leakage~$L_1$ (light-blue symbols) and average gate fidelity~$F_\mathrm{avg}$ (gold symbols) as a function of capacitive quality factor~$Q$. 
            The dashed lines represent the respective results in absence of dissipation.}
\end{figure}
To explore this application, we set the coupler frequency to be~$\omega_c/2\pi=6.0\,\mathrm{GHz}$ (see~\cref{fig:ZZ interaction}\textsf{b}), where the static ZZ interaction is approximately~$-0.7\,\mathrm{MHz}$. 
Because such a large spurious interaction would be highly detrimental, we use a microwave drive on the coupler to counteract the static ZZ coupling. 
We find that a drive tone of frequency~$\omega_0/2\pi=5.836\,\mathrm{GHz}$ and amplitude~$\Omega/2\pi=225\,\mathrm{MHz}$ is a possible condition for zero ZZ coupling, as shown in~\cref{fig:Frequency adiabatic gate 0ZZ}\textsf{a}. 
Because this relatively strong drive is always-on, the coupled logical qubits are better thought of as Floquet qubits~\cite{huang2021engineering,mundada2020floquet}.
We note that ZZ cancellation using a microwave-driven coupler has also been considered in Ref.~\cite{ni2021scalable}. 

The ZZ coupling has a strong dispersion against drive frequency.
This is mainly due to the anticrossing between the computational state~$|\Phi_{110}\rangle$ and the noncomputational state~$|\Phi_{013}\rangle$, shown in~\cref{fig:Frequency adiabatic gate 0ZZ}\textsf{b}.
Using~\cref{eq:gap derivative}, we determine the frequency~$\omega_\times/2\pi=5.870\,\mathrm{GHz}$ and the size~$\Delta_\times/2\pi\approx 65.8\,\mathrm{MHz}$ of the anticrossing.

We leverage the dispersion of the ZZ coupling with respect to drive frequency to engineer a fast two-qubit gate using frequency modulation only.
We design the frequency chirp to obey the boundary conditions~$\omega_0$ and~$\omega_1/2\pi=5.865\equiv\omega_{\times} + \delta_{\times}$, for which the ZZ interaction reaches about~$-22.4\,\mathrm{MHz}$.
Here, we incorporate a small detuning~$\delta_\times=5\,\mathrm{MHz}$ with respect to~$\omega_\times$, to be used as an additional parameter to optimize the pulse.
We derive a LIME pulse schedule for the drive-frequency chirp~$\omega_0\to\omega_1$, which we concatenate with its time-reverse version to obtain the complete~$\omega_\mathrm{eff}(t)$ pulse shown in~\cref{fig:Frequency adiabatic gate 0ZZ}\textsf{c}.  
We use~$\delta_\times$ to adjust the total conditional phase accumulated during the pulse to~$\varphi=\pi$, while ensuring the leakage-cancellation condition in~\cref{eq:Chirp interference condition}.
We show the instantaneous drive frequency according to~\cref{eq:Effective frequency} in panel~\textsf{d}.
Thanks in part to the large ZZ interaction, and the fact that no amplitude modulation is needed in this case, the resulting LIME pulse schedule is much shorter than the one shown in~\cref{fig:Frequency adiabatic gate}. 

Next, we perform time-domain simulations of the pulse schedule, finding a unitary average-gate-fidelity of~$F_\mathrm{avg}=99.92\%$, limited by leakage~$L_1=0.07\%$. 
Because the coupler frequency is no longer fine-tuned to achieve zero static ZZ, a frequency-tunable transmon coupler is no longer necessary.
For this reason, we now assume that the coupler mode is a~$T_1$-limited fixed-frequency transmon.

\Cref{fig:Frequency adiabatic gate}\textsf{e} shows the leakage and the average gate fidelity as a function of capacitive quality factor~$Q$.
According to our Lindblad master-equation simulations, the gate fidelity reaches~$99.9\%$ for $T_1$-times in the range $200-300\,\mu$s (see~\cref{subsubsec:Modeling dielectric loss}), which are several times longer than the gate time of approximately $55\,\mathrm{ns}$.
This is because the two-qubit gate leverages a two-photon transition between a computational state and the higher excited state~$|\Phi_{013}\rangle$, that involves the third level of the coupler. 
Limitations to the gate fidelity that arise from dissipation can be mitigated choosing other possible transitions to noncomputational states.  
Moreover, the operating condition~$(\Omega_0,\omega_0)$ should ideally take into account the impact of dissipation in the presence of always-on drives, and leverage sweet spots in drive amplitude and frequency, when possible~\cite{didier2019ac,huang2021engineering}.

\textit{Additional remarks--}
We conclude this section by discussing some of the technical and implementation details of frequency-modulated pulses.

Due to spurious~$m$-photon transitions with~$m\gg 1$, frequency chirps spanning a large frequency bandwidth can be challenged by the presence of multiple anticrossings in the range~$[\omega_0,\omega_1]$~\cite{breuer1988role}.
However, because the effective coupling between computational and noncomputational levels quickly decreases as $m$ increases, typical pulse times in the order of 10s of nanoseconds result in largely diabatic transitions across spurious anticrossings.
Indeed, even in such cases, we find that average gate fidelities beyond~$99.9\%$ are still possible, and conclude that these spurious interactions do not significantly impact gate performance in practice.
Ultimately, however, the presence of spurious resonances can be taken into account when choosing the operating frequency and the boundary conditions for the frequency chirp.

In addition, we briefly discuss some of the experimental implementation details of frequency chirps.
Because the frequency modulation in~\cref{fig:Frequency adiabatic gate,fig:Frequency adiabatic gate 0ZZ} is only a few~10's of MHz, these pulses are straightforwardly realizable using single-sideband mixing with a~$1\,\mathrm{GHz}$-bandwidth arbitrary waveform generator (AWG) modulating a microwave source.
This is the current approach to microwave electronics for superconducting-qubit control.
However, direct-digital synthesis using higher-bandwidth AWG's would be a better solution, allowing us to digitally specify the pulse parameters without the need for analog sideband mixing. 
Certain aspects of implementing frequency chirps in circuit QED have been discussed in the context of fundamental studies of dynamic phase-locking~\cite{naaman2008phase,murch2011quantum,murch2012quantum,shalibo2012quantum}, quantum simulation~\cite{salis2020time} and the realization of a quantum perception~\cite{pechal2021direct}.

It is worth noticing that the proposed swept-frequency adiabatic gates can be viewed as the driven counterpart of conventional two-qubit gates based on adiabatic swept-flux control~\cite{dicarlo2009demonstration}.
In other words, frequency chirps are the baseband flux-control analogue for Floquet qubits.
Extensions of the proposed two-qubit gate leveraging simultaneous drive amplitude and frequency modulation and nonzero geometric phases are also possible.
We speculate that multi-photon processes and frequency chirps could become an increasingly useful tool to mitigate frequency crowding and coherent errors in future quantum processors with steadily increasing qubit coherence~\cite{kjaergaard2020superconducting,gyenis2021moving}.

\section{Extensibility analysis}
\label{sec:Extensibility analysis}

Finally, we address the extensibility of the proposed two-qubit interactions and controls in a multi-qubit setup.
In particular, can our circuit-QED architecture accommodate a large number of qubits without sacrificing average two-qubit gate fidelity?
We now briefly investigate this question, dividing the subject into two parts. 

First, in~\cref{subsec:Extensible frequency layouts}, we consider frequency allocation within our architecture.
Our approach seeks to determine the frequencies of qubit modes, coupler modes, and driving fields in a manner that suppresses coherent errors and avoids leakage in multi-qubit lattices.
Second, in~\cref{subsec:Adiabatic microwave multi-qubit control}, we qualitatively discuss quasiadiabatic multi-qubit microwave control. 

\subsection{Frequency allocation}
\label{subsec:Extensible frequency layouts}

Frequency allocation in a multi-qubit chip with driven interactions has been theoretically studied for the crossresonance~\cite{brink2018device,hertzberg2021laser,malekakhlagh2020first} and the controlled-phase~\cite{morvan2021optimizing} gates based on direct capacitive coupling. 
These works use a predetermined list of frequency collisions and an optimizer to distribute qubit frequencies such that unwanted resonances are avoided to a desired tolerance. 
These requirements in conjunction with limited control bandwidth rarely lead to optimal solutions that can satisfy all constraints. 
This is due in part to the competing requirements for fast two-qubit gates and reduced leakage. 

Using perturbation theory, we develop an alternative approach that systematically accounts for similar frequency constraints, with and without a microwave drive.
We take advantage of the approximate decoupling between static and driven interactions in the proposed architecture, and optimize the parameters using a two-step process.
The frequencies of the qubit and coupler modes are chosen first to minimize undesired static ZZ couplings.  
The drive frequencies for the two-qubit gates are selected in a second step, maximizing desired gate interactions over undesired drive-activated spectator-qubit effects.

\begin{figure*}[hbt!]
    \includegraphics[scale=1]{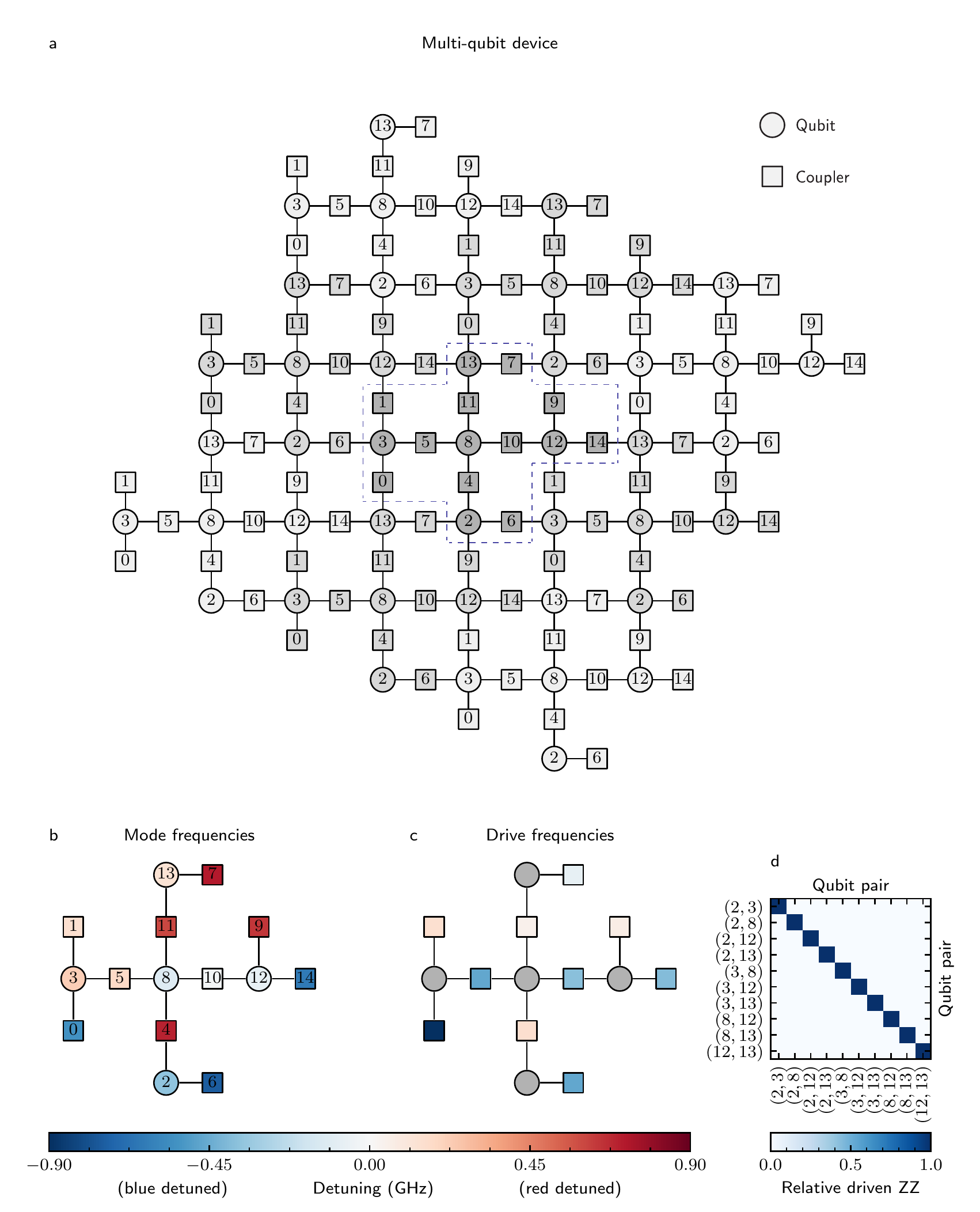}
    \caption{\label{fig:extensibility} Frequency allocation. See~\cref{subsubsec:Multi-qubit setup} for a complete list of circuit parameters.
            \textsf{a} Device model. 
            The unit cell ($d_\mathrm{UC}=3$) is enclosed by dashed lines.
            \textsf{b} Mode frequencies (w.r.t. a reference) minimizing the static ZZ interaction between all pair of qubits.
            \textsf{c} Coupler-drive frequencies (w.r.t. the respective mode frequency) minimizing spectator-qubit effects.
            \textsf{d} Relative two-qubit gate (diagonal) and spurious (off-diagonal) drive-activated ZZ interactions for all pair of qubits.}
\end{figure*}

To simplify the problem, we define a unit cell with a fixed number of qubits and couplers that can be tiled to realize a larger-scale quantum processor. 
An example is shown in~\cref{fig:extensibility}\textsf{a}, where qubits (couplers) are represented by circles (squares), and labelled according to their frequency.
The unit cell is defined by the minimum Manhattan distance between two qubits or couplers with the same frequency, $d_\mathrm{UC}$. 
This parameter determines the number of qubits ($\lceil d_{\mathrm{UC}}^2/2 \rceil$) couplers ($2 \lceil d_{\mathrm{UC}}^2/2 \rceil$), and total modes ($3\lceil d_{\mathrm{UC}}^2/2 \rceil$) in the unit cell. 

We consider direct coupling between adjacent qubits and couplers, and spurious next-nearest-neighbor couplings.
Furthermore, our unit-cell Hamiltonian includes fictitious couplings which model the additional interactions that appear at the boundary of the unit cell when tiled (see~\cref{fig:extensibility}\textsf{a}).
Optimizing the unit-cell frequency layout with these additional links makes the result in principle extensible to a device with an arbitrary number of qubits. 
For concreteness, we now focus on~$d_\mathrm{UC}=3$ with a total of 15 modes. 

In~\cref{sec:Two-step frequency allocation}, we describe our frequency-allocation approach in detail. 
We work with smaller coupling and drive strengths than those considered for the two-qubit gates in previous sections to ensure the validity of our perturbative treatment.
We do this because benchmarking the general case in the multi-qubit limit would require substantial computing resources.
Nonetheless, this approach is informative for our purposes here, and it provides a meaningful starting point from which one can further optimize the multi-qubit system, \textit{e.g.}, increasing the gate speed. 
We also introduce simplifications to the circuit model and work with a multi-qubit KNO Hamiltonian under a rotating-wave approximation.

The result of a typical optimizer run for the qubit and coupler frequencies is shown in~\cref{fig:extensibility}\textsf{b}.
The mode frequencies are shown as detunings with respect to an arbitrary common frequency. 
We find that the static ZZ interaction lies below the desired bound of~$20\,\mathrm{kHz}$ for all pairs of qubits.

With the mode frequencies determined, we run the optimizer a second time to determine suitable coupler-drive frequencies (see~\cref{fig:extensibility}\textsf{c}) such that spectator qubit effects are minimized.
There, the drive frequency is shown as a detuning with respect to the respective coupler frequency defined in the previous step.
Panel~\textsf{d} shows the relative driven ZZ interaction for all pair of qubits.
A diagonal matrix element in this plot corresponds to the normalized gate interaction between a pair of qubits~$(i,j)$ connected by a driven coupler mode. 
In the row associated with~$(i,j)$, off-diagonal matrix elements represent undesired two-qubit couplings between other pair of qubits~$(i',j')$, activated by the drive on the coupler mode that connects~$(i,j)$.
These spurious couplings are shown normalized with respect to the~$(i,j)$ interaction rate.

In practice, the results obtained using perturbation theory should be complemented by numerical simulations of a more elaborate circuit model, including simultaneous drives and potentially larger couplings and drive strengths. 
While a side-by-side comparison against other transmon-based architectures is outside the scope of this work, we expect our setup to allow for comparable or larger extensibility with respect to other all-microwave architectures.
The results of this section are a first step toward building a full-scale processor based on our architecture. 

\subsection{Quasiadiabatic microwave multi-qubit control}
\label{subsec:Adiabatic microwave multi-qubit control}

We conclude our extensibility analysis with a qualitative description of quasiadiabatic multi-qubit microwave control.
To this end, we consider a system of~$K$ superconducting qubits subject to~$d$ microwave drives. 
Following the derivation in~\cref{subsec:Response to slow drive-parameter changes}, we group the drive phases and parameters in the vectors~$\boldsymbol{\theta}(t)=(\theta_1,\theta_2,\dots,\theta_d)^T$.

The expanded-space wavefunction in~\cref{eq:unsqueezed initial condition} and the prescription in~\cref{eq:wavefunction prescription} can be straightforwardly expanded to~$d$ dimensions.
Using that~$(2\pi)^{-d}\smallint d^d\vartheta|\boldsymbol{\vartheta}\rangle=|\boldsymbol{0}\rangle$ and~$\langle\boldsymbol{\vartheta}|=\sum_{\boldsymbol{m}}e^{i\boldsymbol{\vartheta}\cdot\boldsymbol{m}}\langle\boldsymbol{m}|$, where~$|\boldsymbol{m}\rangle$ are the eigenstates of~$\hat{\boldsymbol{m}}$, we arrive at the propagator
\begin{equation}
    \begin{split}
        \mathcal{U}(t) = \sum_{\alpha,\boldsymbol{m},\boldsymbol{m'}}& \langle\boldsymbol{m'}-\boldsymbol{m}|\Psi_\alpha^{\boldsymbol{0}}[\boldsymbol{\lambda}(t)]\rangle\langle\Psi_\alpha^{\boldsymbol{0}}[\boldsymbol{\lambda}(0)]|-\boldsymbol{m}\rangle \\
        & \times e^{i\boldsymbol{\theta}(t)\cdot\boldsymbol{m'}}e^{-i\int_0^t  \varepsilon_\alpha^{\boldsymbol{m}}[\boldsymbol{\lambda}(t')]dt'},
    \end{split}
    \label{eq:Adiabatic Floquet propagator bold}
\end{equation}
where~$|\Psi_\alpha^{\boldsymbol{m}}[\boldsymbol{\lambda}(t)]\rangle$ and~$\varepsilon_\alpha^{\boldsymbol{m}}[\boldsymbol{\lambda}(t)]$ are the parametric eigenstates of the expanded-space Hamiltonian provided in~\cref{sec:Expanded-space Hamiltonian for multiple drives}, and~$\boldsymbol{\lambda}(t)=[\boldsymbol{\Omega}(t),\boldsymbol{\omega}_{\mathrm{eff}}(t)]^T$ groups the drive-parameter vectors~$\boldsymbol{\Omega}(t)=[\Omega_1(t),\dots,\Omega_d(t)]^T$ and~$\boldsymbol{\omega}_\mathrm{eff}(t)=[{\omega}_{\mathrm{eff}_1}(t),\dots,{\omega}_{\mathrm{eff}_d}(t)]^T$. 

The~$K$-qubit Floquet-mode basis~$\{|u_\alpha[\boldsymbol{\theta}(t)]\rangle\}$ with~$\alpha\in\{0,1\}^K$ defines the multi-qubit computational basis. 
A gate operation from time~$t_0$ to~$t_1$ implements a unitary map between the bases~$\{|u_\alpha[\boldsymbol{\theta}(t_0)]\rangle\}$ and~$\{|u_\alpha[\boldsymbol{\theta}(t_1)]\rangle\}$.
If such an operation is performed in a nonadiabatic fashion, then multi-qubit control must explicitly account for the instantaneous phases of the microwave drives.
This would potentially require to introduce phase delays (frequency chirps) to synchronize the logical subspace with independently calibrated waveforms for single- and two-qubit gates.

\begin{figure}[t!]
    \includegraphics[scale=1]{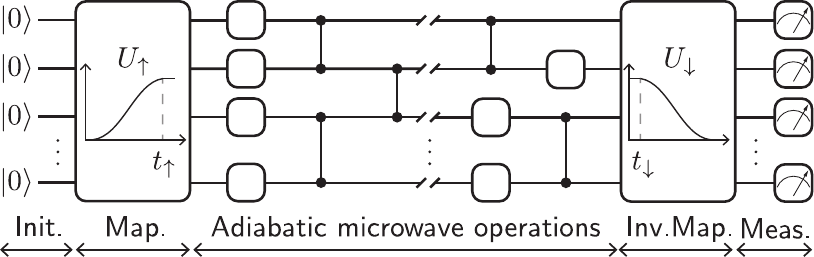}
    \caption{\label{fig:Multiqubit processor} Quantum computation with several driven qubit modes and quasiadiabatic microwave operations.
            For initialization (Init. and Map.), an eigenstate of the static Hamiltonian is adiabatically connected to a Floquet mode of the driven Hamiltonian by a slow rise of the drive amplitudes. 
            For readout (Inv.Map. and Meas.), a state specified in the Floquet-mode basis at time~$t$ is adiabatically connected to an equivalent superposition of the static Hamiltonian eigenstates by a slow ramp down of the drive amplitudes.
            }
\end{figure}

At the expense of longer gate times, quasiadiabatic microwave control (see~\cref{fig:Multiqubit processor}) could help mitigating these fine-tuning conditions and be less prone to noise in the control parameters. 
In a full-scale chip, adiabatic initialization and readout would also be necessary~\cite{huang2021engineering}.  

One can also picture the adiabatic limit using the framework of analog quantum computation: The multi-qubit dynamics follow the slow change of the drive parameters in the expanded-space Hamiltonian (see~\cref{sec:Expanded-space Hamiltonian for multiple drives}), leading to digital gate operations on the multi-qubit state at specific times during the pulse schedule.

\section{Conclusion}
\label{sec:Conclusion}

In summary, we introduced a circuit-QED architecture that leverages microwave drives of variable amplitude and frequency to perform two-qubit gates and suppress coherent errors by driving a coupler mode. 
Our frequency-modulated two-qubit gates expand the microwave-control toolbox for superconducting qubits.  

From a technical standpoint, we introduced a number of developments to qualitatively understand and precisely engineer driven dynamics in circuit-QED setups.
Using a parametric extension of Floquet theory, we derived a pulse-envelope parametrization for low-leakage microwave control and used it to engineer high-fidelity controlled-phase gates with little added control complexity. 
Using our Floquet-theory framework, we derived semi-analytical expressions for the conditional phase and leakage-cancellation conditions. 
We performed numerical simulations of the proposed two-qubit gates, taking into account the full-circuit Hamiltonian and including dissipation.
In particular, we showed that high-fidelity controlled-phase gates can be achieved through multiple approaches, including both resonantly and off-resonantly driving multi-photon transitions with modulated amplitude and/or chirped frequency. 
Indeed, we showed that chirped-frequency control is the driven analog to baseband flux control and carries many of the same advantages. 
The average gate fidelity for all of these gates can exceed~99.9\% for realistic circuit parameters.

To understand the origin of the driven two-qubit interaction, we developed a version of perturbation theory that accurately captures drive-induced frequency shifts on the two-qubit states.
We moreover used our perturbative approach to analyze the extensibility of the proposed architecture, addressing the allocation of qubit and drive frequencies within a systematic framework. 

Finally, many of the techniques introduced here are applicable to other circuit-QED architectures (\textit{e.g.}, bosonic qubits) and more generally cavity-QED modalities based on drive-activated interactions and/or stabilized Floquet modes.
 
\section{Acknowledgements}
\label{sec:Acknowledgements}

We are grateful to Junyoung An, Jeffrey Grover, Patrick Harrington, Amir Karamlou and Sarah Muschinske for insightful conversations, and to Chihiro Wantanabe for administrative assistance.
This research was funded in part by the U.S. Army Research Office Grant W911NF-18-1-0411; the U.S. Department of Energy, Office of Science, National Quantum Information Research Centers, Co-design Center for Quantum Advantage (C2QA) under Contract No. DE-SC0012704, Quantum Systems Accelerator (QSA); NSERC; the Canada First Research Excellence Fund; the Minist\`ere de l’\'Economie et de l'Innovation du Qu\'ebec; and the Under Secretary of Defense for Research and Engineering under Air Force Contract No. FA8702-15-D-0001. Any opinions, findings, conclusions or recommendations expressed in this material are those of the author(s) and do not necessarily reflect the views of the Under Secretary of Defense for Research and Engineering.

\appendix

\section{Perturbation theory calculations}
\label{sec:Perturbation theory calculations}

\subsection{Derivation of the self-energy expression}
\label{subsec:Derivation of the self-energy expression}

Here, we provide the derivation of the self-energy equation. 
We write the eigenstates of the complete Hamiltonian~$\hat{H} = \hat{H}^0+\eta\hat{V}$ as~$|\Phi_\alpha\rangle=|\Phi_\alpha^0\rangle + \eta |d\Phi_\alpha\rangle$, where~$|\Phi_\alpha^0\rangle$ is an eigenstate of~$\hat{H}^0$.
Since the perturbation~$\eta\hat{V}$ is off-diagonal in the basis~$\{|\Phi_\alpha^0\rangle\}$, one has~$\langle \Phi_\alpha^0|d\Phi_\alpha\rangle=0$.
From the Schr\"odinger equation~$\hat{H}|\Phi_\alpha\rangle = ({\epsilon}_\alpha^0 + \Sigma_\alpha)|\Phi_\alpha\rangle$, it follows that 
\begin{equation}
    \langle\Phi_{\alpha'}^0|\Phi_\alpha\rangle = \frac{\langle\Phi_{\alpha'}^0|\eta\hat{V}|\Phi_\alpha\rangle}{{\epsilon}_\alpha^0 + \Sigma_\alpha - {\epsilon}_{\alpha'}^0}.
    \label{eq:dot product}
\end{equation} 
For~$\alpha'=\alpha$, \cref{eq:dot product} reduces to 
\begin{equation}
    \Sigma_\alpha = \langle\Phi_\alpha^0|\eta\hat{V}|\Phi_\alpha\rangle = \sum_{\alpha_1}\langle\Phi_\alpha^0|\eta\hat{V}|\Phi_{\alpha_1}^0\rangle\,\eta\langle\Phi_{\alpha_1}^0|d\Phi_\alpha\rangle,
    \label{eq:self-energy from dot product diagonal}
\end{equation} 
For~$\alpha'=\alpha_1\neq\alpha$, we instead find
\begin{equation}
    \begin{split}
        \eta\langle\Phi_{\alpha_1}^0|d\Phi_\alpha\rangle &= \frac{\langle\Phi_{\alpha_1}^0|\eta\hat{V}|\Phi_\alpha^0\rangle}{{\epsilon}_\alpha^0 + \Sigma_\alpha - {\epsilon}_{\alpha_1}^0} \\
        &+ \sum_{\alpha_2} \frac{\langle\Phi_{\alpha_1}^0|\eta\hat{V}|\Phi_{\alpha_2}^0\rangle}{{\epsilon}_\alpha^0 + \Sigma_\alpha - {\epsilon}_{\alpha_1}^0}\,\eta\langle\Phi_{\alpha_2}^0|d\Phi_\alpha\rangle.
    \end{split}
    \label{eq:self-energy from dot product off-diagonal}
\end{equation} 
Inserting~\cref{eq:self-energy from dot product off-diagonal} in~\cref{eq:self-energy from dot product diagonal}, we arrive at the self-consistent expression
\begin{equation}
    \begin{split}
    \Sigma_\alpha = \sum_{k=1}^{\infty}\sum_{\alpha_1,\dots,\alpha_k}&\langle\Phi_\alpha^0|\eta\hat{V}|\Phi_{\alpha_1}^0\rangle\frac{\langle \Phi_{\alpha_1}^0|\eta\hat{V}|\Phi_{\alpha_2}^0\rangle}{{\epsilon}_\alpha(\Sigma_\alpha)-{\epsilon}_{\alpha_1}^0}\dots\\
    &\dots\times\frac{\langle \Phi_{\alpha_k}^0|\eta\hat{V}|\Phi_{\alpha}^0\rangle}{{\epsilon}_\alpha(\Sigma_\alpha)-{\epsilon}_{\alpha_k}^0}.
    \end{split}
    \label{eq:self-energy appendix}
\end{equation}
Here, $k+1$ represents the order of perturbation theory and the sum over~$\alpha_{1\dots k}$ accounts for all possible processes of order~$k+1$.

\subsection{Self-energy resummation technique}
\label{subsec:Graph-based self-energy resummation technique}

Our Self-Consistent Perturbation Theory (SCPT) technique yields equations for the computational-state self-energies with bounded order.
Furthermore, the implicit form of~\cref{eq:self-energy appendix} prevents divergences due to degeneracies of~$\hat{H}^0$.

As an example, let us consider the subspace spanned by the two bare states~$\{|\Phi_\alpha^0\rangle,|\Phi_\beta^0\rangle\}$ that we assume are coupled by a nearly resonant drive. 
Recasting~\cref{eq:self-energy appendix} as a geometric series, we find
\begin{equation}
    \Sigma_{\alpha,\beta}=\pm\frac{{\epsilon}_\beta^0-{\epsilon}_\alpha^0}{2}\left(1-\sqrt{1+\frac{4|\langle \Phi_\beta^0|\eta\hat{V}|\Phi_\alpha^0\rangle|^2}{({\epsilon}_\beta^0-{\epsilon}_\alpha^0)^2}}\right),
    \label{eq:self-energy two-state}
\end{equation}
which does not diverge for~${\epsilon}_\beta^0-{\epsilon}_\alpha^0\to 0$.
We have shown that multi-photon resonances of the form~$|{\epsilon}_\alpha^0-{\epsilon}_\beta^0|\approx m\omega$, with~$m$ an integer, are central to realize fast two-qubit gates.
The regular behavior of~\cref{eq:self-energy two-state} for exact resonance conditions makes of SCPT a useful tool in this context.

Two-state approximations are, however, not enough in most cases. 
For this reason, we now generalize our approach to include more than two states with the help of a graph-based algorithm next.

\subsubsection{Graph definition}
\label{subsubsec:Graph problem definition}

\Cref{eq:self-energy appendix} is exact but has an infinite number of terms.
We thus need a method to efficiently truncate the number of terms controlling the precision of such an approximation. 
To this end, we reinterpret~\cref{eq:self-energy appendix} with a graph.
The vertices of the graph are the eigenstates of~$\hat{H}^0$.
An edge between two vertices represents the matrix element of~$\eta\hat{V}$ with respect to the two nodes that are connected. 
With this definition, we rewrite the self-energy expression as 
\begin{equation}
    \begin{split}
    \Sigma_\alpha &= \sum_{\boldsymbol{\alpha}\in\{\mathcal{G}_\alpha^{\mathrm{irr.}}\in\mathcal{G}_\alpha\}}\langle\Phi_\alpha^0|\eta\hat{V}|\Phi_{\alpha_1}^0\rangle\dots\\
    &\dots\times\frac{\langle\Phi_{\alpha_k}^0|\eta\hat{V}|\Phi_\alpha^0\rangle}{\left[1-f_{\alpha,\alpha_k}\left(\{\boldsymbol{\alpha}\}\right)\right]\left[{\epsilon}_\alpha(\Sigma_\alpha)-{\epsilon}_{\alpha_k}^0\right]},
    \end{split}
    \label{eq:self-energy graph}
\end{equation}
where
\begin{equation}
    \begin{split}
    f_{\alpha,\alpha_i}(\mathcal{T}) &= \sum_{\boldsymbol{\alpha}\in\{\mathcal{G}_{\alpha_i}^{\mathrm{irr.}}\in\mathcal{G}_\alpha\setminus\mathcal{T}\}}\frac{\langle\Phi_{\alpha_i}^0|\eta\hat{V}|\Phi_{\alpha_1}^0\rangle}{{\epsilon}_\alpha(\Sigma_\alpha)-{\epsilon}_{\alpha_i}^0}\dots\\
    &\dots\times\frac{\langle\Phi_{\alpha_k}^0|\eta\hat{V}|\Phi_{\alpha_i}^0\rangle \left[{\epsilon}_\alpha(\Sigma_\alpha)-{\epsilon}_{\alpha_k}^0\right]^{-1}}{\left[1-f_{\alpha,\alpha_k}\left(\mathcal{T} \cup\{\boldsymbol{\alpha}\}\right)\right]}.
    \end{split}
    \label{eq:fofalpha}
\end{equation}
Here, $\mathcal{T}$ is the set of all nodes defining the spanning tree, $\mathcal{G}_\alpha$ is the undirected graph from node~$\alpha$ and~$\{\mathcal{G}_{\alpha_i}^{\mathrm{irr.}}\in\mathcal{G}_\alpha\setminus\mathcal{T}\}$ is the set of all irreducible (loopless) directed graphs starting at node~$\alpha_i$ in the complete graph~$\mathcal{G}_\alpha$ after removing~$\mathcal{T}$.
Moreover, $\boldsymbol{\alpha}=(\alpha_1, \alpha_2,\dots,\alpha_k)$ is a vector containing the ordered~$k$ nodes inside some irreducible directed graph~$\mathcal{G}^{\mathrm{irr.}}_{\alpha_i}$. 

It is possible to show that~\cref{eq:self-energy graph} reduces to~\cref{eq:self-energy appendix} by induction. 
Importantly, $[1-f_{\alpha,\alpha_i}(\mathcal{T})]^{-1}=\sum_{l=0}^{\infty}[f_{\alpha,\alpha_i}(\mathcal{T})]^l$ for $|f_{\alpha,\alpha_i}(\mathcal{T})|<1$ generates an infinite number of loops inside the irreducible diagrams defining the spanning trees. 
This expression captures the infinite number of times that a given cycle is consecutively repeated in the perturbation theory expansion in \cref{eq:self-energy appendix}, such that the infinite-order expression in~\cref{eq:self-energy appendix} can be reduced to finite-order in~\cref{eq:self-energy graph}.

\subsubsection{Self-consistent perturbation-theory (SCPT) algorithm}
\label{subsubsec:SCPT algorithm}

Our SCPT algorithm relies on finding irreducible cycles inside the graph that represents the perturbation in a truncated Hilbert space. 
Two parameters determine the accuracy of this technique: the maximum path-length~$L_{\mathrm{max}}$ of all cycles of the perturbation graph, and the maximum tree-depth~$D_\mathrm{max}$. 
The latter is defined as the longest length of a path between two vertices in the graph that includes such vertices only once.
The recursive algorithm estimates the self-energy~$\Sigma_\alpha$ associated with a system eigenstate~$|\Phi_\alpha\rangle$, by following the steps:
\begin{enumerate}
    \item Define the graph~$\mathcal{G}_\alpha$ starting at vertex~$|\Phi_0\rangle=|\Phi_\alpha\rangle$.
    \item Define a set~$\{L_1, L_2, \dots\}$ of cycle lengths and the maximum allowed tree-depth~$D^*\leq D_\mathrm{max}$.
    \item Find all irreducible cycles of the perturbation graph of maximum length~$L=L_1$ starting at~$|\Phi_\alpha\rangle$. 
          Set~$D=1$.
    \item Initialize the total weight~$W_\alpha=0$.
    \item For each cycle~$|\Phi_0\rangle\to|\Phi_1\rangle \to\dots|\Phi_J\rangle\to|\Phi_0\rangle$:
            \begin{enumerate}
                \item Initialize the cycle weight~$w=\langle\Phi_J|\eta\hat{V}|\Phi_0\rangle$.
                \item For each state~$|\Phi_{j>0}\rangle$ in the cycle:
                \begin{enumerate}
                    \item If $D=D^*$, set~$w=0$ and break.
                    \item Update the length variable as~$L=L_2$ and the current tree-depth as~$D=D+1$.
                    \item Update the perturbation graph~$\mathcal{G}_\alpha$ by removing the states~$\{|\Phi_{j'<j}\rangle\}$, i.e. define the graph~$\mathcal{G}_\alpha\setminus\{|\Phi_0\rangle,|\Phi_1\rangle,\dots,|\Phi_{j-1}\rangle\}$.
                    \item Repeat steps 3-5 recursively to find the total weight~$W_j$ associated with~$|\Phi_j\rangle$.
                    \item Update the cycle weight according to~$$w= w\times \frac{\langle\Phi_{j-1}|\eta\hat{V}|\Phi_j\rangle}{{\epsilon}_\alpha(\Sigma_\alpha) - {\epsilon}_j^0+W_j}.$$
                \end{enumerate}
                \item Update the total weight as~$W_\alpha=W_\alpha+w$.
            \end{enumerate}
    \item Set~$\Sigma_\alpha=W_\alpha$.
\end{enumerate}

We use this algorithm to derive symbolic expressions for the self-energies using the Python package~\texttt{Sympy}. 
Because of the recursion, we obtain a self-consistent expression that we then solve numerically using a root-finder routine. 
The order of the self-energy equation is determined in part by~$D^*$ and the maximum length~$L$ at each recursion level. 
We discuss how we select these parameters below. 

\subsubsection{Estimating the ZZ interaction}
\label{subsubsec:Estimating the ZZ interaction}

The ZZ interaction is estimated by first computing the self-energies of the computational states.
Given the multiplicity of the numerical roots, we pick the roots that are closest to the self-energy obtained using full numerical diagonalization. 
Alternatively, we simply pick the smallest root found for each state, thus minimizing the self-energy. 
This second approach is stand-alone, but it can, in cases, differ from the numerical estimation at anticrossings where the labelling of states can be done in multiple ways.

\textit{Static case}--
Because the Hamiltonian in~\cref{eq:KNO Hamiltonian} conserves the number of excitations for~$\Omega=0$, the perturbation graph is clustered. 
This property is important, as the self-energies can then be computed exactly using our resummation technique, where~$D_\mathrm{max}=6$ and~$L_\mathrm{max}=12$. 

\textit{Driven case}--
For~$\Omega\neq 0$, the inclusion of the the coupler drive leads to a perturbation graph that is no longer clustered.
In other words, $D_\mathrm{max}$ and~$L_\mathrm{max}$ are infinite for all computational states. 
This situation makes the resummation exponentially harder in graph size. 

An approximate clustering of the perturbation graph is therefore needed, at the cost of rendering the theory only approximate. 
For the numerical simulations in this work, we consider~$D_\mathrm{max}=6$ and~$L_\mathrm{max}=12$.
However, for each computational state, we truncate the perturbation graph to include states with up to two additional excitations with respect to the computational state in consideration.
We also truncate the path-length at each level of recursion, following the rule~$L_k = L_\mathrm{max} - 2(k-1)$. 
This reduces the total number of cycles that need to be found recursively in the graph, which is a NP-hard problem, and is motivated by energy scales, as the self-energy renormalization decreases with increasing tree-depth. 

\subsubsection{Low-order estimations}
\label{subsubsec:Low-order estimations}

Next, we illustrate how our perturbation theory can be used to understand the processes that dominantly contribute to the ZZ interaction. 
This section expands on the details provided in~\cref{subsubsec:System parameters and ZZ interaction}.
First, we find an effective coupling~$J$ between the two states~$|\Phi_{100}\rangle$ and~$|\Phi_{020}\rangle$ introduced in~\cref{subsubsec:System parameters and ZZ interaction}.
We approximate~\cref{eq:self-energy appendix} by dropping the self-energies in all denominators that do not include~${\epsilon}_{100}^0-{\epsilon}_{020}^0$. 
This is a good approximation if~$|\Sigma_\alpha|\ll |{\epsilon}_{100}^0-{\epsilon}_{\alpha\neq 020}^0|$. 
To determine the coupling, we consider only those terms that connect~$|\Phi_{100}\rangle$ and~$|\Phi_{020}\rangle$ up to fourth order and in a frequency bandwidth of~$1\,\mathrm{GHz}$ centered in~${\epsilon}_{100}^0/h$, such that~$J\to J^{(4)}$.
Accordingly, we find that the self-energy of the computational state can approximated as
\begin{equation}
    \Sigma_{100}^{(8)}\approx \frac{|J^{(4)}|^2}{{\epsilon}_{100}(\Sigma_{100}^{(8)})-{\epsilon}^0_{020}-\Lambda_{020}},
    \label{eq:self-energy 100}
\end{equation}
where 
\begin{equation}
    \begin{split}
        J^{(4)} & = \sqrt{2} J_\mathrm{bc} \frac{J_\mathrm{ab}}{{\epsilon}^0_{100}-{\epsilon}^0_{011}}\frac{\Omega/2}{{\epsilon}^0_{100}-{\epsilon}^0_{101}} \\
        & + \sqrt{2}J_\mathrm{ab}\frac{J_\mathrm{bc}}{{\epsilon}^0_{100}-{\epsilon}^0_{110}}\frac{\Omega/2}{{\epsilon}^0_{100}-{\epsilon}^0_{101}} \\
        & + \sqrt{2}J_\mathrm{bc}\frac{J_\mathrm{ac}}{{\epsilon}^0_{100}-{\epsilon}^0_{011}}\frac{J_\mathrm{bc}}{{\epsilon}^0_{100}-{\epsilon}^0_{110}}\frac{\Omega/2}{{\epsilon}^0_{100}-{\epsilon}^0_{101}} \\
        & + \sqrt{2}J_\mathrm{bc}\frac{\sqrt{2}J_\mathrm{bc}}{{\epsilon}^0_{100}-{\epsilon}^0_{011}}\frac{\sqrt{2}J_\mathrm{ac}}{{\epsilon}^0_{100}-{\epsilon}^0_{002}}\frac{\Omega/2}{{\epsilon}^0_{100}-{\epsilon}^0_{101}} \\
        & + \sqrt{2}J_\mathrm{bc}\frac{\sqrt{2}J_\mathrm{bc}}{{\epsilon}^0_{100}-{\epsilon}^0_{011}}\frac{\Omega/\sqrt{2}}{{\epsilon}^0_{100}-{\epsilon}^0_{002}}\frac{J_\mathrm{ac}}{{\epsilon}^0_{100}-{\epsilon}^0_{001}} \\
        & + \sqrt{2}J_\mathrm{bc}\frac{\Omega/2}{{\epsilon}^0_{100}-{\epsilon}^0_{011}}\frac{J_\mathrm{ab}}{{\epsilon}^0_{100}-{\epsilon}^0_{010}} \\
        & + \sqrt{2}J_\mathrm{bc}\frac{\Omega/2}{{\epsilon}^0_{100}-{\epsilon}^0_{011}}\frac{J_\mathrm{bc}}{{\epsilon}^0_{100}-{\epsilon}^0_{010}}\frac{J_\mathrm{ac}}{{\epsilon}^0_{100}-{\epsilon}^0_{001}},
    \end{split}
    \label{eq:G4}
\end{equation}
and~$\Lambda_{020}$ is an energy shift induced by the drive on the noncomputational state~$|\Phi_{020}\rangle$.
In particular, we consider the approximation~$\Lambda_{020}\to\Lambda^{(2)}_{020}=(\Omega/2)^2/({\epsilon}^0_{100}-{\epsilon}^0_{021})$, which is second-order in the drive amplitude. 

Two terms belonging to~\cref{eq:G4} are illustrated in~\cref{fig:ZZ interaction}\textsf{c-d}.
The impact of~$\Lambda_{020}$, which includes the processes that connect~$|\Phi_{020}\rangle$ to itself via states that are not~$|\Phi_{100}\rangle$, can be significant in the limit~${\epsilon}^0_{100}-{\epsilon}^0_{020}\to 0$, and should therefore be included. 
Crucially, this frequency shift can also be used as a resource to maximize the ZZ interaction at finite coupling and drive power.

Using~\cref{eq:self-energy two-state}, we arrive at the expression
\begin{equation}
    \Sigma_{100}^{(8)}\approx \frac{\Delta}{2}\left(1-\sqrt{1+\left|\frac{2 J^{(4)}}{\Delta}\right|^2}\right),
    \label{eq:self-energy 100 2 app}
\end{equation}
where~$\Delta={\epsilon}_{100}^0-{\epsilon}^0_{020}-\Lambda_{020}^{(2)}$.
The driven ZZ interaction can then be approximated as~$-\Sigma_{100}^{(8)}$.
In the dispersive limit where~$J^{(4)}\ll \Delta$, \cref{eq:self-energy 100 2 app} reduces to the simpler form~$\Sigma_{100}^{(8)}\approx -|J^{(4)}|^2/\Delta$.
According to this expression, and assuming~$J_\mathrm{ab}\neq 0$, the ZZ interaction is at least sixth order: second order in~$\Omega$ and fourth order in~$J_{\mu\nu}$. 

\section{Expanded Hilbert space Hamiltonian}
\label{sec:Extended Hilbert space Hamiltonian}

Here, we present a derivation of~\cref{eq:Extended Hilbert space Hamiltonian}. 
We start with the Schr\"odinger equation
\begin{equation}
    i\hbar\partial_t|\psi(\boldsymbol{{\phi}},t)\rangle = \hat{H}(\boldsymbol{\hat{\phi}},t)|\psi(\boldsymbol{{\phi}},t)\rangle,
    \label{eq:Original Schrodinger equation}
\end{equation}
and define the expanded-Hilbert-space wavefunction by
\begin{equation}
    |\psi(\boldsymbol{{\phi}},t)\rangle = |\Psi({{\vartheta}}, \boldsymbol{{\phi}},t)\rangle|_{\vartheta=\theta(t)}.
    \label{eq:Extended Hilbert space wavefunction definition}
\end{equation}
Accordingly, we have 
\begin{equation}
\begin{split}
    \partial_t|\psi(\boldsymbol{{\phi}},t)\rangle &= \partial_t|\Psi({{\vartheta}}, \boldsymbol{{\phi}},t)\rangle|_{\vartheta=\theta(t)}\\
    &+\dot{\theta}(t){\partial}_{\vartheta}|\Psi({{\vartheta}}, \boldsymbol{{\phi}},t)\rangle|_{\vartheta=\theta(t)}.
    \label{eq:Derivatives extended Hilbert space}
\end{split}
\end{equation}
Next, we rewrite~\cref{eq:Original Schrodinger equation} as 
\begin{equation}
    i\hbar\partial_t|\psi(\boldsymbol{{\phi}},t)\rangle = \left[\hat{H}(\hat{\vartheta},\boldsymbol{\hat{\phi}},t)|\Psi({\vartheta},\boldsymbol{{\phi}},t)\rangle \right]_{\vartheta=\theta(t)}.
    \label{eq:Original Schrodinger equation 2}
\end{equation}
Replacing the r.h.s. of~\cref{eq:Derivatives extended Hilbert space} in~\cref{eq:Original Schrodinger equation 2}, we arrive at
\begin{widetext}
\begin{equation}
    i\hbar\partial_t|\Psi({{\vartheta}}, \boldsymbol{{\phi}},t)\rangle|_{\vartheta=\theta(t)}+i\hbar\dot{\theta}(t){\partial}_{\vartheta}|\Psi({{\vartheta}}, \boldsymbol{{\phi}},t)\rangle|_{\vartheta=\theta(t)} = \left[\hat{H}(\hat{\vartheta},\boldsymbol{\hat{\phi}},t)|\Psi({\vartheta},\boldsymbol{{\phi}},t)\rangle \right]_{\vartheta=\theta(t)}.
    \label{eq:Original Schrodinger equation 3}
\end{equation}
\end{widetext}
This equation can be rewritten as 
\begin{equation}
    \left[i\hbar\partial_t|\Psi({{\vartheta}}, \boldsymbol{{\phi}},t)\rangle= \hat{H}_\mathrm{eff}(t)|\Psi({\vartheta},\boldsymbol{{\phi}},t)\rangle \right]_{\vartheta=\theta(t)}.
    \label{eq:Original Schrodinger equation 4}
\end{equation}
where
\begin{equation}
    \hat{H}_\mathrm{eff}(t)=\hat{H}(\hat{\vartheta},\boldsymbol{\hat{\phi}},t)-i\hbar\dot{\theta}(t){\partial}_{\vartheta}.
    \label{eq:Heff definition}
\end{equation}
This expression is equivalent to~\cref{eq:Extended Hilbert space Hamiltonian} written in terms of~$\hat{m}\to-i{\partial}_\vartheta$ in the phase representation.

\section{Full-circuit numerical simulations}
\label{sec:Full-circuit numerical simulations}

In this section, we provide details regarding the full-circuit model and numerical simulations. 
The full-circuit parameters are provided in~\cref{tab:Full-circuit enegry parameters}.
\begin{table*}[t]
    \begin{ruledtabular}
    \begin{tabular}{cccccccccc}
    Parameter set & $E_{\mathrm{C}_\mathrm{a}}/h$ & $E_{\mathrm{C}_\mathrm{b}}/h$ & $E_{\mathrm{C}_\mathrm{c}}/h$ & $E_{\mathrm{J}_\mathrm{a}}/h$ & $E_{\mathrm{J}_\mathrm{b}}/h$ & $E_{\mathrm{J}_\mathrm{b}}/h$ & $g_{\mathrm{ac}}/2\pi$ & $g_{\mathrm{bc}}/2\pi$ & $g_{\mathrm{ab}}/2\pi$ \\
    \hline
    Zero static ZZ & 0.2315 & 0.2499 & 0.2947 & 15.414 & 17.189 & 14.152 & 0.0752 & 0.0825 & 0.0072 \\
    \hline
    Nonzero static ZZ & 0.2315 & 0.2499 & 0.2987 & 15.414 & 17.189 & 16.687 & 0.0723 & 0.0792 & 0.0072 \\
    \end{tabular}
    \end{ruledtabular}
    \caption{\label{tab:Full-circuit enegry parameters} Energy parameters for the full-circuit Hamiltonian. 
    All values are provided in GHz.
    `Zero static ZZ' specifies the parameters used in~\cref{fig:Amplitude adiabatic gate},
    \cref{fig:Amplitude adiabatic resonant gate},
    \cref{fig:Frequency adiabatic gate},
    and~\cref{fig:ZZ interaction appendix}.
    `Nonzero static ZZ' reports the circuit parameters used in~\cref{fig:Frequency adiabatic gate 0ZZ}.}
\end{table*}

\subsection{KNO and full-circuit model comparison}
\label{subsec:KNO vs. full-circuit comparison}

In~\cref{subsubsec:System parameters and ZZ interaction}, we use a Kerr-nonlinear-oscillator to understand the drive-activated ZZ interaction using perturbation theory.
Here, we numerically compare the ZZ coupling obtained for the KNO model against that predicted for the full-circuit Hamiltonian. 

\begin{figure}[t!]
    \includegraphics[scale=1]{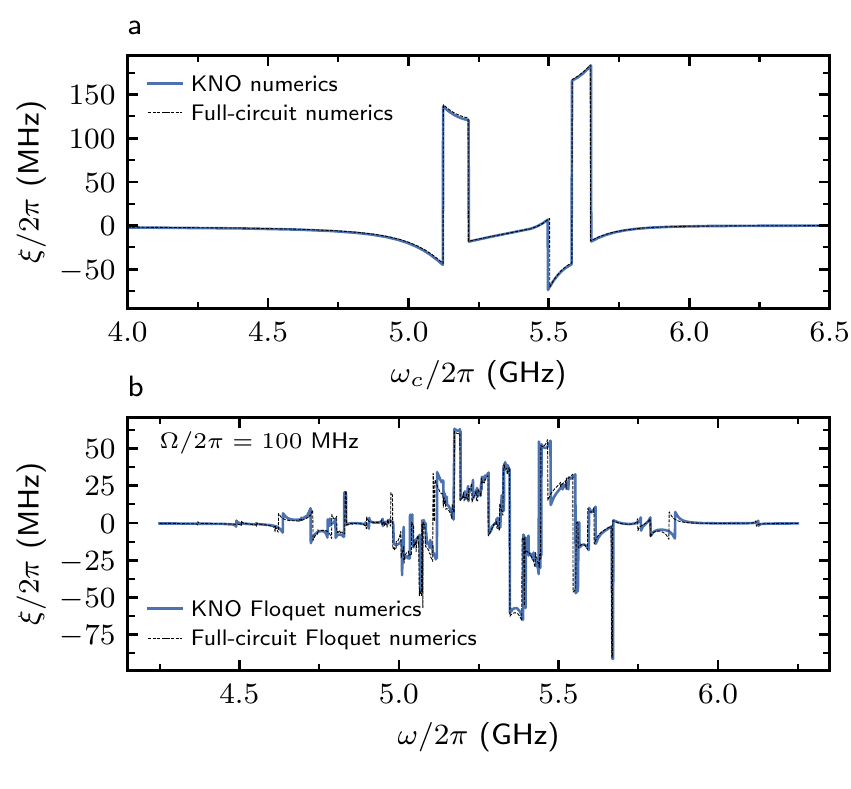}
    \caption{\label{fig:ZZ interaction appendix} Two-qubit ZZ interaction. 
            \textsf{a} Static ZZ interaction as a function of coupler frequency.
            Comparison between the KNO model numerics against full-circuit numerics.
            \textsf{b} Driven ZZ interaction for~$\Omega/2\pi=100\,\mathrm{MHz}$ as a function of drive frequency. 
            The coupler frequency is chosen to minimize the static ZZ interaction.
            Comparison between KNO Floquet numerics against full-circuit Floquet numerics. 
            }
\end{figure}

\cref{fig:ZZ interaction appendix} shows the numerical ZZ interaction for the KNO and full-circuit models (see also~\cref{fig:ZZ interaction}).
\cref{fig:ZZ interaction appendix}\textsf{a} shows the static ZZ interaction as a function of coupler frequency. 
In addition, \cref{fig:ZZ interaction appendix}\textsf{b} shows the driven ZZ interaction as a function of drive frequency, for~$\Omega/2\pi=100\,\mathrm{MHz}$. 
In panel~\textsf{b}, the coupler frequency is chosen such that the static ZZ coupling is zero. 

These results demonstrate that the KNO model is overall a very good approximation to the full-circuit Hamiltonian, even for the large coupling strengths and drive amplitudes considered in this work. 
While the results in panel~\textsf{b} differ quantitatively at specific drive frequencies, presumably due to the full-transmon nonlinearity and counter-rotating terms that are not present in the KNO Hamiltonian, the driven ZZ interaction is qualitatively accurate.
This makes it possible to understand the driven ZZ coupling in the KNO limit, justifying our perturbation-theory approach (SCPT) to this task. 

\subsection{Lindblad master equation and noise model}
\label{subsec:Lindblad master equation}

This section discusses details of the Lindblad master-equation simulations and dissipation rates in~\cref{eq:Master equation}. 

\subsubsection{Incoherent relaxation and excitation processes}
\label{subsubsec:Modeling dielectric loss}

We model dissipation processes assuming capacitive loss for the qubit and coupler modes and following Ref.~\cite{pop2014coherent}.
For simplicity, we assume the same frequency-independent capacitive quality factor~$Q$ for all circuit modes.  
The incoherent transition rates take the form 
\begin{equation}
    \gamma^{1}_{\alpha\beta} = \sum_{{\mu}}\frac{8E_{\mathrm{C}_{\mu}}}{Q}|\langle\Phi_\alpha|\hat{n}_{\mu}|\Phi_\beta\rangle|^2\Big|1+\coth\Big(\frac{\hbar\omega_{\alpha\beta}}{2k_B T}\Big)\Big|,
    \label{eq:gamma_1}
\end{equation}
where~$\omega_{\alpha\beta}=\omega_\beta-\omega_\alpha$ is the transition frequency between two eigenstates~$|\Phi_\alpha\rangle$ and~$|\Phi_\beta\rangle$ of the full-device Hamiltonian, and~$T=10\,\mathrm{mK}$ is the base temperature.

\begin{figure}[t!]
    \includegraphics[scale=1.]{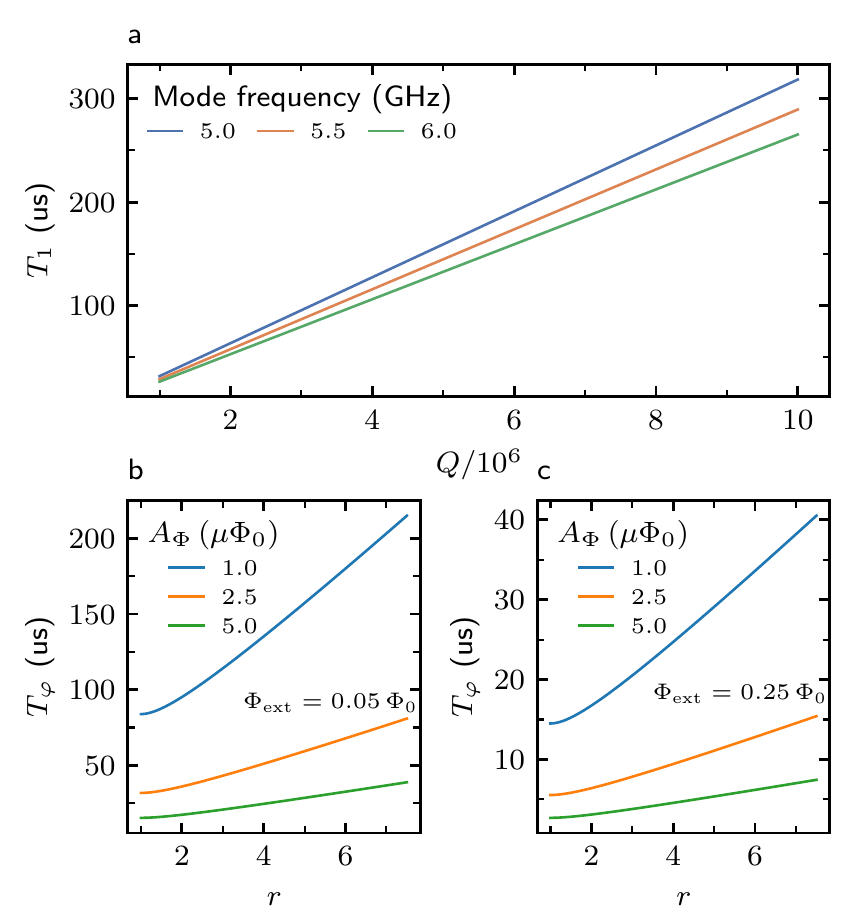}
    \caption{\label{fig:Dissipation information} Main dissipation channels. 
            \textsf{a} $T_1$-times for a typical transmon qubit with frequency in the range~$5-6\,\mathrm{GHz}$ and anharmonicity~$\alpha/2\pi=-300\,\mathrm{MHz}$, as a function of the capacitive quality factor~$Q$.
            \textsf{b-c} Coherence-time estimations according to~\cref{eq:implicit equation coherence times} as a function of the split-transmon junction-asymmetry parameter~$r$. 
            \textsf{b} Flux-bias~$\Phi_\mathrm{ext}=0.05\,\Phi_0$.
            \textsf{c} Flux-bias~$\Phi_\mathrm{ext}=0.25\,\Phi_0$.}
\end{figure}

The sign of~$\omega_{\alpha\beta}$ determines whether~$\gamma^{1}_{\alpha\beta}$ models relaxation~$(\omega_{\alpha\beta}>0)$ or excitation~($\omega_{\alpha\beta}<0$). 
Note that the sum over the circuit modes~${\mu}=(\mathrm{a,b,c})$ in~\cref{eq:gamma_1} adds up the contribution of each circuit component to the total rate~$\gamma^1_{\alpha\beta}$.
\Cref{fig:Dissipation information}\textsf{a} shows single-mode~$T_1$ estimations based on~\cref{eq:gamma_1} as a function of~$Q$.

\subsubsection{Pure-dephasing processes}
\label{subsubsec:White-noise modeling of coupler dephasing}

We consider pure-dephasing processes due to $1/f$ flux noise affecting the coupler flux bias.
However, modeling~$1/f$ noise using a Lindblad master equation is rather inconvenient because it involves time-dependent rates that also depend on the details of the noise~\cite{di2021efficient}. 
To study the gate fidelity using a simpler metric, we derive a multi-level white-noise approximation to the pure-dephasing rates~$\gamma^{\varphi}_{\alpha}$, parametrized by the coupler's pure-dephasing time~$T_\varphi$.

\textit{Multi-level white-noise approximation--}
 We consider a split-transmon model for the coupler and we account for flux noise via the effective Josephson energy
\begin{equation}
    E_{\mathrm{J}_\mathrm{c}}(\phi_\mathrm{ext}) = E_{\mathrm{J}_\Sigma}\cos\left(\frac{\phi_\mathrm{ext}}{2}\right)\sqrt{1+d_r^2\tan^2\left(\frac{\phi_\mathrm{ext}}{2}\right)},
    \label{eq:coupler Josephson energy}
\end{equation}
where~$E_{\mathrm{J}_\Sigma}=E_{\mathrm{J}_1} + E_{\mathrm{J}_2}$ adds the Josephson energies of the junctions that form the SQUID loop, 
\begin{equation}
    d_r = \frac{r-1}{r+1},
    \label{eq:junciton difference}
\end{equation}
and~$r=E_{\mathrm{J}_1}/E_{\mathrm{J}_2}> 1$ is the junction assymmetry parameter~\cite{hutchings2017tunable}. 

Carrying out a derivation similar to the one outlined in Ref.~\cite{di2021efficient}, but assuming a noise-spectral density~$S[\omega]$ that is nonsingular at zero frequency, we arrive at the multi-level dephasing rates
\begin{equation}
    \gamma^\varphi_{\alpha \beta} = \hbar^{-2}S[0]\langle\Phi_\alpha|\partial_{\phi_\mathrm{ext}}\hat{H}_\mathrm{s}|\Phi_\alpha\rangle \langle\Phi_\beta|\partial_{\phi_\mathrm{ext}}\hat{H}_\mathrm{s}|\Phi_\beta\rangle,
    \label{eq:multi-level dephasing rates}
\end{equation}
where~$\{|\Phi_\alpha\rangle\}$ are the eigenstates of the full device Hamiltonian. 
In the two-state subspace spanned by~$\{|\Phi_\alpha\rangle,|\Phi_\beta\rangle\}$, the coherence~$\rho_{\alpha\beta}$ evolves in the interaction frame according to 
\begin{equation}
    \dot{\rho}_{\alpha\beta} = -\frac{1}{2}\left(\gamma^\varphi_{\alpha \alpha}+\gamma^\varphi_{\beta \beta}-2\gamma^\varphi_{\alpha \beta}\right)\rho_{\alpha\beta}.
    \label{eq:differential equation density matrix coherences}
\end{equation}
Therefore, the coupler coherence time under pure-dephasing noise is given by
\begin{equation}
    T_\varphi = \frac{1}{\frac{\gamma^\varphi_{\alpha \alpha}+\gamma^\varphi_{\alpha \beta}}{2}-\gamma^\varphi_{\alpha \beta}},
    \label{eq:coupler coherence time}
\end{equation}
where~$\alpha=000$ and~$\beta=001$ correspond to the hybridized ground and first-excited states of the coupler mode. 

In our simulations, we consider~$T_\varphi$ as a parameter and use~\cref{eq:coupler coherence time} to infer the multi-level pure-dephasing rates. 
We do so by first approximating the matrix elements in~\cref{eq:multi-level dephasing rates} for the case of a transmon coupler, arriving at the expression
\begin{equation}
    \frac{1}{T_\varphi} = S[0]\times\frac{E_{\mathrm{C}_\mathrm{c}}}{E_{\mathrm{J}_\mathrm{c}}(\phi_\mathrm{ext})}\times[\partial_{\phi_\mathrm{ext}}E_{\mathrm{J}_\mathrm{c}}(\phi_\mathrm{ext})/\hbar]^2,
    \label{eq:approximate expression for Tvarphi}
\end{equation}
which we rewrite as
\begin{equation}
    S[0] = \frac{1}{T_\varphi}\times \frac{E_{\mathrm{J}_\mathrm{c}}(\phi_\mathrm{ext})}{E_{\mathrm{C}_\mathrm{c}}}\times \frac{1}{[\partial_{\phi_\mathrm{ext}}E_{\mathrm{J}_\mathrm{c}}(\phi_\mathrm{ext})/\hbar]^2}.
    \label{eq:approximate expression for S0}
\end{equation}
We use~\cref{eq:approximate expression for S0} to compute the rates in~\cref{eq:multi-level dephasing rates} for all pair of device eigenstates in the model.

\textit{Comparison against 1/f noise estimations--}
We now consider a model of flux noise where the spectral noise density takes the more realistic form 
\begin{equation}
    S[f] = \frac{A_\Phi^2}{|f|}.
    \label{eq:1/f spectral noise density}
\end{equation}
Here,~$A_\Phi$ quantifies the flux noise amplitude that is typically in the range~$10^{-6}-10^{-5}\,\mu\Phi_0$, with~$\Phi_0=h/2e$~\cite{koch2007charge}. 
Next, we follow the derivation in Ref.~\cite{di2021efficient}, arriving at an equation similar to~\cref{eq:differential equation density matrix coherences}.
However, because of the singular behavior of~\cref{eq:1/f spectral noise density} at low frequency, the pure-dephasing rates are now time-dependent. 
Taking this fact into consideration, we define the coherence time~$T_\varphi$ according to the relation~$\rho_{\alpha\beta}(T_\varphi)=\rho_{\alpha\beta}(0)/e$, arriving at the implicit equation  
\begin{widetext}
\begin{equation}
    2\times(2\pi)^2\times\left(\frac{A_\Phi}{\Phi_0}\right)^2\times T_\varphi^2\left[\left(\frac{3}{2}-\gamma\right)-\log(\omega_\mathrm{ir}T_\varphi)\right]\times\frac{E_{\mathrm{C}_\mathrm{c}}}{E_{\mathrm{J}_\mathrm{c}}(\phi_\mathrm{ext})}\times[\partial_{\phi_\mathrm{ext}}E_{\mathrm{J}_\mathrm{c}}(\phi_\mathrm{ext})/\hbar]^2 = 1,
    \label{eq:implicit equation coherence times}
\end{equation}
\end{widetext}
where we have introduced the infrared cutoff~$\omega_\mathrm{ir}/2\pi\simeq 1$~Hz and the Euler constant $\gamma\approx 0.58$.

\Cref{fig:Dissipation information}\textsf{b} and~\textsf{c} show the coherence times obtained by numerically solving~\cref{eq:implicit equation coherence times} for a split-transmon qubit of frequency~$5\,\mathrm{GHz}$ and~$-300\,\mathrm{MHz}$ anharmonicity.
The coherence times are shown as a function of the SQUID-junction assymmetry parameter~$r$, for different values of~$A_\Phi$ that are typical in experiments. 
Panel~\textsf{b} shows the result for the external flux bias~$\Phi_\mathrm{ext}=0.05\,\Phi_0$, while panel~\textsf{c} considers the case of~$\Phi_\mathrm{ext}=0.25\,\Phi_0$. 
We observe that coherence times as large as~$100\,\mu$s are in principle possible for split-transmon couplers with large junction assymmetry that opperate closer to their upper sweet spot. 
We consider the SQUID-junction asymmetry ratio~$r=7$ and~$\Phi_\mathrm{ext}=0.25$ for all simulations in the main text.

\section{Two-step frequency allocation}
\label{sec:Two-step frequency allocation}

In this section, we describe additional details of the frequency allocation method outlined in~\cref{subsec:Extensible frequency layouts}.

\subsection{Multi-qubit Hamiltonian}
\label{subsubsec:Multi-qubit setup}

As in~\cref{subsec:Stationary two-qubit interaction rates}, we model all circuit modes as Kerr nonlinear oscillators with frequencies $\{\omega_{{\mu}}\}$ and anharmonicities $\{\alpha_{{\mu}}\}$. 
For simplicity, we assume uniform first-neighbor (qubit-coupler) coupling~$J_1$ and next-neighbor (qubit-qubit and coupler-coupler) coupling~$J_2$. 
We also consider a single coupler drive at a time.
The drive amplitude is~$\Omega$, and we work in a frame rotating at the drive frequency~$\omega$.
In what follows, we assume that the couplings~$J_1$ and~$J_2$ are given, and optimize over the mode parameters and the drive frequencies.
To make sure that the perturbative result is accurate, we work with moderate coupling parameters~$J_1/2\pi=25\,\mathrm{MHz}$, $J_2/2\pi=2\,\mathrm{MHz}$, and drive strength~$\Omega/2\pi=50\,\mathrm{MHz}$.
We consider a total frequency bandwidth for the qubit and coupler modes of~$1.5\,\mathrm{GHz}$, and anharmonicities in the range of~$[-350,-250]\,\mathrm{MHz}$.

We write the unit-cell Hamiltonian~$\hat{H} = \hat{H}^0+\eta\hat{V}$ as in~\cref{subsubsec:Perturbation theory}, where~$\hat{H}^0$ denotes the noninteracting part.
Once the interaction~$\eta\hat{V}$ is turned on, the self-energies associated with the computational eigenstates are approximated to second (leading) order in the couplings (drive amplitude). 
In the limit where the self-energy~$\Sigma_\alpha$ of a computational state~$|\Phi_\alpha\rangle$ is much smaller than the detunings~${\epsilon}_\alpha^0-{\epsilon}_\beta^0$, where~$\beta$ labels a noncomputational state, the self-consistent nature of~\cref{eq:self-energy appendix} can be simplified by omitting the self-energy from all denominators in that expression. 
In this approximation, a two-level truncation near the resonance~${\epsilon}_\alpha^0 - {\epsilon}_\beta^0 \to 0$ leads to the dispersive form of the self-energy~$\Sigma_\alpha \approx |G_{\alpha\beta}|^2/({\epsilon}_\alpha^0-{\epsilon}_\beta^0)$, where
\begin{equation}
    \begin{split}
    G_{\alpha\beta}  = \sum_{k=1}^{\infty}\sum_{\alpha_1,\dots,\alpha_k}&\langle\Phi_\beta^0|\eta\hat{V}|\Phi_{\alpha_1}^0\rangle\frac{\langle \Phi_{\alpha_1}^0|\eta\hat{V}|\Phi_{\alpha_2}^0\rangle}{{\epsilon}_\alpha^0-{\epsilon}_{\alpha_1}^0}\dots\\
    &\dots\frac{\langle \Phi_{\alpha_k}^0|\eta\hat{V}|\Phi_{\alpha}^0\rangle}{{\epsilon}_\alpha^0-{\epsilon}_{\alpha_k}^0},
    \end{split}
    \label{eq:virtual-interaction-strength}
\end{equation}
is the virtual interaction rate between the eigenstates~$|\Phi_\alpha^0\rangle$ and~$|\Phi_{\beta}^0\rangle$ of the bare Hamiltonian~$\hat{H}^0$. 
Here, we assume that~$|G_{\alpha\beta}|$ is small compared to $|{\epsilon}_\alpha^0-{\epsilon}_\beta^0|$. 
We use this simplified expression to compute the self-energies of computational states in the unit cell. 

\subsection{Cost function and optimization}
\label{subsubsec:Two-step frequency allocation}

Our two-step frequency optimization seeks to: \textit{i}) minimize the static ZZ interaction between all pairs of qubits, 
\textit{ii}) minimize drive-activated ZZ interactions on spectator qubits,
\textit{iii}) maximize desired two-qubit gate rates.
The cost function also incorporates the dispersive-coupling ratios~$\nu_{\alpha\beta}=|G_{\alpha\beta}/({\epsilon}_\alpha^0-{\epsilon}_\beta^0)|$, where~$\alpha$ and~$\beta$ are computational and noncomputational states, respectively.
Because the excitation number is constant in the absence of a drive and under a rotating-wave approximation, we estimate the static ZZ interaction taking into account states with up to two excitations. 
However, we consider computational (noncomputational) states with up to two (three) excitations in the presence of a drive.

\textit{Static ZZ interaction}\textit{--}
We begin by generating several frequency layouts as independent initial conditions for the optimizer. 
At this point, the qubits and couplers anharmonicities are~$-300\,\mathrm{MHz}$ and~$-350\,\mathrm{MHz}$, respectively.
Next, we optimize these frequency layouts independently using a least-squares algorithm.
The optimization is done over the mode frequencies and anharmonicities.
The cost function incorporates penalties to ensure that both the dispersive-coupling ratios and the the ZZ interaction between any pair of qubits are below the chosen bounds of 0.05 and~$20\,\mathrm{kHz}$, respectively.

\textit{Two-qubit gate rate}\textit{--}
Each pair of qubits offers many potential operating points for the two-qubit gate, depending on the drive frequency. 
However, not all resonances are equivalent when it comes to minimizing the impact of the coupler drive on neighboring qubits.
Thus, we take into account all possible single- and two-photon resonances for a given pair of qubits, and computing the driven ZZ interaction for any other qubit pair in the device.

More precisely, our cost function incorporates: 
\textit{i}) the driven ZZ interaction between the qubits that participate of the two-qubit gate, 
\textit{ii}) the ratio between the driven ZZ interaction for other pair of qubits and that calculated in~\textit{i}), and 
\textit{iii}) the ratios~$\nu_{\alpha\beta}$ for qubits that do not participate of the gate. 
The cost is evaluated for all coupler-drive frequencies corresponding to a transition between a computational and a noncomputational state of a specific two-qubit pair, and then added up for all pair of neighboring qubits in the device.
While information in \textit{i-iii}) is somewhat redundant, we find that partitioning the cost function in this way gives us enough flexibility for the optimization.
Only the coupler-drive frequencies are optimized during this second step.

\section{Expanded-space Hamiltonian for multiple drives}
\label{sec:Expanded-space Hamiltonian for multiple drives}

With the definitions introduced in~\cref{subsec:Adiabatic microwave multi-qubit control}, the system Hamiltonian [see~\cref{eq:complete Hamiltonian}] takes the form
\begin{equation}
    \hat{H}(\hat{\boldsymbol{\phi}},\hat{\boldsymbol{n}},t) = \hat{H}_\mathrm{s}(\hat{\boldsymbol{\phi}},\hat{\boldsymbol{n}}) + \hat{H}_\mathrm{drive}[\hat{\boldsymbol{\phi}},\hat{\boldsymbol{n}},\boldsymbol{\Omega}(t),\boldsymbol{\theta}(t)].
    \label{eq:complete Hamiltonian bold}
\end{equation}
Next, we promote the drive-phase variables to a vector operator~$\boldsymbol{\theta}(t)\to\hat{\boldsymbol{\vartheta}}$, with conjugate momenta~$\hat{\boldsymbol{m}}\to-i{\partial}_{\boldsymbol{\vartheta}}$ in the phase representation. 
Under this transformation, the effective Hamiltonian that extends~\cref{eq:Extended Hilbert space Hamiltonian} to $d$ microwave drives takes the form 
\begin{equation}
    \hat{H}_\mathrm{eff}(\hat{\boldsymbol{\vartheta}},\hat{\boldsymbol{m}},\hat{\boldsymbol{\phi}},\hat{\boldsymbol{n}},t) = \hat{H}(\hat{\boldsymbol{\phi}},\hat{\boldsymbol{n}},t) + \hbar
    \boldsymbol{\omega}_{\mathrm{eff}}(t)\cdot \hat{\boldsymbol{m}},
    \label{eq:Extended Hilbert space Hamiltonian bold}
\end{equation} 
where~$\boldsymbol{\omega}_{\mathrm{eff}}(t)=\dot{\boldsymbol{\theta}}(t)$ groups the drive frequencies. 

\bibliography{library}

\end{document}